# Bound states of solitons in fiber lasers

Yudong Cui[1,2,†], Tianchang Lu[1,†], Yusheng Zhang[3,*], Dong Mao[4], and Boris A. Malomed[5,*]

[1]*State Key Laboratory of Extreme Photonics and Instrumentation, College of Optical Science and Engineering, Zhejiang University, Hangzhou 310027, China*

[2]*ZJU-Hangzhou Global Scientific and Technological Innovation Center, No.733 Jianshe San Road, Xiaoshan District, Hangzhou 311200, Zhejiang, China*

[3]*Hangzhou Institute of Advanced Studies, Zhejiang Normal University, Hangzhou 311231, China*

[4]*Key Laboratory of Light Field Manipulation and Information Acquisition, Ministry of Industry and Information Technology, Shaanxi Key Laboratory of Optical Information Technology and School of Physical Science and Technology, Northwestern Polytechnical University, Xi'an 710129, China*

[5]*Instituto de Alta Investigación, Universidad de Tarapacá, Casilla 7D, Arica, Chile*

[†]*These authors contributed equally to this work.*

*Corresponding author: yszhang@zjnu.edu.cn and malomed@tauex.tau.ac.il

**Abstract:** Bound states of solitons, alias "soliton molecules", in fiber-laser cavities is one of central topics in the ongoing experimental and theoretical work in nonlinear optics. This topic has drawn much interest as a unique platform for the studies of dynamics of dissipative solitons, and also due to its vast potential for various applications in optics and photonics. This article presents a systematic review of theoretical and experimental findings for bound states of two and several dissipative solitons in fiber lasers. The theoretical basis underlying the formation and stabilization of soliton molecules in the fibers, which is provided by the complex Ginzburg-Landau equations and bound states of such equations, is presented in necessary detail, which is followed by a detailed presentation of experimental findings, including very recent ones. In particular, included are the results for the multi-soliton bound states in the fibers, as well as for the bound states in the temporal and frequency domains, single-component (scalar) and two-component (vector), two- and multi-soliton modes, as well as for bound states of spatiotemporal dissipative solitons in the lasers based on multimode fibers.

**Keywords:** Interactions of solitons; soliton molecules; laser cavities; mode-locking; Ginzburg-Landau equations; nonlinear dynamics

## Table of Contents

## 1 Introduction

Solitons are commonly known self-trapped states of wave fields which play a profoundly important role in diverse areas of physics, including optics, photonics, and magnetism, hydrodynamics and plasmas, quantum matter (Bose-Einstein condensates), etc. In their original classical form, the solitons were found as solutions of exactly integrable conservative equations. In particular, this is the celebrated nonlinear Schrödinger equation (NLSE), with has a great number of physical realizations[1]. The fundamental solitons predicted by NLSE have been observed in nonlinear optical fibers[2] and other optical and photonic media[1], plasmas, water waves, Bose-Einstein condensates, etc. Once the existence of stable fundamental solitons is firmly established, a natural question is if it may be possible to predict and observe stable physical states that represent bound states of two or several solitons – in particular, pairs of interacting solitons which maintain a constant separation between them, or stationary multisoliton clusters. However, the integrable equations, such as NLSE, do not produce solutions of such types. These equations provide exact solutions for collisions between moving fundamental solitons, and also oscillatory bound states (breathers) of solitons with coinciding centers [3–5] (in particular, see Fig. 1(a) below), but stationary solutions for bound solitons do not exist.

On the other hand, the integrable NLSE is not a fully realistic model for fiber optics and other physical realizations, as it does not include terms which account for various important features of the underlying physical system, such as loss and gain, higher-order dispersion, nonlinearity saturation, etc. (see details below). Adding terms representing these features makes the equation an essentially more accurate model of the physical system but simultaneously breaks its integrability. This means that soliton solutions are not available in an exact analytical form, although in many relevant cases they may be obtained in an approximate analytical form by means of the perturbation theory, provided that the integrability-breaking terms may be considered as small perturbations [6] (conservative non-integrable equations, such as NLSE with the combination of cubic and quintic nonlinear terms[7], may admit exact solutions for fundamental solitons). In the general case, the perturbations may be naturally classified as conservative and dissipative ones. In practical terms, dissipative terms (such the loss and gain in optical fibers) added to NLSE or other unperturbed integrable equations produce strongest effects. In particular, while the classical NLSE solitons are maintained by the balance between the linear dispersion (or diffraction, depending on the physical realization) and the self-focusing cubic nonlinearity (which represents the Kerr effect in fiber optics), in the presence of dissipative terms one must simultaneously take care of the balance between the loss and compensating gain. The solitons which simultaneously satisfy both balance conditions are often called dissipative solitons[8,9]. Their radical difference from solitons in conservative models (both integrable and non-integrable ones) is the fact that, in the absence of the dissipative effects, fundamental solitons exist in continuous families, which are parametrized by the value of the soliton's norm (also called energy in fiber optics, being, however, different from the underlying Hamiltonian, see below). On the other hand, because dissipative solitons must maintain the balance between loss and gain, this additional condition selects one or several (most frequently, two) discrete values of the norm. Therefore, the dissipative solitons exist not in families, but as isolated states (if they are stable, they actually play the role oof attractors, in the framework of the dissipatively perturbed equations).

While the dissipative solitons lose many sophisticated features of the solitons in conservative models, they acquire some new properties, that may be quite relevant in terms of underlying physics and for potential applications. In particular, such soliton metamorphoses were studied in detail in the framework of the complex Ginzburg-Landau equations (CGLEs), which are introduced, roughly speaking, as NLSEs (usually ones including the quintic nonlinear term) with real coefficients replaced by complex ones[8,10]. In this context, it was found that the CGLE, which includes the linear and nonlinear (quintic) loss terms and a cubic gain term, gives rise not only to isolated dissipative-soliton solutions, but also to their stationary bound states, that may include two or several individual solitons (as mentioned above, such stationary bound stated cannot be produced by the classical integrable equations). This possibility was first predicted theoretically in 1991[11], in terms of an effective potential of the interaction between separated dissipative solitons with equal amplitudes. The potential was derived as a function of the distance between the solitons and phase shift between them, see Eq. (23) below. The bound states were thus predicted as a series of local minima of the effective potential (see details below). For the first time, the existence of the robust bound states of solitons in optical fibers (actually, in fiber lasers, which may be modeled by the cubic-quintic CGLE) was demonstrated experimentally in 2001[12]. Such bound states of solitons in lasers were later termed "soliton molecules" (SMs) in Ref. [13] (the same term was earlier applied to bound states of spatiotemporal (multidimensional) optical solitons [14]). While the fundamental interaction mechanisms for optical solitons are often rooted in the (1+1) dimensional framework, this review extends significantly beyond this limit. We provide a comprehensive analysis of soliton bound states across multiple degrees of freedom, including polarization (vector solitons), frequency (multi-color compounds), and transverse spatial modes. Furthermore, we discuss how the interplay of these dimensions forms higher-dimensional structures, such as the 3D spatiotemporal soliton molecules.

The topic of SMs in fiber lasers has gradually drawn a great deal of interest for two reasons: (i) they offer a unique platform for theoretical and experimental studies of soliton dynamics in nonlinear dissipative systems, which are scarcely available for experimental realization in other physical settings; (ii) robust bound states of solitons in the fiber-laser cavities may find various practical applications (see details below). The present article provides a systematic up-to-date review of basic theoretical and experimental results on this topic, including very recent ones. The presentation starts with the summary of the theoretical basis, i.e., approximate analytical results which predict single-component (scalar) and two-component (vector) SMs as stationary two-soliton solutions of the respective CGLEs and systems of coupled CGLE, respectively. Various mechanisms underlying the formation and stabilization of SMs are considered in detail. The larger subsequent part of the review presents a detailed summary of experimental results, including very recent ones, which have been obtained for bound states of two and several dissipative solitons in various setups implemented in fiber lasers. The experimental findings are reported for the bound states in the temporal and frequency domains, as well as for both scalar and vector SM states. The soliton bound states in the time domain have been reported in mode-locked fiber lasers under the dispersion conditions ranging from anomalous to normal, as well as using various mode-locking techniques, demonstrating the widespread existence of soliton bound states. In recent years, the transient vibration and evolution dynamics of bound states stay in the spotlight, with the help of the dispersion Fourier- transform technique. Furthermore, efforts have been made to produce on-demand soliton bound states, to address various applications. The solitons with different wavelengths and polarizations can be bound by the cross-phase modulation (XPM), offering the potential to manipulate the electromagnetic field structure of the laser pulses. Also included are recently reported results for bound states of spatiotemporal solitons in lasers based on multimode fibers (unlike the single-mode ones which are used in traditional fiber-laser setups). In that case, the bound states feature their specific structure not only in the longitudinal direction, but also in the fiber's transverse cross-section.

## 2 Bound states of solitons: the theoretical basis

### 2.1 Bound-state solutions of the nonlinear Schrödinger equation

In the basic approximation, the propagation of solitons in single-mode optical fibers is governed by the basic NLSE [2]:

$$i\frac{\partial u}{\partial \xi}+\frac{1}{2}\frac{\partial^2 u}{\partial \tau^2}+|u|^2 u=0\ , \tag{1}$$

where $u(\xi,\ \tau)$ is the normalized envelope of the light pulse, with $\xi$ and $\tau$ being the normalized propagation distance and reduced time, respectively. As its name implies, this equation is similar to the quantum-mechanical Schrödinger equation, including the nonlinear potential term, $U = -|u|^2$, which represents the Kerr self-focusing in silica. In its basic form, Eq. (1) neglects gain, loss, and higher-order dispersion and nonlinearity. Its commonly known stable single (fundamental)-soliton solution is:

$$u_s(\xi,\tau)=\eta\,\mathrm{sech}(\eta\tau)\cdot e^{i\eta^2\xi/2}\ . \tag{2}$$

where real parameter $\eta$ represents the amplitude and inverse width of the soliton. The dynamics of the fundamental solitons may be significantly affected by interactions between them. The evolution of two-soliton states is governed by the exact two-soliton solution of Eq. (1), which are available due to the exact integrability of NLSE by means of the inverse-scattering transform[1]. In the context of fiber optics, Gordon had derived the exact solution for two counterpropagating solitons and analyzed their interaction in terms of the corresponding effective equations of motion[2]. Chu and Desem later extended this approach, proposing a more comprehensive two-soliton model[5]. Accordingly, the exact two-soliton solution can be expressed in the following form:

$$u_d(\xi,\tau)=W\left\{\eta_1\,\mathrm{sech}[\eta_1(\tau+\gamma_0/2)]\cdot e^{i\eta_1^2\xi/2}+\eta_2\,\mathrm{sech}[\eta_2(\tau-\gamma_0/2)]\cdot e^{i\eta_2^2\xi/2}\right\}. \tag{3}$$

It is composed of two factors: the one combined of the two sech terms, which represents the propagation of the individual solitons, and factor $W$ which quantifies the interaction between them:

$$W=\frac{\eta_2^2-\eta_1^2}{(\eta_1^2+\eta_2^2)-2\left[\sqrt{\eta_1^2-P_1^2}\sqrt{\eta_2^2-P_2^2}+P_1P_2\cos\psi\right]}\ , \tag{4}$$
$$P_{1,2}=\eta_{1,2}\,\mathrm{sech}[\eta_{1,2}(\tau\pm\gamma_0/2)],\qquad \psi=(\eta_2^2-\eta_1^2)\xi/2,$$

where $\gamma_0$ is the initial separation between the solitons. In the configurations with both solitons having equal amplitudes and sharing an initial phase, they periodically merge and separate, forming a bound state whose oscillation frequency is $4\pi/(\eta_2^2-\eta_1^2)$ , determined by their initial separation (see Fig. 1(a)) [15]. On the other hand, when the solitons are launched with an initial phase difference, their coalescence is inhibited, leading to the repulsion between them (see Fig. 1(b)).

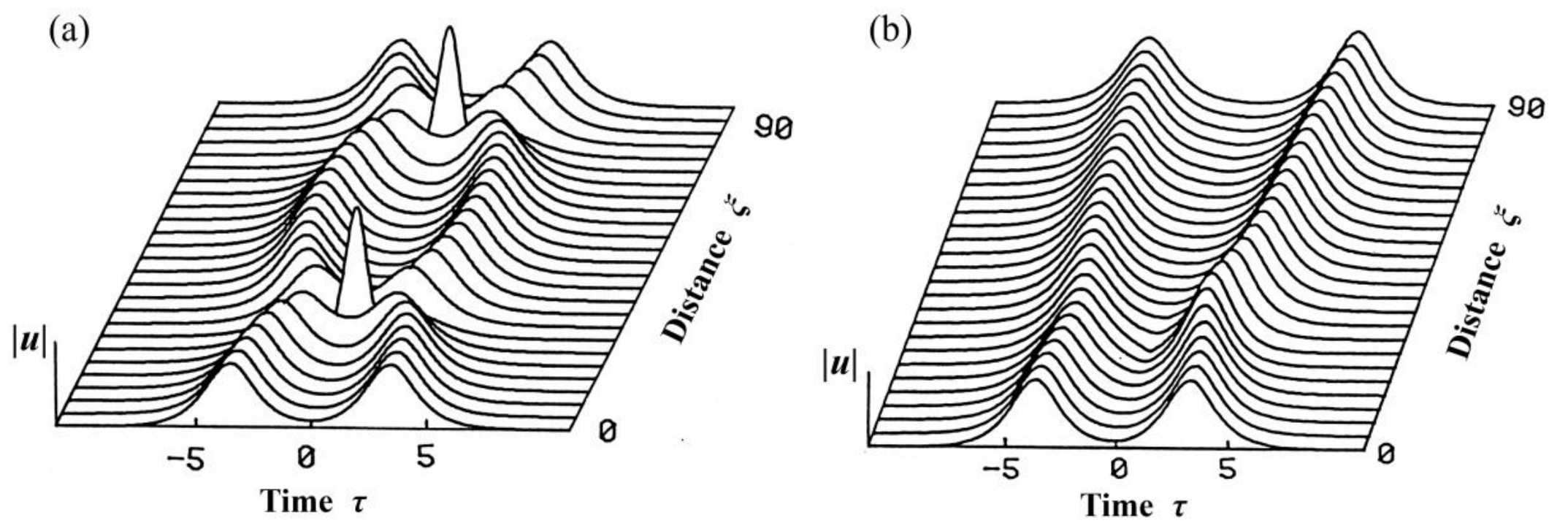

**Fig. 1** The interaction of two solitons with equal amplitudes, $\eta$=1, as per Ref. [15]. **(a)** The bound state formed by the solitons with identical initial phases. **(b)** The repulsion between the solitons with the initial phase difference π/4.

It is relevant to mention another class of exact solutions of NLSE for bound states of solitons, in the form of breathers, alias *N*-solitons, which are produced by the fundamental soliton as described in Eq. (2) initially multiplied by an integer factor *N*[3]. They are construed as complexes of overlapping fundamental solitons with a common central point and amplitudes $\eta(2n\text{-}1)$, $n = 1, 2, \ldots, N$.

An alternative approach to studying the interactions between separated fundamental solitons treats the interaction as a perturbation [11,16–19]. Initially implemented by Karpman and Solov'ev, this approach postulates that the weak interactions arise from the overlap of the solitons' "tails" [16]. A similar tail-mediated interaction has also been studied in two-dimensional spatial systems, where the overlap of oscillatory transverse tails can stabilize bound clusters and other two-dimensional localized structures[20,21]. Accordingly, NLSE was written separately for each soliton as

$$i\frac{\partial u}{\partial \xi}+\frac{1}{2}\frac{\partial^2 u}{\partial \tau^2}+|u|^2 u=i\kappa(u)\ , \tag{5}$$

with term $\kappa(u)$ accounting for the perturbation effects induced by neighboring solitons. It is assumed that the interacting solitons propagate in the optical fiber keeping their unperturbed form similar to that given by Eq. (2),

$$u(\xi,\tau)=\eta(\xi)\operatorname{sech}\{\eta(\xi)[\tau-q(\xi)]\}\cdot\exp[i\delta(\xi)\tau+i\phi(\xi)] \tag{6}$$

while parameters $\eta$, $q$, $\delta$, and $\phi$ of this *ansatz*, which define the amplitude, position, frequency and phase of the soliton, respectively, evolve as functions of the propagation distance $\xi$.

In the framework of this approach, the two-soliton solution is expressed as the superposition of two solitons, $u(\xi,\tau) = u_1(\xi,\tau) + u_2(\xi,\tau)$. Each one is perturbed by the tail of the other, with the respective perturbation term written as $\kappa_{mn}(u_n) = i(u_m^* u_n^2 + 2u_m u_n u_n^*)$, where $m,n$ = 1,2 and $m \neq n$, describing the influence of the $m$-th soliton on its $n$-th counterpart. the asterisk in the expression representing the complex conjugate. By incorporating this perturbation term, the evolution of parameters $\eta$, $q$, $\delta$, and $\phi$ can be derived by means of the adiabatic perturbation theory[22], inverse scattering perturbation theory[23], or the Lagrangian perturbation theory[24] (alias the variational approximation[25]). The corresponding evolution equations are obtained in the following form[26]:

$$\frac{d\eta}{d\xi}=\operatorname{Re}\int_{-\infty}^{\infty}\varepsilon(u)u^*(\tau)\ \mathrm{d}\tau\ , \tag{7}$$

$$\frac{d\delta}{d\xi} = -\operatorname{Im}\int_{-\infty}^{\infty} \varepsilon(u)\tanh[\eta(\tau-q)]u^*(\tau)\ \mathrm{d}\tau\ , \tag{8}$$

$$\frac{dq}{d\xi} = -\delta + \frac{1}{\eta^2}\operatorname{Re}\int_{-\infty}^{\infty} \varepsilon(u)(\tau-q)u^*(\tau)\ \mathrm{d}\tau\ . \tag{9}$$

$$\frac{d\phi}{d\xi} = \operatorname{Im}\int_{-\infty}^{\infty} \varepsilon(u)\left\{1/\eta-(\tau-q)\tanh[\eta(\tau-q)]\right\}u^*(\tau)\ \mathrm{d}\tau + \frac{1}{2}(\eta^2-\delta^2) + q\frac{d\delta}{d\xi}\ . \tag{10}$$

Then, the evolution equations can be derived from Eqs. (7) – (10) for variables $\eta_d = (\eta_1+\eta_2)/2$ , $\delta_d = (\delta_1+\delta_2)/2$ , $\Delta\eta = (\eta_2-\eta_1)/2$ , $\Delta\delta = (\delta_2-\delta_1)/2$ , $\Delta q = q_2 - q_1 > 0$ , and $\Delta\phi = \phi_2 - \phi_1 - \delta\Delta q$ , which are introduced to describe the interaction per se:

$$\frac{d\eta_d}{d\xi} = 0, \qquad \frac{d\delta_d}{d\xi} = 0\ . \tag{11}$$

$$\frac{d\Delta\eta}{d\xi} = 4\eta_d^3 e^{-\eta_d\Delta q}\sin(\Delta\phi)\ . \tag{12}$$

$$\frac{d\Delta\delta}{d\xi} = -4\eta_d^3 e^{-\eta_d\Delta q}\cos(\Delta\phi)\ , \tag{13}$$

$$\frac{d\Delta q}{d\xi} = 2\Delta\delta, \qquad \frac{d\Delta\phi}{d\xi} = 2\eta_d\Delta\eta\ . \tag{14}$$

When the perturbation coefficients are sufficiently small, the following integral of motion can be obtained from Eq. (11)-(14):

$$Y^2 - 16\eta_d^2\exp(-\eta_d\Delta q + \Delta\phi) \equiv \Lambda^2\ . \tag{15}$$

Here, $Y \equiv 2(\Delta\delta + \mathrm{i}\Delta\eta)$ and $\Lambda^2$ remains constant. Equation (15) can be solved for $Y$:

$$Y = \Lambda\tanh(2\eta_d\Lambda\xi + \theta) \tag{16}$$

where $\theta$ is a complex integration constant. Using Eq. (15) and (16), an explicit expression for the soliton separation $\Delta\gamma$ can be obtained. Karpman and Solov'ev examined the evolution of $\Delta q$ [16]. In the case of two solitons with equal amplitudes and identical phases, the following expression can be derived[27]:

$$\Delta\gamma = \Delta\gamma_0 + 2\ln\left|\cos(\operatorname{Im}[\Lambda]\xi/2)\right| \tag{17}$$

From Eq. (17), it is evident that the soliton separation oscillates with the period $\arccos(e^{-\Delta q_0/2})/(\eta_d\,\Delta\eta)$ . Notably, for large $\Delta q_0$, the resulting period aligns with that obtained from the inverse-scattering-transform method[5].

However, NLSE lacks stationary solutions in which two or more solitons with identical amplitudes maintain a fixed separation. Many works, chiefly based on numerical simulations, have addressed the problem of the soliton-soliton interactions in more realistic models of fiber optics (in particular, fiber lasers), including additional physically relevant terms, such as higher-order dispersion[5,28], loss[29,30], and gain[29,31]. In these studies, initial soliton parameters (the amplitude[32,33] and phase difference[34–36]) were varied to assess the feasibility of the formation of stable bound states of solitons in a stationary form (alias "soliton molecules" (SMs) [14]). From the standpoint of the fiber transmission link, the third-order dispersion can weaken the effective attractive force between the solitons by disrupting their phase coherence[28]. In driven Kerr cavities and all-fiber resonators, higher-order dispersion can markedly reshape soliton tails via Cherenkov radiation and thereby modify the binding forces between temporal localized states. For example, Vladimirov et al. showed that third-order dispersion can break the interaction symmetry and extend the interaction range, facilitating the formation of temporal bound states[37], whereas positive fourth-order dispersion can further enhance tail-mediated coupling and stabilize multiple equidistant bound states; moreover, spectral filtering may damp the radiative tails and affect bound-state stability[38]. This fact, in turn, attenuates the oscillatory dynamics and prevents their merging. Additionally, the bandwidth-limited amplification can also reduce soliton interaction[31]. In contrast, incorporating fiber loss into Eq. (1) tends to exacerbate the instability of soliton bound states[29,30]. As concerns the conditions, an increase in the pulse amplitude leads to an enhanced oscillatory tail, thereby enhancing the instability[32]. However, assigning unequal amplitudes to the two solitons can effectively maintain a constant separation between them[5,33]. Notably, a greater amplitude difference results in a more stable bound state of solitons. Additionally, introducing an initial phase difference between the solitons may attenuate the interaction (as it allows for the avoidance of pulse coalescence in this case, see Fig. 1(b)), although the same factor may cause the solitons to propagate at different velocities[35,36]. Furthermore, if the initial pulse is formed as a Gaussian rather than a soliton-like sech (see Eq. (2)), the oscillatory behavior may also be significantly attenuated, resulting in a potential increase of the transmission bandwidth by at least 50%[32].

The previous investigations were primarily focused on the region of anomalous group-velocity dispersion (GVD) in optical fibers, while solitons and SMs can also be observed in dispersion-managed (DM) systems[39–49] and all-normal-GVD regimes[50–53]. A DM system is built of alternating segments of fibers exhibiting anomalous and normal GVD as shown in Fig. 2, with fiber Bragg gratings serving as a quintessential example[43,45]. In such systems, the pulses featuring stable transmission are termed DM solitons. They can achieve a larger peak power than those found in the anomalous-GVD regime[39,41] and are less susceptible to various detrimental effects, including the Gordon-Haus timing jitter, induced by the amplifier noise[54], and four-wave mixing[55]. Additionally, bound states of DM solitons can also form in these systems, taking into regard effects of the third-order dispersion, filtering, and amplification[10,45]. Similar to the properties of the soliton interactions outlined above, appropriately controlling the initial phase difference and unequal initial amplitudes of the pulses can effectively reduce the instability of bound states of the DM solitons, leading to the formation of a bound state with the fixed separation[44,46–48].

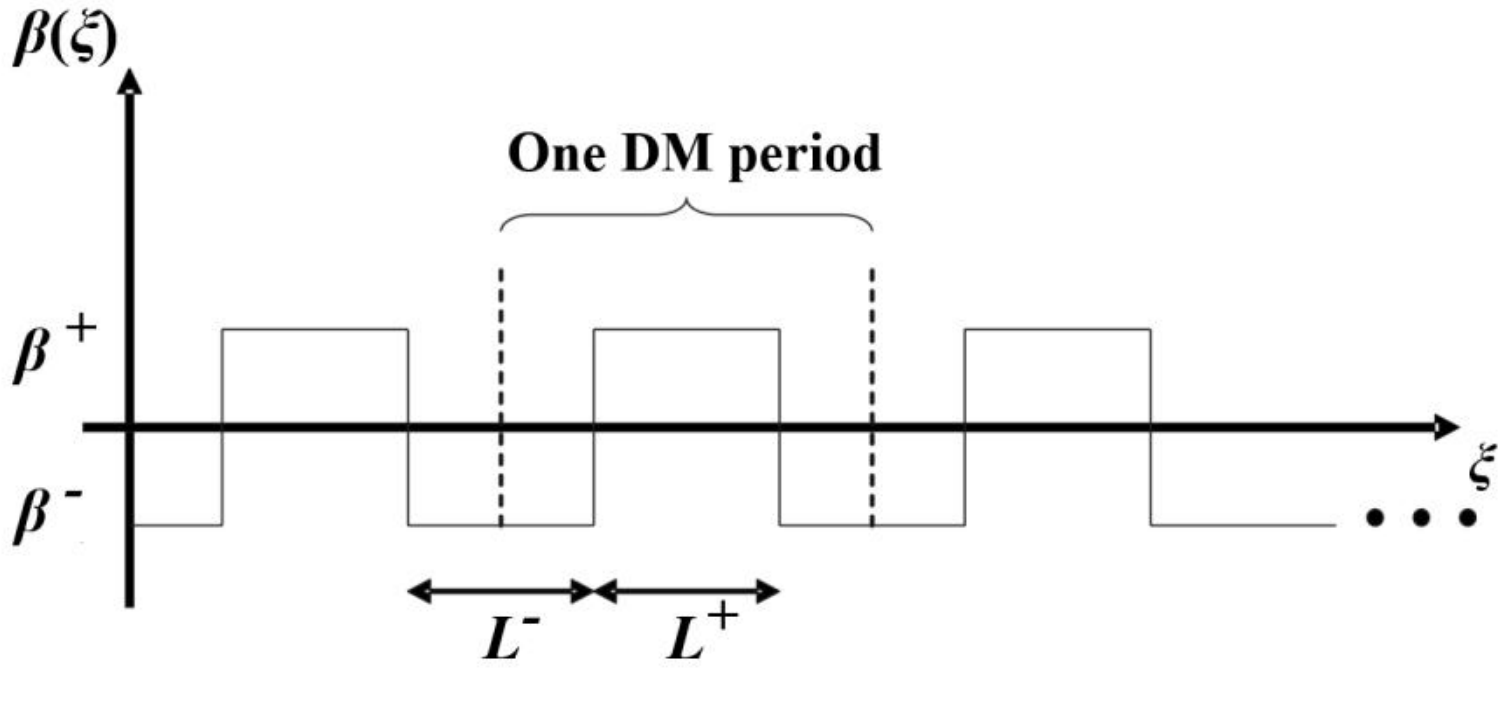

**Fig. 2** The scheme of the periodic dispersion management (DM), as per Ref. [48,49].

In the normal-GVD regime of the light propagation in optical fibers, another type of solitons emerges, known as dark solitons, which exhibits an intensity profile with a dip superimposed on a uniform background [52]. This solution is obtained as the solution of Eq. (1) with the opposite sign in front of the second term

(GVD):

$$u(\xi,\tau)=\eta\left[B\tanh(\zeta)-i\sqrt{1-B^2}\right]\exp(i\eta^2\xi)\ , \tag{18}$$

where $\zeta=\eta B(\tau-\tau_s-\eta B\sqrt{1-B^2})$ , parameters $\eta$ and $\tau_s$ denote the background amplitude and the location of the dip, respectively, and $B$ determines the depth of the dip. Specifically, $|B|=1$ indicates that the intensity at the center of the dip is reduced to zero, qualifying it as the black soliton, while $|B|<1$ characterizes it as a gray soliton, as shown in Fig. 3. Dark solitons are particularly effective in mitigating dispersion effects in the course of the propagation and demonstrate the enhanced resistance to perturbations compared to bright solitons [51,56,57]. Furthermore, they can interact with multiple modes, facilitating multimodal signal transmission in a singular system[52]. The dynamics of the interaction between dark solitons have also been extensively investigated. Unlike bright solitons, dark ones exhibit a direct correlation between the amplitude and velocity, which simplifies the solution processing, reducing the number of free parameters[58]. The studies have demonstrated that the interaction force between adjacent dark solitons is inherently repulsive, leading to a monotonic increase in their separation rather than periodic oscillations[50]. Notably, this repulsive force exponentially decays with the increase of the initial separation, the decay rate being twice as large as that for bright solitons.

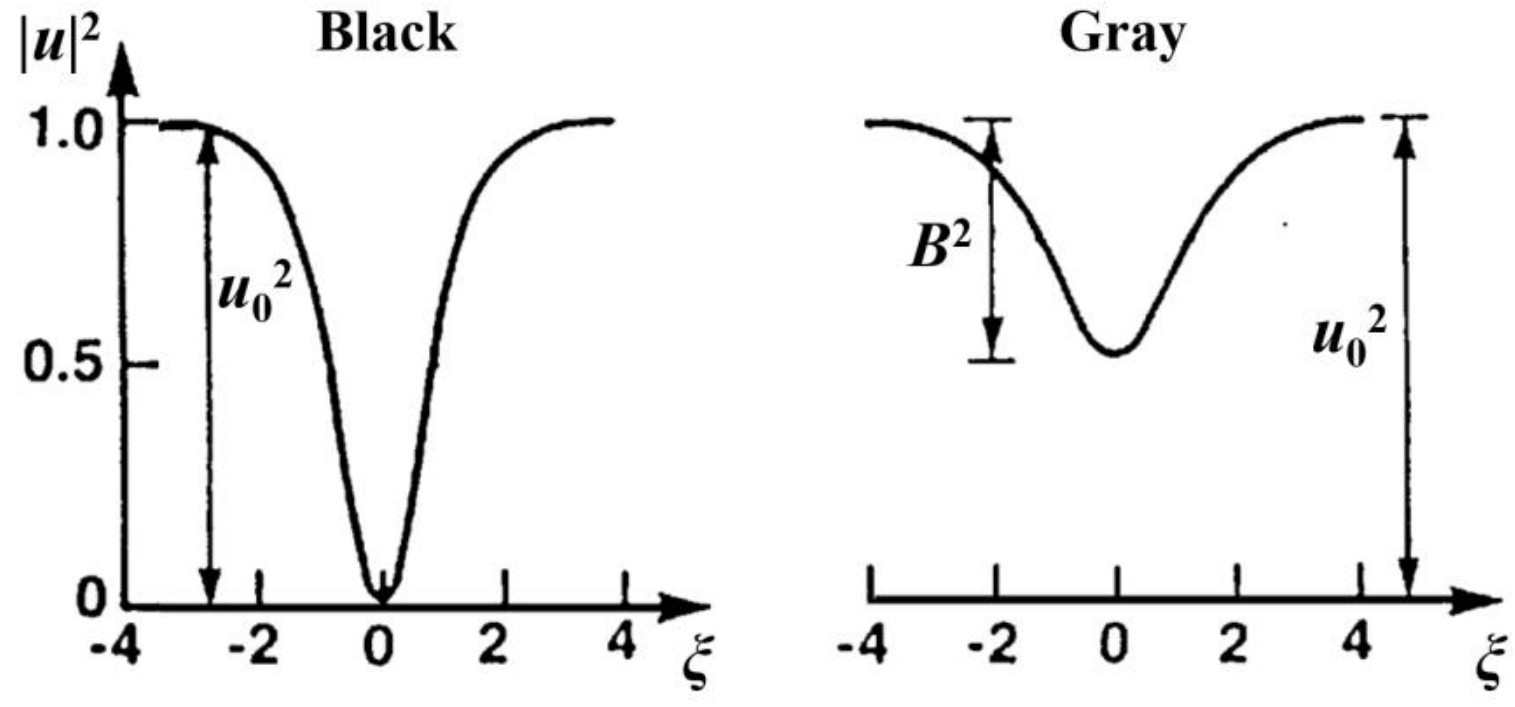

**Fig. 3** The intensity profiles for black and gray solitons, as per Ref. [52].

### 2.2 Bound-state solutions of the complex Ginzburg-Landau equation (CGLE)

In fiber laser and ultra-long transmission fiber links, the soliton dynamics is not governed by conservative NLSEs as described in Eqs. (3) and (6) [59,60]. The realistic model must include gain, loss, dispersion, and nonlinear effects, which transforms the model into a dissipative one, dramatically altering the solitons and their interactions[61,62]. Accordingly, the governing equation replacing Eq. (1) should incorporate the gain and loss terms[63,64]

$$i\frac{\partial u}{\partial\xi}+\frac{1}{2}\frac{\partial^2 u}{\partial\tau^2}+|u|^2u=-i\delta u+i\beta\frac{\partial^2 u}{\partial\tau^2}+i\varepsilon|u|^2u-i\mu|u|^4u\ . \tag{19a}$$

As mentioned above, this equation is often referred to as CGLE. Here, $\delta$ denotes the linear gain or loss coefficient, $\beta$ accounts for spectral filtering, $\varepsilon$ represents the nonlinear gain, and $\mu$ accounts for the quintic nonlinear loss (actually, it represents the saturation of the cubic gain). It is relevant to mention that CGLEs form a vast class of models with diverse realizations in physics (and even in chemistry, in the form of chemical waves)[10]. As concerns the interpolation of the optical losses and nonlinearity, an important model of externally driven laser cavities is provided by the Lugiato-Lefever equation[65],

$$i\frac{\partial u}{\partial \xi}+\frac{1}{2}\frac{\partial^2 u}{\partial x^2}+|u|^2 u+\left(i\delta-\Delta\right)u=iE, \tag{19b}$$

where $x$ is the transverse coordinate in the cavity, real $\Delta$ is the mismatch parameter, and $E$ is the external pump. Notably, driven–damped nonlinear Schrödinger–type equations closely related to (and sharing the same mathematical structure as) the Lugiato–Lefever equation were derived and studied earlier in other physical contexts, including radiofrequency-driven plasmas [66,67], and condensates under applied alternating-current (AC) fields [68], highlighting the broad theoretical foundations of this framework beyond nonlinear optics. In particular, Turaev et al. [69] studied the interaction of oscillating dissipative solitons in the Lugiato–Lefever model and demonstrated that long-range coupling, mediated by dispersive waves, can lead to harmonic synchronization and stable breather bound states, thus providing a dynamical route to breather 'molecules' in driven–dissipative systems.

The quintic CGLE significantly enriches the complexity of the soliton interactions, facilitating the formation of stable bound states of solitons across a wider parameter space[70,71]. Generally, the CGLE does not admit analytical soliton solutions, but an approximate solution can be produced by the perturbative method, when both the gain and loss coefficients are sufficiently small[11]. To this end, a solution to Eq. (19a) is looked for as

$$u_g = u_s\cdot\exp(ik\,|\tau-q_0|)\ . \tag{20}$$

In this expression, $\exp(ik|\tau-q_0|)$ is the phase factor with the wave number determined by the loss parameters and soliton's amplitude $\eta$ as $k=-\delta/\eta+\beta\eta$, $q_0$ being the central position of the soliton. This factor represents the internal phase chirp of the CGLE soliton, resulting in solitons' "tails" that decay exponentially with oscillations, in contrast to the smooth decay featured by the classical NLSE solitons (see Eq. (6)). Similar to the analysis of soliton interactions in the NLSE framework, the interaction between two solitons in the CGLE model can also be represented by the respective term in the Hamiltonian term accounting for the cubic nonlinearity[11],

$$H=-\int_{-\infty}^{+\infty}|u(\tau)|^4\,d\tau\ . \tag{21}$$

Using the perturbation method based on the tail overlapping, the evolution equations for the separation and phase caused by the interactions of the solitons can be derived[11]:

$$\frac{d^2\Delta\tau}{dz^2}+\frac{\sqrt{2}}{3}\beta\frac{d\Delta\tau}{dz}+e^{-\Delta\tau}[\cos(k\Delta\tau)+b\sin(k\Delta\tau)]\cos\Delta\phi=0\ , \tag{22a}$$

$$\frac{d^2\Delta\phi}{dz^2}+\frac{p}{\sqrt{2}}\frac{d\Delta\phi}{dz}-e^{-\Delta\tau}\cos(k\Delta\tau)\sin\Delta\phi=0\ . \tag{22b}$$

Here, the normalized propagation distance is $z=2\sqrt{2}\eta^2\xi$, while the normalized separation and phase difference between the interacting solitons are $\Delta\tau = \eta(q_1 - q_2)$ and $\Delta\phi = \phi_1 - \phi_2$. Coefficient $p = \sqrt{25(2\varepsilon-\beta)^2 - 480\mu\delta}/15$ appears as a combination of all the gain and loss terms from Eq. (18). Note that $\cos(k\Delta\tau)$ and $\sin(k\Delta\tau)$ are induced by the oscillatory shape of the tails. Equations (22) correspond to the equations of motion for a mechanical system with two degrees of freedom and potential

$$U(\Delta\tau,\Delta\phi) = -e^{-\Delta\tau}\cos(k\Delta\tau)\cos\Delta\phi \,. \tag{23}$$

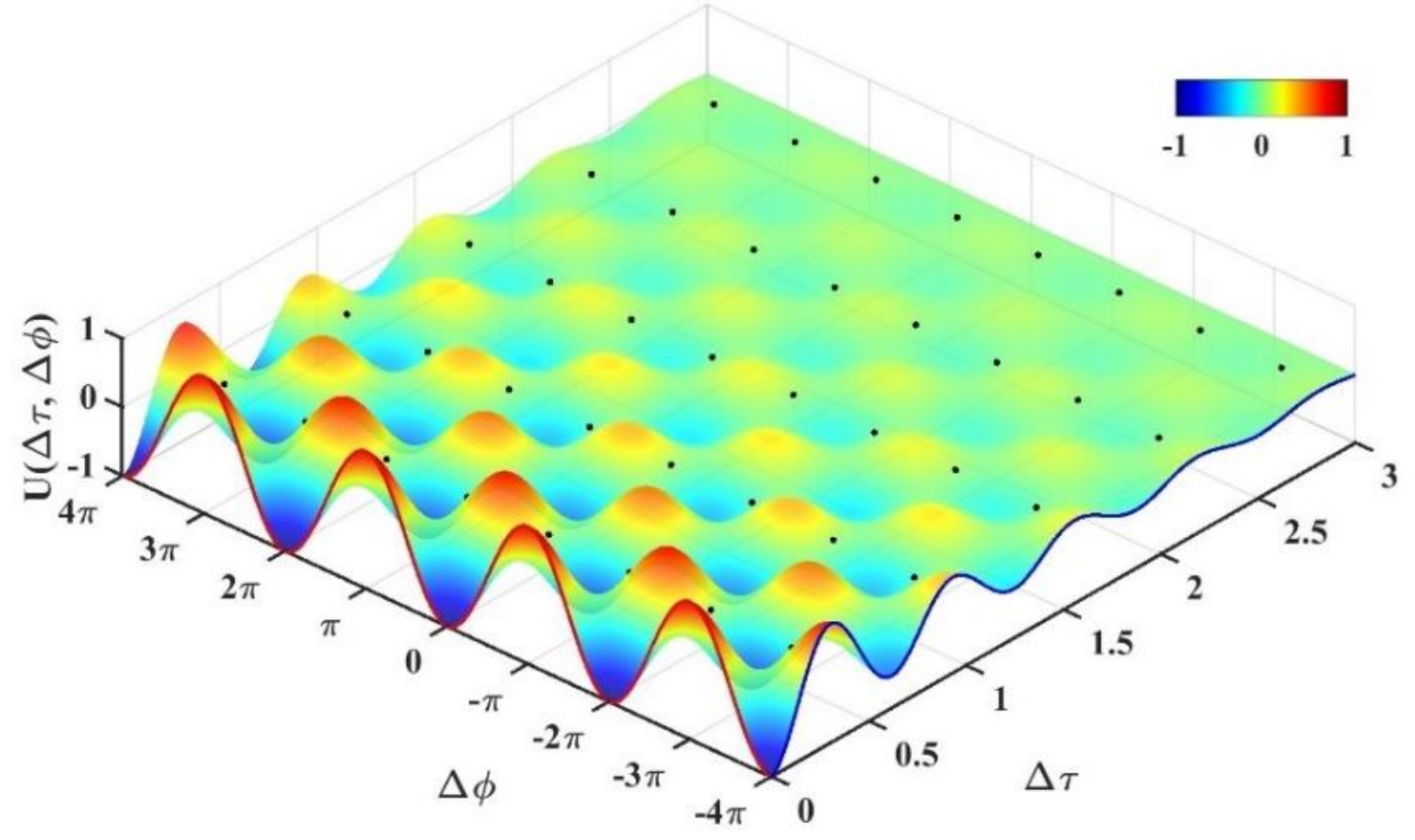

**Fig. 4** Original "piece of art" for the potential Eq. (23) as the function of $\Delta\tau$ and $\Delta\phi$ (for $k$ = 10). Positions of all peaks and troughs (local extrema) represent the fixed points corresponding to Eq. (25), while all black dots indicate the fixed points corresponding to Eq. (26).

The potential Eq. (23), whose evolution is illustrated in Fig. 4, possesses two sets of stationary (or fixed) points:

$$\sin\Delta\phi = 0, \quad \cos(k\Delta\tau) + k\sin(k\Delta\tau) = 0 \,. \tag{24a}$$

$$\cos\Delta\phi = 0, \quad \cos(k\Delta\tau) = 0 \,. \tag{24b}$$

Thus, the separation values between bound solitons, which correspond to the local extrema of the Hamiltonian (21) (indicated by the peaks and troughs in Fig. 4), are determined by Eq. (24a) as

$$\Delta\tau = -\frac{1}{k}\arctan\frac{1}{k} + \frac{m\pi}{k}, \quad m = 0,1,2,... \tag{25a}$$

$$\Delta\phi = n\pi, \quad n = \pm 1, \pm 2,... \tag{25b}$$

This type of soliton bound states sustains a positive binding energy, due to the fact that $\cos(k\Delta\tau)\cos\Delta\phi>0$. Accordingly, the full analysis, corroborated by the numerical results, reveals that these states are unstable saddle points, with a negative effective mass associated with the degree of freedom $\Delta\phi$[11].

Another set of fixed points, represented by the black dots in Fig. 4, can be derived from Eq. (24b):

$$\Delta\tau = \frac{\pi}{2k}(2m+1), \quad m = 0,1,2,... \tag{26a}$$

$$\Delta\phi = \frac{\pi}{2}(2n+1), \quad n = \pm 1, \pm 2,... \tag{26b}$$

The eigenvalue analysis of the system indicates that, although these states correspond to unstable spirals, their instability is relatively weak for small perturbation parameters[11]. The respective instability growth rate is proportional to the square of the exponentially small factor $e^{-\Delta\tau}$, while the instability growth rate of the bound states (25) grows linearly with this factor.

The qualitative phase portrait for the bound states, based on Eqs. (25) and (26) in the framework of Eq. (22), are depicted in Fig. 5. Looking at it, one concludes that the spirals, except for those corresponding to $n$ = 0 in Eq. (26), give rise, at $x\rightarrow\infty$, to infinite-period limit cycles coinciding with an elementary cell of the separatrix grid. The spirals corresponding to $n$=0, i.e., to the bound state with the smallest possible separation between the solitons, formally give rise to a similar cycle, which, however, including a segment at $r$ = 0, implies a collision between the two solitons.

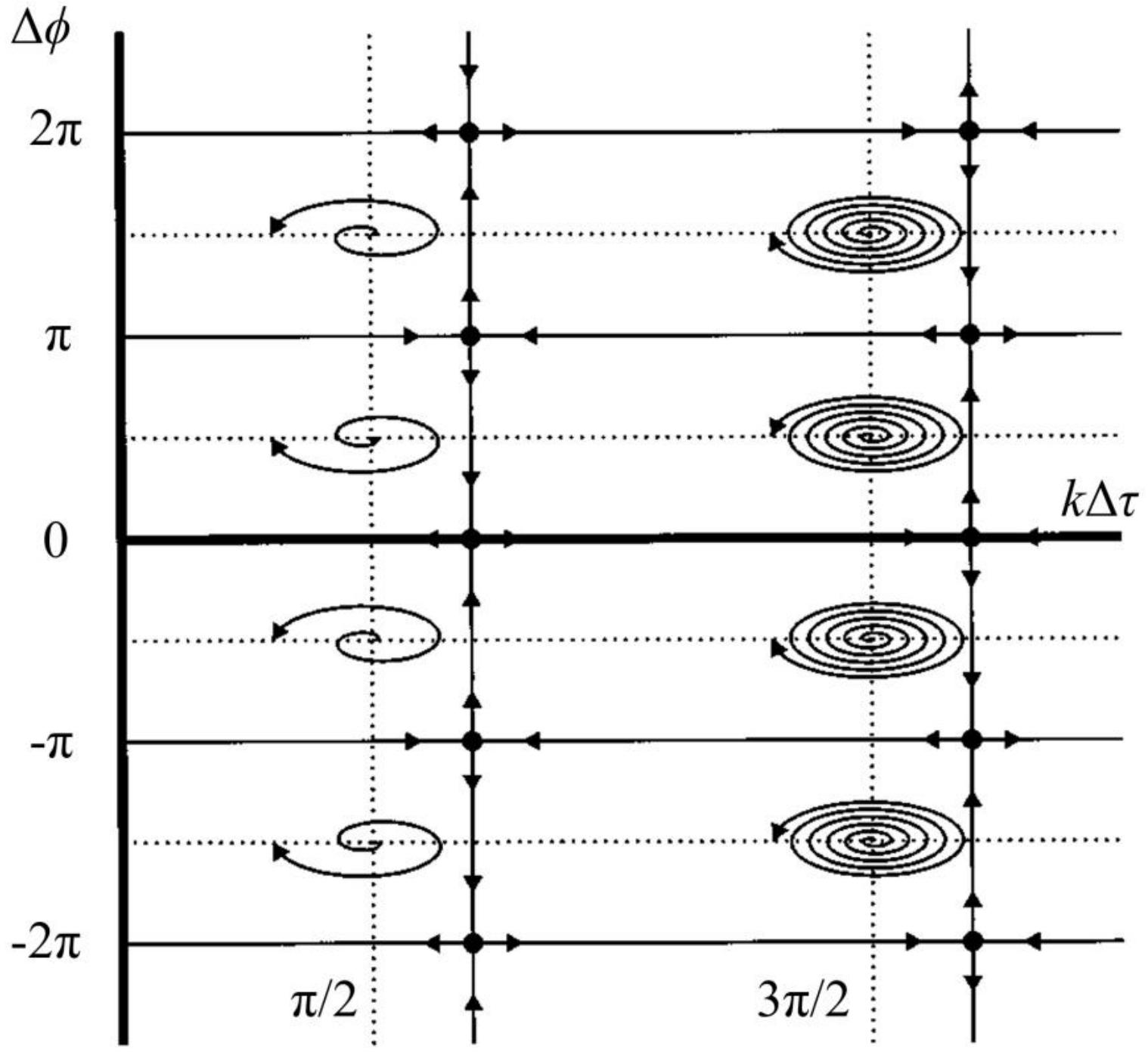

**Fig. 5** The phase portrait of the reduced dynamical system for the two-soliton bound states corresponding to Eqs. (25) and (26), as per Ref.[11].

This characteristic of CGLE implies that only a single or a limited number of soliton solutions exist for a given set of parameters, including stable soliton bound states[11]. Subsequently, efforts have been focused on employing numerical calculations to identify stable bound states under varying parameters. Prior to conducting numerical studies, the existence of stable bound states can be analyzed in terms of the system's energy and momentum[72,73]. In the dissipative system described by CGLE, the rates of change of the energy, $Q=\int_{-\infty}^{+\infty}|u|^2\,d\tau$, and momentum, $M=\mathrm{Im}\left(\int_{-\infty}^{+\infty}u_\tau^* u\,d\tau\right)$, are derived as follows:

$$\frac{dQ}{d\xi}=F[u],\qquad \frac{dM}{d\xi}=J[u]\,, \tag{27}$$

where

$$F[u]=2\int_{-\infty}^{+\infty}\left[\delta|u|^2+\varepsilon|u|^4+\mu|u|^6-\beta\left|\frac{\partial u}{\partial\tau}\right|^2\right]d\tau \tag{28}$$

$$J[u]=2\,\mathrm{Im}\int_{-\infty}^{+\infty}\left[(\delta+\varepsilon|u|^2+\mu|u|^4)+\beta\frac{\partial^2 u}{\partial\tau^2}\right]\left(\frac{\partial u}{\partial\tau}\right)^* d\tau \tag{29}$$

In terms of Eq. (27), stationary points must simultaneously satisfy the conditions that both energy and momentum are constant, i.e., $F[u]=0$ and $J[u]=0$. Akhmediev and Ankiewicz *et al.* conducted the numerical calculations of the energy and momentum, illustrating their findings on a two-dimensional interaction phase plane, as shown in Fig. 6 [72]. In this case, the perturbation parameters in the CGLE are no longer small ones. With the nonlinear gain coefficient $\varepsilon = 1.8$, the curves of $F$ and $J$ in Figs. 5(c,d) reveal the existence of five sets of stationary points. Here, $S_1$, $S_2$, and $S_3$ represent in-phase or out-of-phase soliton bound states, corresponding to the solutions of Eq. (25). In contrast, bound states $F_1$ and $F_2$ represent solitons with the $\pm\pi/2$ phase difference, corresponding to the solutions in Eq. (26). Building on this foundation, the trajectories of the bound states, which are shown in Fig. 6(b), indicate that the first set of the stationary points represent unstable saddle points, whereas the second set corresponds to stable spirals. This outcome is consistent with the above-mentioned theoretical predictions. Notably, the formation of a stable bound state requires an extended interval of evolution. As demonstrated in Figs. 5(c) and (d), both $J$ and $F$ oscillate with decaying amplitudes before gradually decaying to zero, marking the convergence to stability. However, when parameter changes occur, stable solutions may not be maintained. As shown in Figs. 5(e) and 5(f), when the nonlinear parameter $\varepsilon$ is reduced to 0.4, $J$ and $F$ undergo significant changes, preventing the realization of stable solutions $F_1$ and $F_2$, and resulting solely in the formation of the saddle points $S_1$, $S_2$.

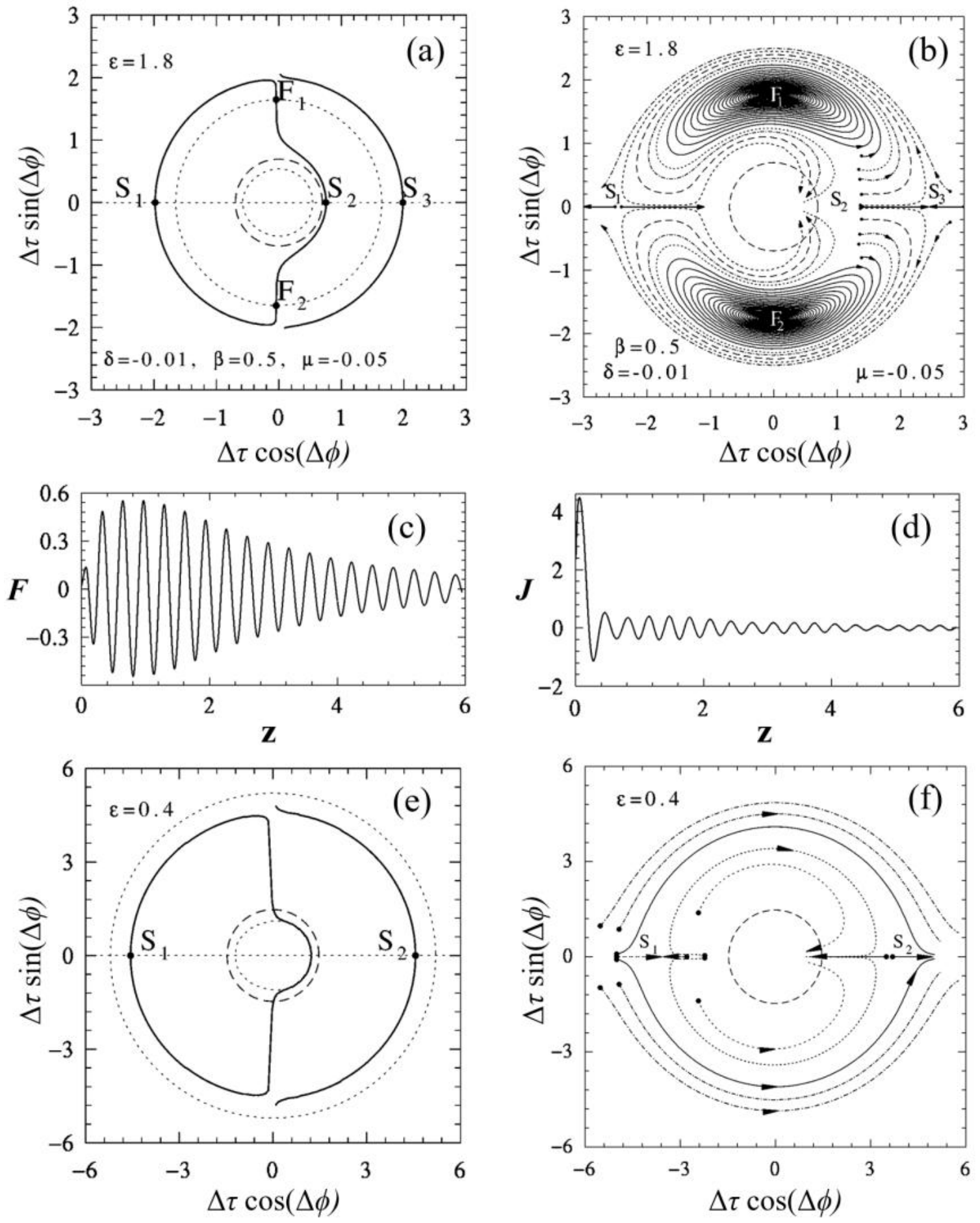

**Fig. 6** The evolutions of the energy and momentum of the two-soliton system, as per Ref. [72]. **(a)** Zeros of $F$ (solid lines) and $J$ (dotted lines) from Eqs. (28) and (29) on the interaction plane for $\varepsilon$ = 1.8. **(b)** The evolution of two-soliton solutions in the phase plane for the same parameters as in (a). **(c)** Oscillations of function $F$ in the process of the convergence. **(d)** Oscillations of function $J$ in the process of the convergence. **(e)** Zeros of $F$ (solid lines) and $J$ (dotted lines) on the phase plane for $\varepsilon$ = 0.4. **(f)** The evolution of two-soliton solutions on the phase plane for the same parameters as in (e).

The in-phase and out-of-phase unstable bound states (saddle points) have been observed in both the normal- [63,74] and anomalous-GVD [63,64] regimes. However, stable bound states (spirals) are typically achievable only in the anomalous-GVD[72] regime, unless a more realistic lumped model is considered[74], where the saturable absorbers exhibit non-instantaneous recovery. These solutions are determined entirely by the system's parameters rather than the initial conditions[62,75], which is one of the primary distinctions between dissipative systems and Hamiltonian systems. This characteristic allows soliton bound states in dissipative systems to exhibit more complex behaviors. Recent numerical investigations have revealed novel entities in the family of bound-state solutions of CGLE under varying parameters [75–80]. Among these, a distinct structure known as the vibrating bound state exhibits an evolution trajectory that converges to a limit cycle, rather than the fixed point illustrated in Fig. 6(b). An example of this behavior is presented in Fig. 7(a), where the trajectory is depicted in the phase plane, with a clockwise orientation indicated by the arrow [76,77]. Notably, both the separation and phase difference between the two solitons exhibit oscillatory behavior.

Additionally, the peak amplitude of the soliton on the right slightly exceeds that of the soliton on the left, which is attributed to the nonsymmetric phase relationship between the two entities. It is observed that, for specific parameter sets, only solitary and vibrating solutions are produced, suggesting that the existence of the vibrating solution is not contingent upon the presence of a stable bound state. The characteristics of the vibration are solely determined by the interactions between the two solitons[76]. Furthermore, some studies have introduced an additional perturbation into the system by injecting a weak signal into the laser, which results in similar vibrating solutions[78].

A two-soliton system characterized by a more intricate trajectory, termed the shaking bound state [76,79], has also been identified, as depicted in Fig. 7(b). In this configuration, the soliton pairs approach the fixed point of the spiral type through two distinct evolutionary scenarios. On the one hand, they spiral inward (as indicated by the arrow), reminiscent of the stable solution's evolution illustrated in Fig. 7(b). Conversely, they may simultaneously wind outward from the fixed point, manifesting a chaotic state. This phenomenon embodies a four-dimensional or higher-dimensional dynamical process, wherein the spiral convergence and chaotic shaking correspond to two separate evolutionary paths in the phase plane, alternating between them. Various amplitudes and periods of shaking can be achieved by modifying parameters $\beta$, $\varepsilon$, and $\mu$.

Another notable case, known as mixed bound states, is also possible[76,79]. In contrast to the previously described soliton pairs comprised of identical pulses, mixed bound states are characterized by the periodic motion of two solitons exhibiting distinct shapes, amplitudes, and phases, as shown in Fig. 7(c). The periodic attraction and repulsion observed in this case imply that the mixed solution constitutes a specific type of vibrating solutions, with the oscillation frequency intrinsically linked to the difference in propagation constants of the individual solitons.

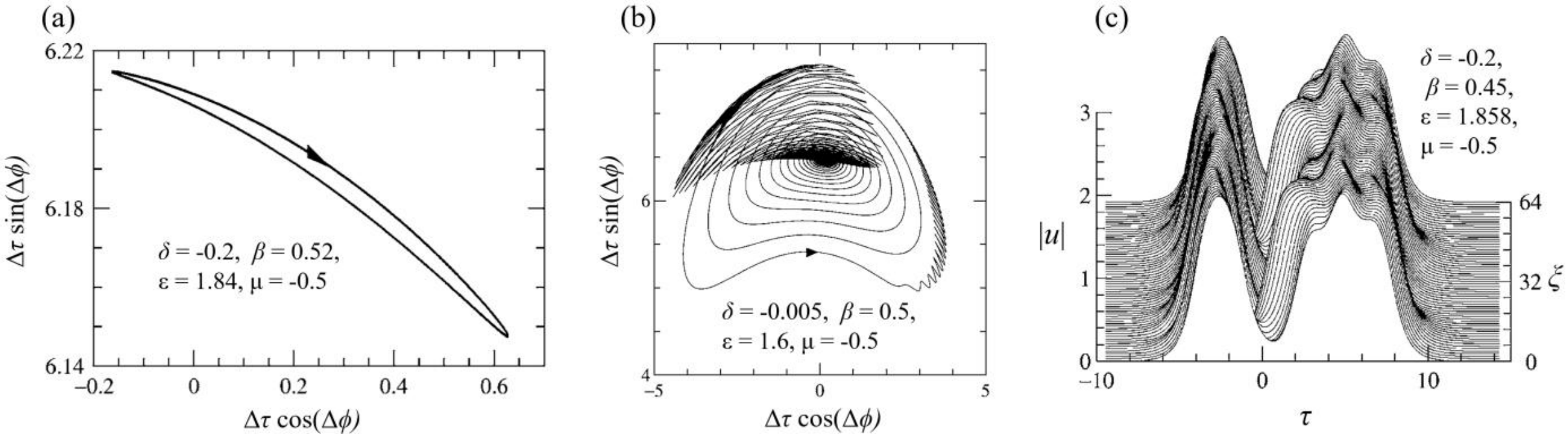

**Fig. 7** New objects in the family of bound state-solutions of CGLE (19a), as per Ref. [76]. **(a)** A vibrating solution. **(b)** A shaking solution. **(c)** A mixed solution. The corresponding numerical parameters are shown in each panel.

The existence of two-soliton bound states may be naturally extended to multi-soliton ones [72,75,81–85]. For the simplest case of a three-soliton bound state, the interactions can be analyzed by considering this state as a combination of a two-soliton bound state and a single soliton [75]. Due to the phase difference between the solitons, the two-soliton bound state constitutes an asymmetric solution of the CGLE and moves relative to the single soliton with a non-zero group velocity, resulting in collisions. There are three possible scenarios of the soliton collision. In the first scenario, the bound state is completely disrupted, leading to soliton fusion (Fig. 8(a)). Depending on the initial conditions, the collision yields one or two solitons that are sufficiently separated to interact weakly. In the second scenario, the single soliton and the bound state undergo an "elastic collision", forming a new soliton pair through an exchange of the middle soliton, as depicted in Fig. 8(b). The binding energy of the solitons is non-zero, and the difference in velocities between the pair and the single soliton is fixed and must be the same before and after collision. In the third scenario, all solitons coalesce into a three-soliton bound state (Fig. 8(c)). In this configuration, the phase differences between the adjacent solitons can be either opposite, *viz*., $-\pi/2$ and $+\pi/2$ [72,83] or identical, both being $+\pi/2$ or $-\pi/2$ [84]. While the separation between the three-soliton solutions remains stable, they exhibit a collective motion with a constant velocity, attributed to the asymmetric phase profile, and is almost equal to the velocity of the soliton pair before the collision. By accurately selecting a symmetric and concave phase profile, zero-velocity evolution of the multi-soliton bound states can be realized too [85], as demonstrated in Fig. 8(d) and 8(e) for the four- and five-solitons, respectively.

A similar mechanism for the formation of bound states of separated dissipative solitons due to the overlapping of their tails was analyzed in the framework of the Lugiato-Lefever equation (see Eq. (19b)) with a randomly varying pump $E(\xi)$. In this case, $E$ is replaced by $E(\xi)$, which represents a randomly varying function with the Gaussian correlator [86]. Thus far, this analytical result has not been verified in numerical simulations.

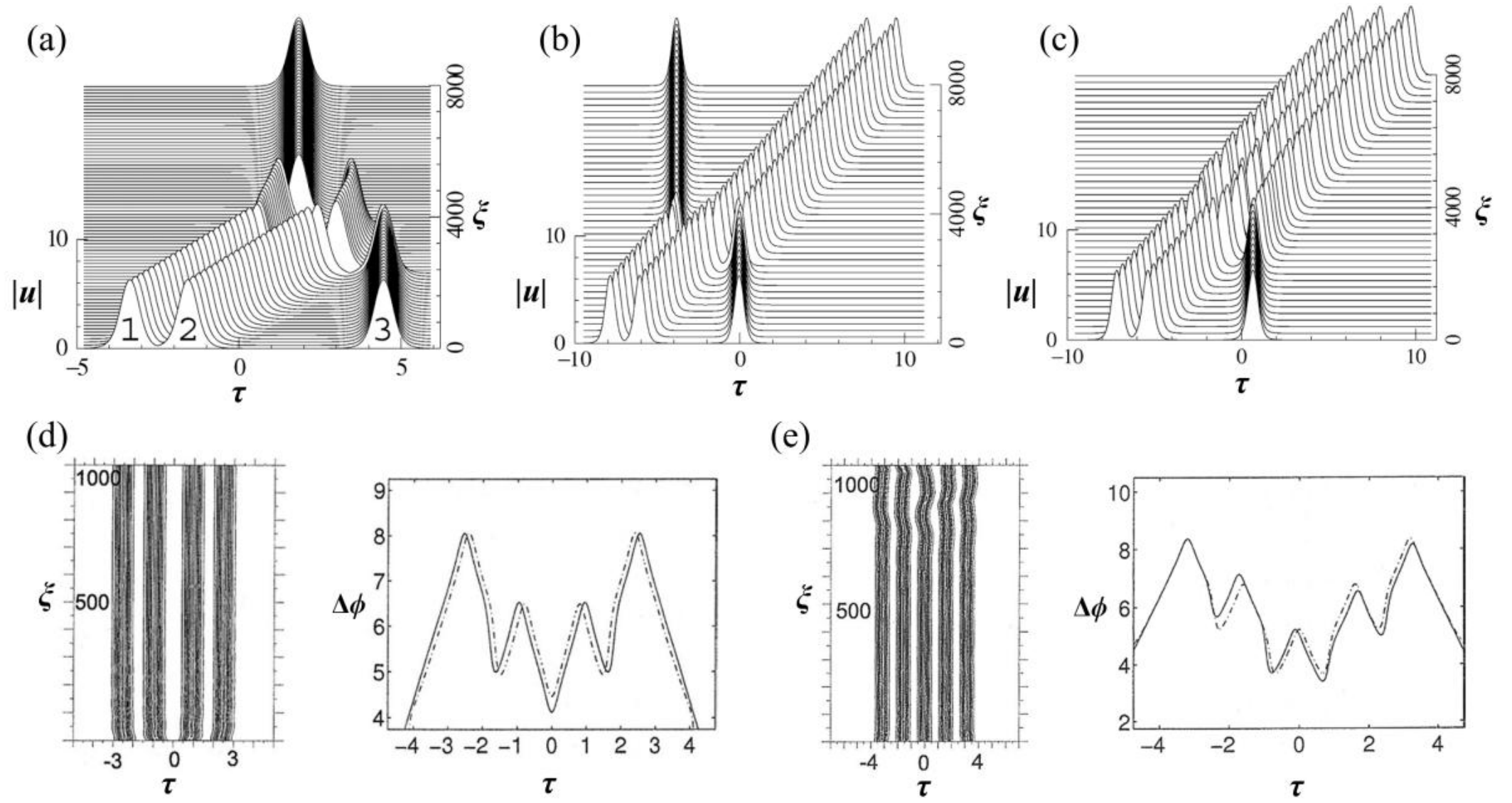

**Fig. 8** Numerical results for multi-soliton solutions, as per Ref. [75,85]. **(a)** The evolution of a three-soliton bound state. **(b)** The formation of a soliton train from a single pulse. **(c)** The coalescence into a three-soliton bound states. Other panels display the zero-velocity evolution of the four-soliton (d) and five-soliton (e) bound states, when choosing a symmetric and concave phase profile.

Stable bound states of dark solitons have been observed through numerical simulations across a wide range of parameters in the cubic-quintic CGLE[50–53,58]. Afanasjev *et al.* found that the bound state of dark solitons can be established for different values of $\varepsilon$ (see Eq. (19a))and remains stable even when the solitons are significantly overlapped (see Fig. 9(a)) [58]. Furthermore, the stability persists even when the two dark solitons possess different initial phases (Fig. 9(b)). Additionally, solutions for the bound states of dark solitons in this model have been analyzed recently [87]. Bound states of two and three dark solitons can also be demonstrated in terms of coupled equations, as illustrated in Figs. 8(c) and 8(d) [53]. Although the propagation direction and periodicity remain unchanged, the depth of the solitons increases as the value of $\varepsilon$ decreases.

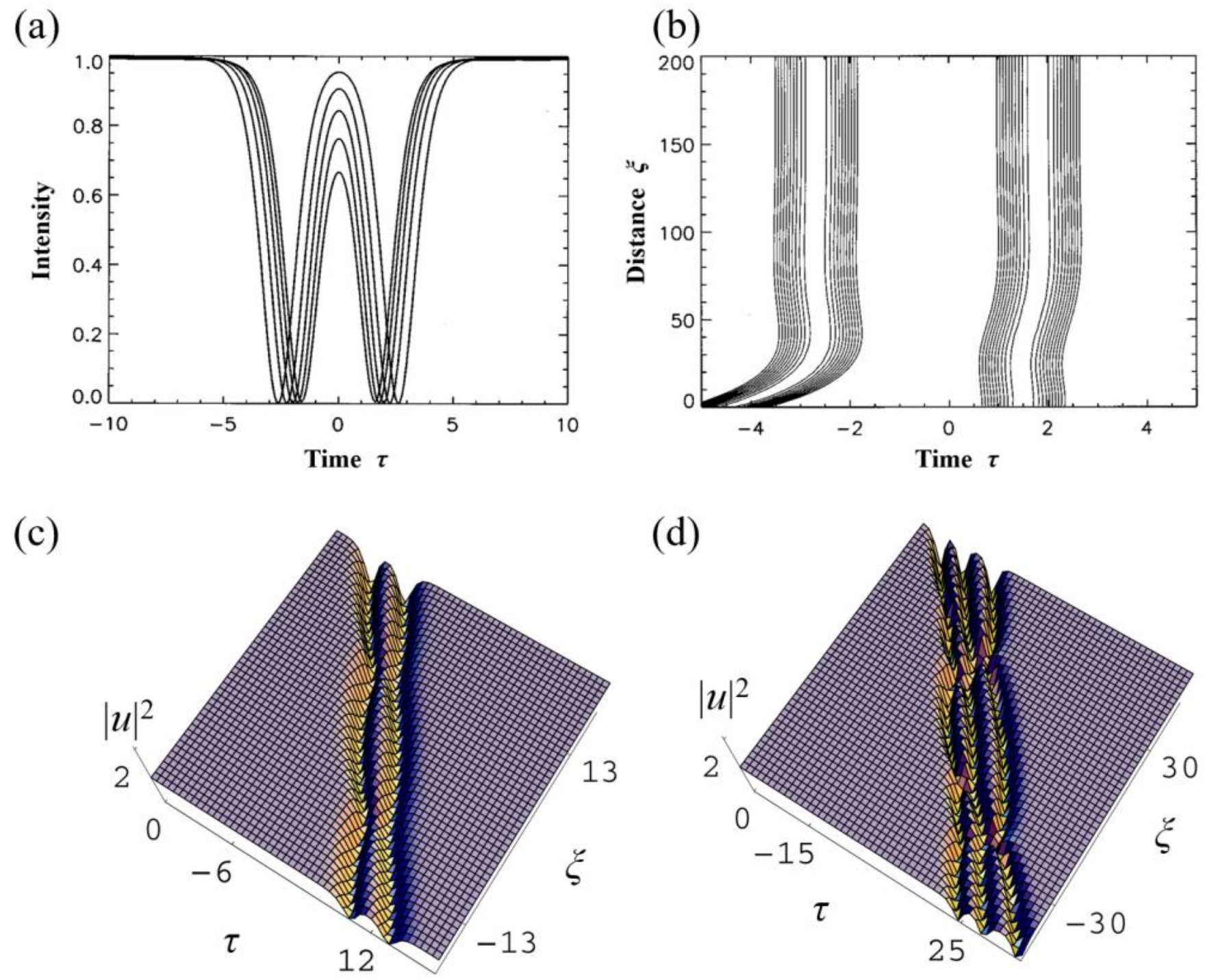

**Fig. 9** Numerical results for bound states of dark solitons, cf. Refs. [53,58]. **(a)** The profiles of the bound state for different values of $\varepsilon$ in Eq. (19a). **(b)** An example of the formation of the bound state from solitons with different initial phases. Other panels display bound states of two **(c)** and three **(d)** dark solitons in the system of coupled equations.

### 2.3 Bound-state solutions of coupled equations

In the theoretical models based on NLSE and CGLE considered above, the soliton interaction is mainly contributed by the self-phase-modulation (SPM) effect, which is related to the phase difference of the solitons, i.e., the coherent interaction between them[2]. However, for the interaction between solitons with different transverse modes, different polarizations and different colors (i.e., different carrier wavelengths), the coupled nonlinear equations govern the incoherent evolution dynamics which is not sensitive to the phase difference between the solitons[88–90]. In this case, solitons with different frequencies, intensities, pulse widths, and spectral profiles can be bound via the cross-phase modulation (XPM), as well as cross gain and loss, forming the bound state of solitons overlapping almost completely in time. This mechanism allows the existence of multi-component soliton bound states, such as dual-color SMs, vector SMs and spatio-temporal SMs[91–98]. Accordingly, richer evolution dynamics can be expected for the XPM-induced soliton bound states.

The fundamental mechanism of the XPM-induced soliton trapping is described by a system of coupled NLSEs, such as the one for the interaction between two orthogonal linear polarizations[2,89,99]:

$$i\frac{\partial u}{\partial \xi}+i\Delta v_g\frac{\partial u}{\partial \tau}+\frac{1}{2}\frac{\partial^2 u}{\partial \tau^2}-\delta u+\left(|u|^2+F|v|^2\right)u+\left(1-F\right)v^2u^*\exp(-2i\Delta\beta\xi)=0\,, \tag{30a}$$

$$i\frac{\partial v}{\partial \xi}-i\Delta v_g\frac{\partial v}{\partial \tau}+\frac{1}{2}\frac{\partial^2 v}{\partial \tau^2}-\delta v+\left(|v|^2+F|u|^2\right)v+\left(1-F\right)u^2v^*\exp(-2i\Delta\beta\xi)=0\,. \tag{30b}$$

where $u$ and $v$ represent the normalized slowly varying envelopes of the two orthogonally polarized states,

while $\xi$ and $\tau$ represent, as in Eq. (1), the normalized propagation distance and time, respectively, and $\delta$ is the same linear gain or loss coefficient as in Eq. (19a). Parameter $\Delta v_g$ denotes the group-velocity difference between the two polarizations, which originates from the fiber's birefringence. The last two terms on the left-hand side in Eqs. (30a) and (30b), account for the incoherent and coherent XPM coupling, quantified by coefficients $F$ and $1-F$, respectively. Values of these coefficients depend on the polarization states. For the coupling between two orthogonal linear polarizations, one has $F$=2/3, which corresponds to the Kerr nonlinearity originating from the electronic response of the dielectric material, while $F$=2 if the coupling occurs between two orthogonal circular polarizations or different carrier wavelengths [89]. Under the phase-matching conditions, the coherent coupling term induces degenerate four-wave mixing, influencing the polarization dynamics[100]. However, in optical fibers, there is always some degree of polarization detuning, dispersion detuning, and other detuning factors between the two polarization components, causing the phase mismatch represented by coefficient $\Delta\beta$ in Eqs. (30).

When studying the interaction between the two polarization components of a soliton pulse, the cases of small and large detuning should be examined separately. Under small detuning, the group-velocity mismatch is relatively small, allowing one to set $\Delta v_g \approx 0$, and, for simplicity, the circular-polarization components are typically used, with the fiber losses neglected. In this case, Eqs. (30a) and (30b) can be rewritten as[2]

$$i\frac{\partial u_+}{\partial \xi}+\frac{i}{2}\frac{\partial^2 u_+}{\partial \tau^2}+bu_-+\left(|u_+|^2+2|u_-|^2\right)u_+=0\,, \tag{31a}$$

$$i\frac{\partial u_-}{\partial \xi}+\frac{i}{2}\frac{\partial^2 u_-}{\partial \tau^2}+bu_++\left(|u_-|^2+2|u_+|^2\right)u_-=0\,, \tag{31b}$$

where $u_+$ and $u_-$ denote the right- and left-hand circular polarization states, and $b = (\Delta\beta)L_D/2$, with $L_D$ being the fiber dispersion length. Taking the polarization detuning as an example, $\Delta\beta$ here is related to the fiber's beat length $L_B$, so that $L_B=2\pi/\Delta\beta$. Numerical results indicate that when the nonlinear length exceeds $L_B$ , i.e., $L_{NL}>L_B$, the polarization components of the soliton along the fast and slow axes can remain stable. Conversely, with $L_{NL}<<L_B$, the component polarized along the fast axis becomes unstable.

Under large-detuning conditions (e.g., in highly birefringent fibers), the group velocity difference between the two polarization components is significant. The phase term in the coherent coupling oscillates rapidly due to large $\Delta\beta$ and averages to zero. In this case, neglecting the coherent coupling term and fiber loss reduces Eqs. (30a) and (30b) to[2]

$$i\frac{\partial u}{\partial \xi}+ic\frac{\partial u}{\partial \tau}+\frac{1}{2}\frac{\partial^2 u}{\partial \tau^2}+\left(|u|^2+2|\upsilon|^2\right)u=0, \tag{32a}$$

$$i\frac{\partial \upsilon}{\partial \xi}+ic\frac{\partial \upsilon}{\partial \tau}+\frac{1}{2}\frac{\partial^2 \upsilon}{\partial \tau^2}+\left(|\upsilon|^2+2|u|^2\right)\upsilon=0\,. \tag{32b}$$

For an input pulse launched at polarization angle $\theta$ (measured from the slow axis), Eqs. (32a) and (32b) should be solved with the following input:

$$u(0,\tau)=N\cos\theta\,\mathrm{sech}(\tau), \qquad \upsilon(0,\tau)=N\sin\theta\,\mathrm{sech}(\tau)\,, \tag{33}$$

where $N$ is the soliton order. In the absence of the XPM-induced coupling, the two polarization components propagate independently and separate under the action of the different group velocities. However, XPM can bind these components together. A numerical example for a fundamental soliton ($N = 1$) is provided in Fig. 10. For $\theta = 45°$ and $c = 0.2$, the polarization components become mutually captured, with overlapping temporal profiles (see Fig. 10(a)). This capture effect results from the frequency-domain shift in the opposite directions, as depicted in Fig. 10(b).

When $\theta$ = 30°, the initial amplitudes of the polarization components are unequal. As shown in Fig. 10(c), the weaker pulse (polarized along the fast axis) remains bound to the stronger one polarized along the slow axis, causing both components to move together. However, the captured pulses are offset from the center, as the slower-moving, stronger pulse initially shifts to the right before capturing the fast-axis counterpart. Figure 10(d) further demonstrates the asymmetric influence of the XPM-induced coupling on the pulse spectra of the two components.

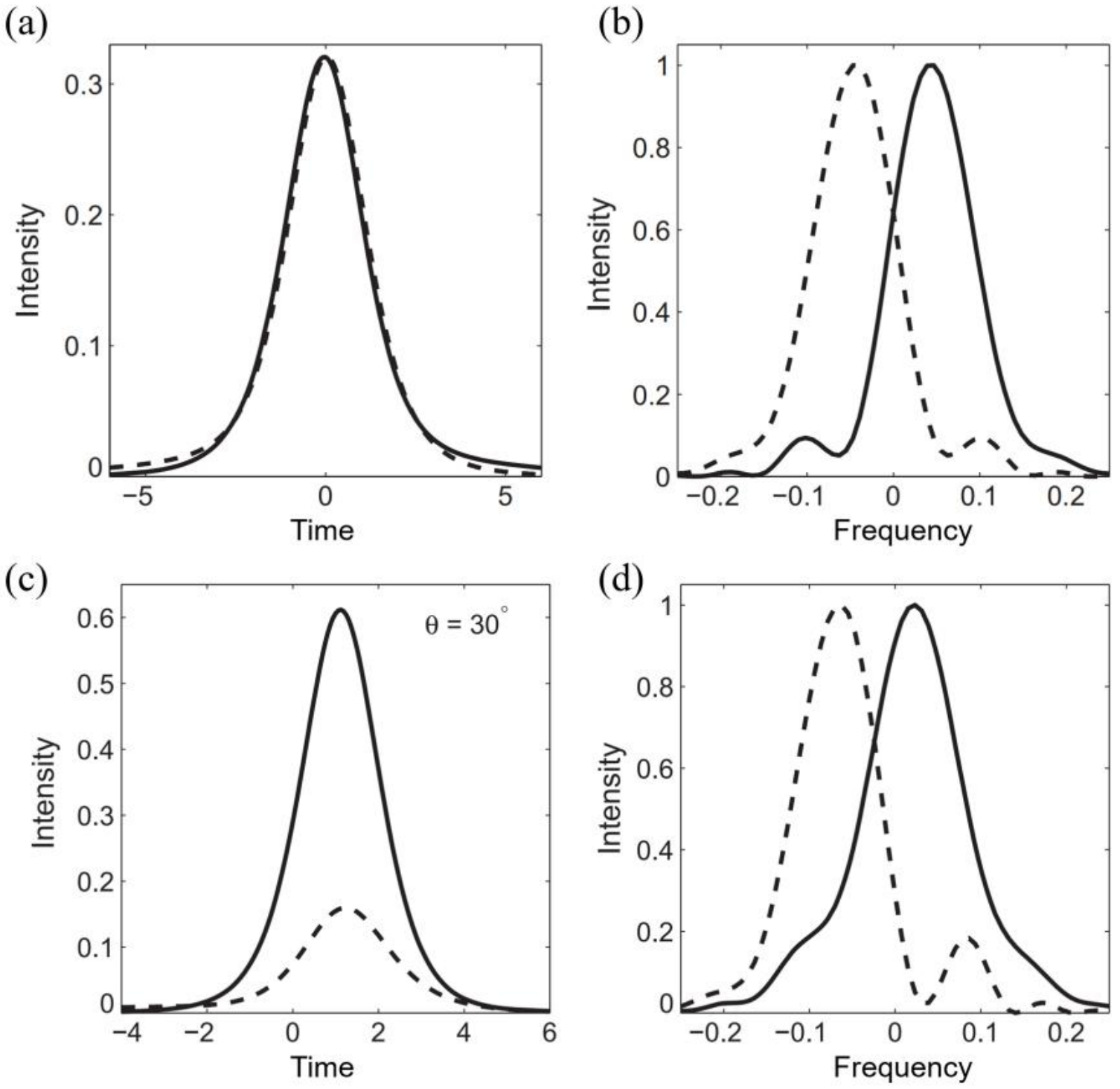

**Fig. 10** The soliton trapping between fast- and slow-axes modes (the dashed and solid lines, respectively), as per Ref. [2]. Panels **(a)** and **(b)** depict the temporal and spectral profiles of the two components when $\theta$ = 45°, while **(c)** and **(d)** show the temporal and spectral profiles of two components when $\theta$ = 30°.

The soliton trapping effect suggests that the coupled equations may possess exact two-component soliton solutions [101,102]. They are referred to as vector solitons, to emphasize the fact that they maintain the polarization state which is not aligned with one of the principal axes of the fiber. Further, the interaction between two orthogonally polarized pulses[95,103–107] can also be investigated in the framework of Eqs. (30a) and (30b), using the ansatz

$$(u,v) = 2i\eta \operatorname{sech}[2\eta(\tau - q_{u,v})]\exp(4i\eta^2\xi + i\phi_{u,v}), \tag{34}$$

where $q_{u,v}$ and $\phi_{u,v}$ represent the temporal positions and phases of the two solitons, respectively. This interaction is driven by the XPM terms, producing two distinct types of bound states[89]. In the first type, where phase mismatch $\Delta\beta$ is small, the soliton centers coincide, leading to strong binding, induced by both the incoherent and coherent coupling terms. This case resembles the one modeled by Eqs. (31a) and (31b), which can form a bound two-soliton state in the form of

$$u = v = 2\sqrt{2} i\eta \,\mathrm{sech}(4\eta\tau)\exp(16 i\eta^2 \xi) \, . \tag{35}$$

The simplest way to analyze the stability of the vector soliton (35) against the splitting into free separated $u$ and $v$ solitons is to compare energies of the initial and final states. Analyzing the system's Hamiltonian and total momentum yields the initial value, $H_0$ = -128$\eta^3$/3+$O(\Delta v_g)$. If the bound state splits into two separated single-component solitons, the total Hamiltonian of the resulting state must be lower than the initial value. Taking into regard the momentum conservation, the net Hamiltonian of the final state is twice that of the free soliton, and reaches a minimum value $(H_f)_{\min}$ = -32$\eta^3$/3+$O(\Delta v_g^2)$ when $\Delta v_g$ is small. Thus, the condition for the energy stability of the two-component soliton against the decay, $E_0 < (E_f)_{\min}$, holds indeed. Note that the respective eigenvalues of the binding energy (Hamiltonian) are

$$H_b \equiv (H_f)_{\min} - H_0 = 32\eta^3 + O(\Delta v_g) \, . \tag{36}$$

In the second type of the bound state, the solitons are separated by a larger distance, with weaker binding primarily determined by the coherent coupling term. In this case, the two-soliton state may be approximated just by a linear superposition of the single solitons. The interaction is governed by the effective potential

$$V \equiv -2B\int_{-\infty}^{+\infty} |u(\tau)|^2 |v(\tau)|^2 \, \mathrm{d}\tau - (1-B)\int_{-\infty}^{+\infty} \left[ u^2 (v^*)^2 + v^2 (u^*)^2 \right] \mathrm{d}\tau \, . \tag{37}$$

A stable bound state of the two weakly overlapping solitons corresponds to a local minimum of the potential, and the respective binding energy (Hamiltonian) can be calculated as

$$H_b \approx 256\eta^3 e^{-2\pi n\eta/\Delta v_g} \, . \tag{38}$$

In Eq. (38) even the binding energy corresponding to the most stable bound state ($n$ = 1) is exponentially small. This smallness, as well as the mechanism for the creation of the bound state of the weakly overlapping solitons, are similar to those reported in Ref. [11] for the strongly separated solitons governed by one equation (not a coupled system) of the NLSE-CGLE type.

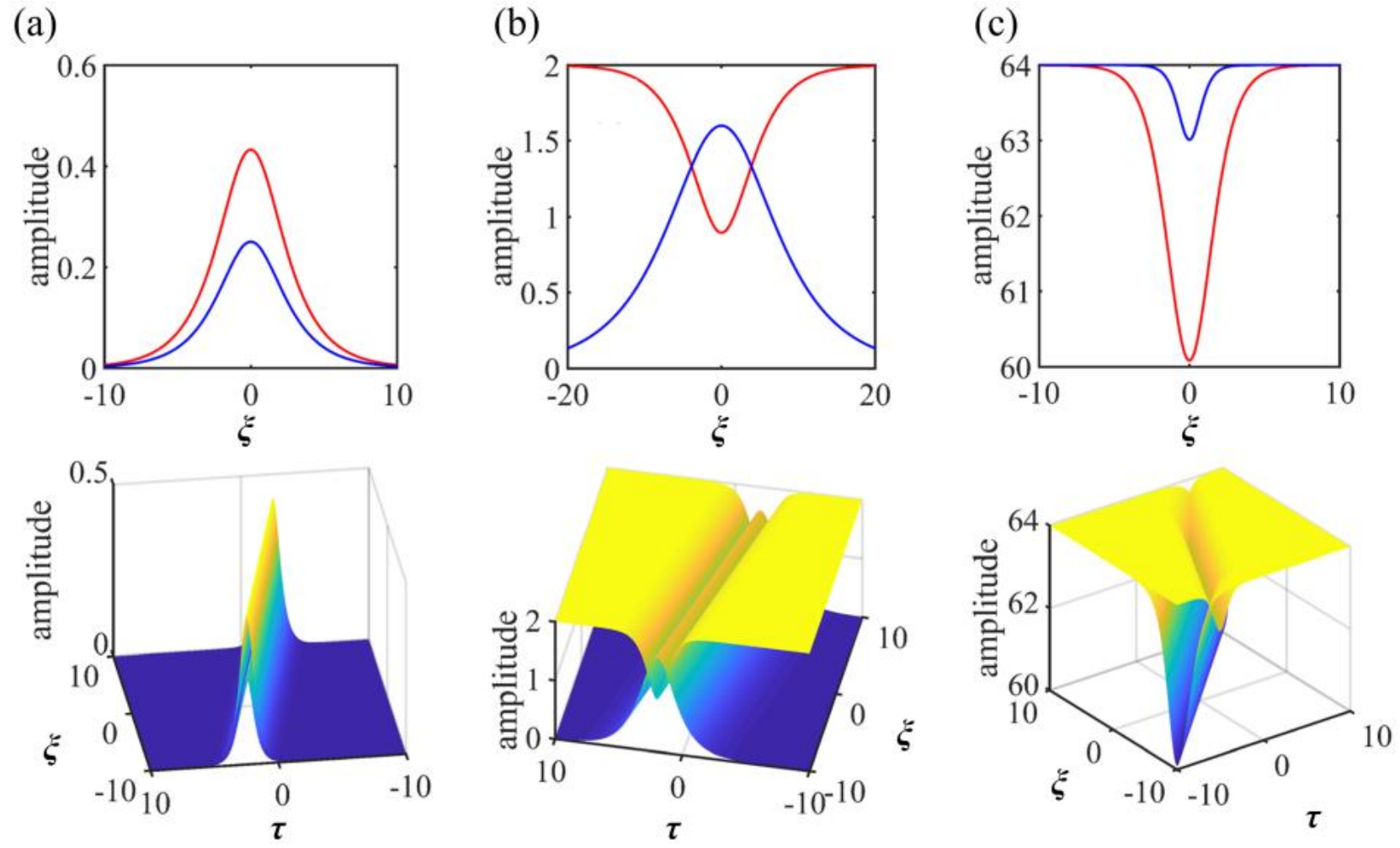

**Fig. 11** Intensity profiles and evolution of three types of vector solitons, as per Ref. [108]. **(a)** A bright-bright vector soliton. **(b)** A dark-bright vector soliton. **(c)** A dark-dark vector soliton.

Based on the effects of the coupling terms in the governing equations, the existence of three distinct types of vector solitons has been predicted: bright-bright[103,104], dark-bright[95,105], and dark-dark solitons[106,107]. A typical example of a bright-bright vector soliton, displayed in Fig. 11(a), occurs when both nonlinear terms are focusing, or when a combination of focusing and defocusing is present[108]. Here, focusing or defocusing refers to the sign of the nonlinear terms in the coupled equations being positive or negative, respectively. In the course of interactions, these vector solitons typically exhibit elastic collisions, with their shapes and velocities remaining the same as prior to the collision. The formation of dark-bright solitons, exemplified in Fig. 11(b), requires at least one nonlinear term to be defocusing, as dark solitons typically exist in the normal-GVD regime. Currently, dark-dark solitons can only be obtained in the numerical form, requiring at least one defocusing nonlinear term, as illustrated by their typical profile in Fig. 11(c). A representative vectorial setting is provided by Kerr resonators described by coupled Lugiato–Lefever equations with nonlinear polarization-mode coupling, where dark vector localized states can emerge[109]. Furthermore, depending on the cavity birefringence and strength of cross-polarization coupling, the resulting vector solitons can be categorized into polarization-locked vector solitons[110,111], polarization-rotation ones[112,113], and group-velocity-locked vector solitons [114,115].

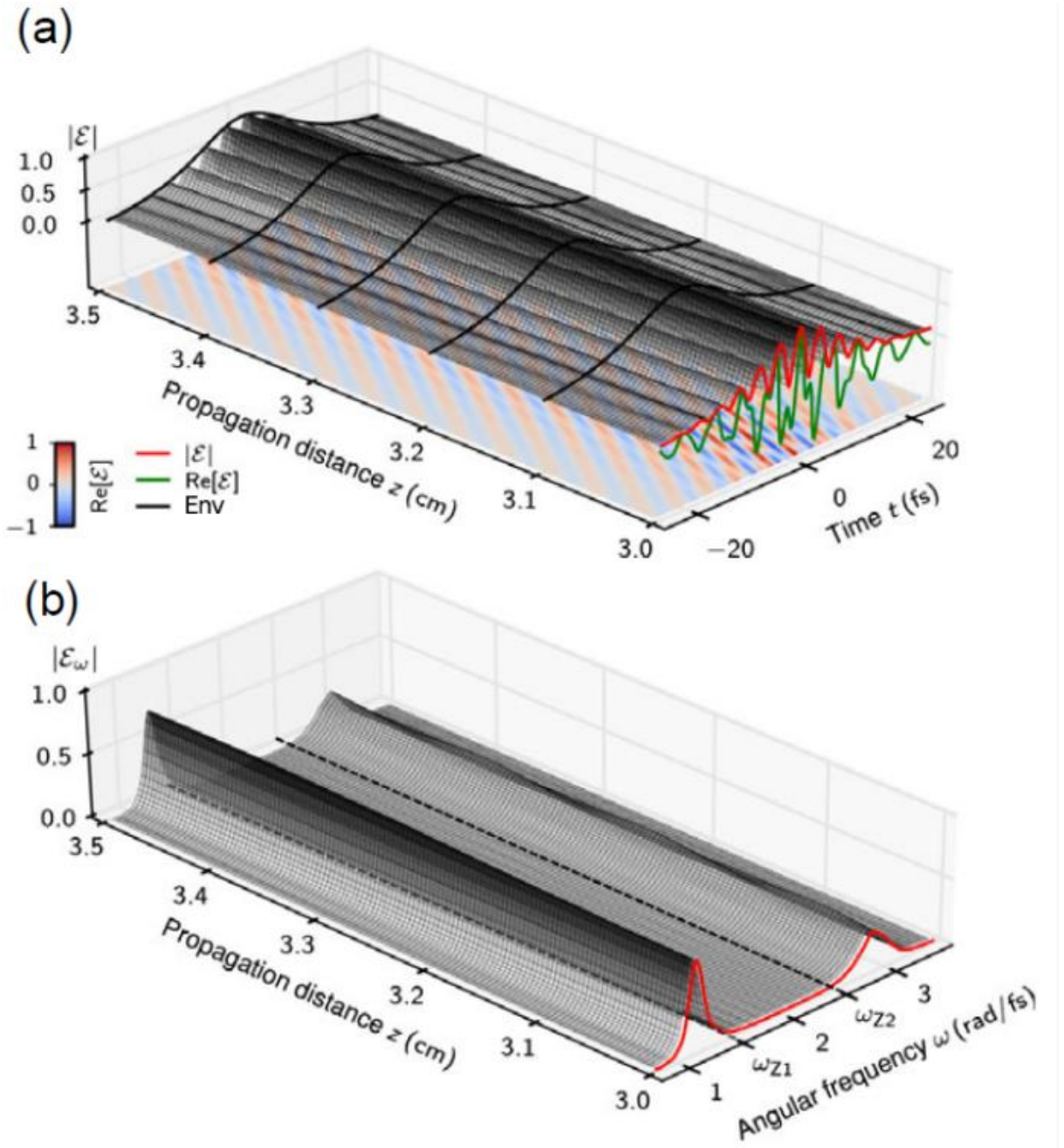

**Fig. 12** The bound state of colored solitons, as per Ref. [92]. Panel **(a)** displays the temporal evolution of the bound state, while panel **(b)** presents its spectral evolution.

Colored solitons, defined as ones with different carrier frequencies, can also be modeled by Eqs. (32) and (33)[90,91]. The threshold values for the frequency difference can be obtained by means of the inverse-scattering-transform technique[91], showing that the soliton modes oscillate between a two-soliton bound state and a single soliton. A typical example of the temporal and spectral evolution of a bound state of the colored solitons is presented in Fig. 12. In the temporal domain, this bound state manifests itself as a solitary pulse characterized by interference fringes, maintaining its shape in the course of the propagation. In the spectral domain, the bound state features distinct peaks in two spatially separated regions. These peaks exhibit minor oscillations around the group velocity associated with the composite state. When the frequency difference is less than the soliton's spectral width, the perturbation theory is applicable, suggesting that the colored solitons will either merge into a single pulse or evolve into two pulses with different velocities, depending on their initial phase difference[93]. In contrast, when the frequency difference is significantly larger, the colored solitons retain a molecule-like binding energy and may exhibit phenomena such as vibrations or radiation. These features may be interpreted as mutual trapping, providing a compelling analogy to quantum mechanics[92].

Spatiotemporal solitons in multimode optical fibers and micro-photonic crystals have emerged as a prominent research frontier[96–98,116–118]. The coupling between spatial and temporal domains can be modeled by coupled equations, where the pulse characteristics transcend the temporal dimension of the longitudinal mode to encapsulate the interplay of the electric field distribution in both temporal and spatial (transverse) dimensions, as shown in Fig. 13(a). In particular, by means of the variational approximation and numerical methods, the existence and stability of spatiotemporal solitons was analyzed in multidimensional optical media with the quadratic nonlinearity[117]. The findings revealed that these solitons, under specific conditions, exhibit stable propagation and are fundamentally distinct from conventional spatial solitons. An illustrative case of the spatiotemporal soliton evolution in (2+1) dimensions, including internal oscillations, is presented in Fig. 13(b). Building on this, Mihalache et.al. explored the dynamical stability of spatiotemporal solitons in quadratic nonlinear media[96]. They observed that, under varying GVD conditions, the transverse spatiotemporal profiles of the solitons are often asymmetric, providing essential insights into the soliton formation and propagation dynamics. Recent investigations have also delved into the spatiotemporal dynamics of breather solitons in micro-photonic crystals[97], highlighting phenomena such as spatial breathers, chaotic transitions, and their implications for nonlinear optics in both experiment and theory.

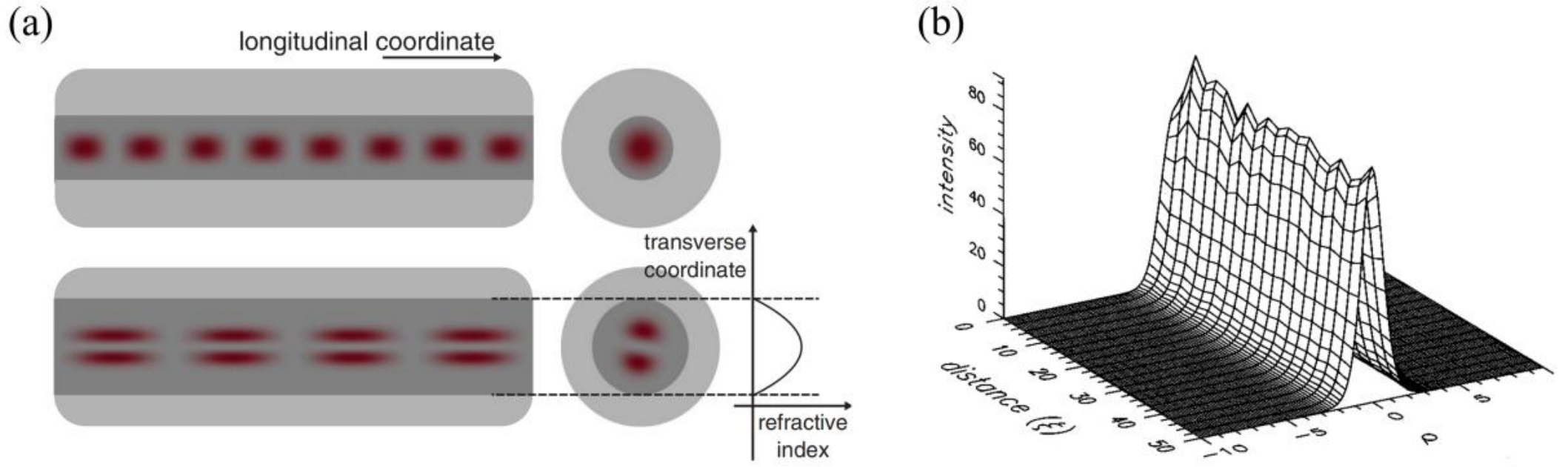

**Fig. 13** Spatiotemporal solitons. **(a)** Lasing occurs in the temporal dimension (longitudinal mode) and spatial dimension (transverse mode), as per Ref. [98]. **(b)** An example of the stable evolution of a spatiotemporal soliton in (2+1) dimensions, as per Ref. [117].

Lastly, the formation of bound states of separated solitons was also analyzed in the model of the mismatched optical coupler, which is represented by a linearly coupled system of NLSEs, with the phase- and group-velocity differences between the coupled equations[119]. Without the inclusion of loss and gain, this conservative system gives rise to effective potential for the interaction between solitons in the two components. The potential features a series of minima which predict bound states of the solitons. This system was not yet studied by means of numerical simulations.

## 3 Binding mechanism for solitons in fiber lasers

Formation of bound state of solitons originates from the interaction between solitons producing the temporal structure via binding forces[27]. The interaction between solitons was firstly studied, in the context of optical telecommunications, in long fiber links [120,121]. The interaction forces were found during the propagation of the solitons in nonlinear fibers, which is an adverse factor for communication and dual-comb sensing system[120,122]. Then, it was observed and widely studied that the soliton interaction can significantly affect the soliton states in mode-locked lasers[13,123–126]. The partial overlapping of adjacent solitons builds the intensity distribution of the interference field, as resulting from the coherent interaction which is related to the solitons' phases [13]. The phase-sensitivity coherent interaction can also occur between soliton and non-soliton component, such as dispersive waves, continuous waves, and even the noise background[120,127,128]. In this case, the NLSE and GLE models are employed to describe the interaction that is mainly induced by the SPM effect. On the other hand, other factors also affect the relative position of the solitons, such as the electrostriction mechanism, gain recovery, and XPM[129–132]. These interactions should be incoherent, as they do not depend on the solitons' phases. These factors always act in the context of long-range interactions[120,133]. Actually, the soliton bound states in fiber lasers are subject to the action of several different binding forces that can produce diverse temporal distributions. In this section, the coherent and incoherent binding mechanisms for solitons in fiber lasers are introduced.

### 3.1 The coherent interaction

The origin of the coherent interaction is discussed above in Sections 2.1 and 2.2 to demonstrate the existence of soliton bound states theoretically. For the standard soliton solution of NLSE, the attractive or repulsive forces can be induced by the SPM effect, which is determined by the initial phase difference and intensity[4,134]. However, the pulse separation varies along with the propagation distance. Stable bound-state solutions in the extended NLSE-based model were found by considering the above-mentioned oscillating factor in the soliton tails[11]. Features of the stable soliton solutions are outlined in above section, the formation mechanism being effective under different dispersion maps (i.e., the different soliton formation scenarios)[135–137]. The stable bound states have also been tested in numerical simulations in which the separation between the interacting solitons always converge to a specific value, starting from different initial parameters[138], as shown in Fig. 14. The stable bound-state solutions act as attractors in mode-locked fiber lasers.

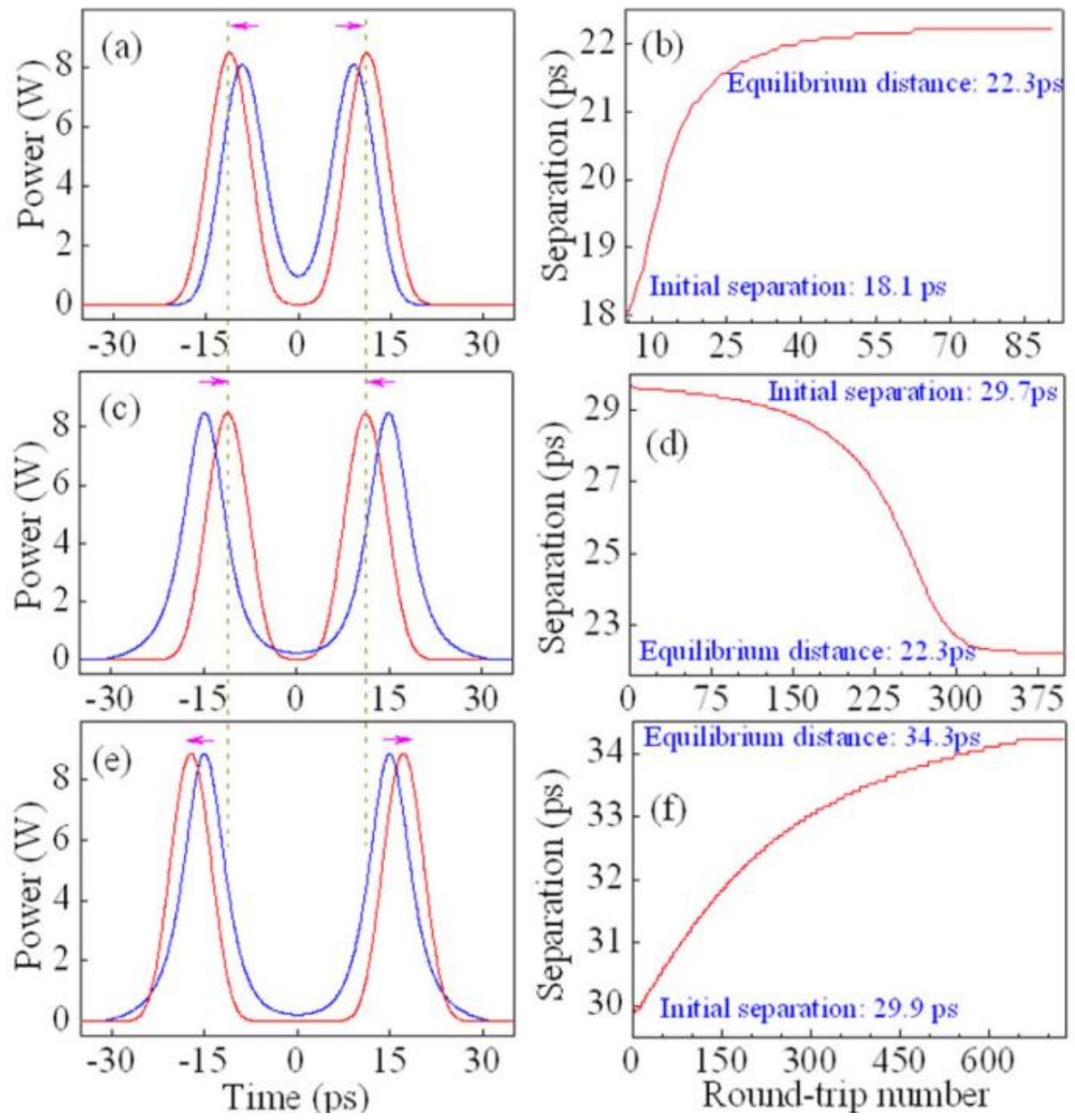

**Fig. 14** The Evolution of soliton bound states produced by numerical simulations. **(a,c,e)** Pulse profiles with different initial separations and the same equilibrium state (the dashed line). **(b,d,f)** The evolution of the pulse separation converging to the equilibrium value. The results were reported in Ref. [138].

Full understanding of the interaction mechanism for the dissipative solitons in fiber lasers, which is determined by several factors, is a challenging objective. The soliton and its oscillatory tails can be affected by the dispersion, nonlinearity, filtering, gain and higher-order effects, leading to the bound states with different separations and phase shifts between the interacting solitons [131,139–143].

It is widely recognized that the formation of stable soliton bound state is facilitated by a non-soliton component, which may be a weak continuous wave (CW) or radiation emitted by the solitons [128,144,145]. Applying the perturbation theory based on the inverse-scattering transform, the initial conditions for NLSE are assumed as a combination of a single-soliton solution $u_s(t)$ and a small finite perturbation $\delta u(t)$, i.e., $u_0(t)= u_s(t)+ \delta u(t)$. The perturbation may be expressed as [144]

$$\delta u\left(t,\tau_{\text{in}}\right)=A\exp\left(i\phi\right)\left\{\exp\left(i\Omega t\right)+\exp\left[i\Omega\left(t-\tau_{\text{in}}\right)-\theta\right]\right\} \tag{39}$$

where $A$ is the CW amplitude and $\Omega$ the frequency difference between the soliton and CW, while $\tau_{\text{in}}$ is the initial separation between the perturbation pulses. The resulting changes of the soliton's amplitude, $\delta\eta$, and velocity, $\delta V$, are

$$\delta\eta\left(\tau_{\text{in}},\Omega\right)=A\pi\,\text{sech}\frac{(\pi\Omega)}{2}\left[\cos\phi+\cos\left(\Omega\tau_{\text{in}}-\phi+\theta\right)\right], \tag{40}$$

$$\delta V\left(\tau_{\text{in}},\Omega\right)=\Omega\,\delta\eta\left(\tau_{\text{in}},\Omega\right). \tag{41}$$

The output soliton separation is obtained as

$$\begin{aligned}\tau_{\text{out}}\left(\tau_{\text{in}}\right)&=\tau_{\text{in}}+\xi\left[\delta V(0)-\delta V\left(\tau_{\text{in}}\right)\right]\\&=\tau_{\text{in}}\pm A\xi\Omega\pi\,\text{sech}\frac{(\pi\Omega)}{2}\sin\left(\Omega\tau_{\text{in}}\right).\end{aligned} \tag{42}$$

From the form of the analytical NLSE soliton it is seen that the frequency-shifted non-soliton component may enhance the interaction of the solitons, resulting in a periodic solution, depending on the amplitude and frequency[144]. The stationary bound states were also found due to mutual trapping of the radiation emitted by them, which can form a standing wave producing local minima in the effective potential of the interaction between neighboring solitons, as shown in Fig. 15(a) [146]. In mode-locked fiber lasers, the effect of the non-soliton components have also been investigated in simulations [128,145]. The wings of the dispersive wave result in strong interaction between the solitons, leading to formation of stable bound states with a large set of energy levels, as shown in Fig. 15(b) [145]. The solitons can remotely interact with each other through dispersive waves [133]. There are several origins for the formation of the non-soliton component in fiber lasers and fiber communication links, such as the periodic amplification and loss, variation of the pulse envelope, and residual CW radiation [147–150].

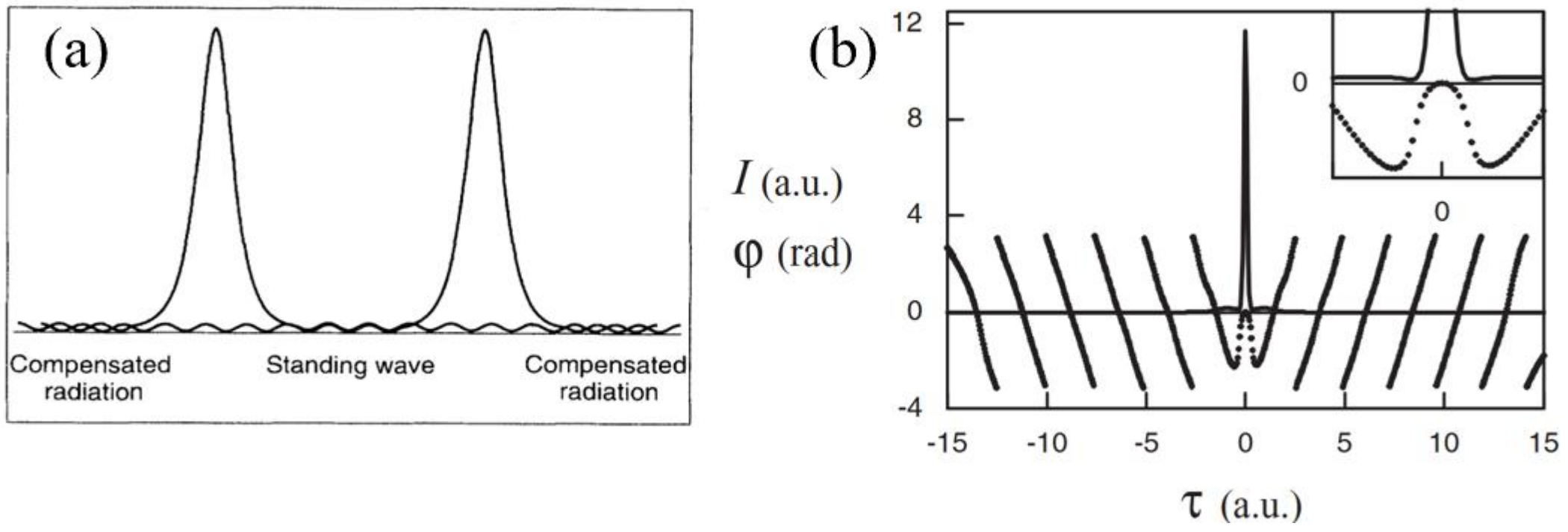

**Fig. 15. (a)** A schematic of the soliton bound state coupled by the background wave, as per Ref. [146]. **(b)** The phase change along with soliton, as per Ref. [145]. The inset zooms the changes of the phase and intensity in the vicinity of the center of the soliton.

A basic type of the nonsoliton components in fiber lasers is widely known as dispersive waves, which can be generated under various dispersion maps and dissipative conditions, facilitating the formation of bound states of conventional, dispersion-managed, and dissipative solitons[150–153]. In particular, the resonant coupling between the dispersive wave and soliton can be observed under the action of the anomalous-GVD map, producing multiple narrow peaks distributed on both sides of the soliton spectrum, which are called Kelly sidebands[150]. The sideband frequencies can be calculated as[2,150]

$$\omega T_S = \pm\sqrt{8m\xi_0/L-1}, \quad \xi_0 = L_D \cdot \pi/2, \quad m = 1, 2, ..., \tag{43}$$

where $T_s$ is the temporal width of the soliton. The features of the enhanced sidebands may significantly affect the soliton interaction, leading to the formation of phase-locked bound states with "exotic" phase values[154,155].

### 3.2 The incoherent interaction

For the long-range interaction between solitons, it was found that the electrostriction mechanism plays an important role [129,156]. Acoustic waves in optical fibers are excited by the optomechanical coupling between cavity modes through the stimulated forward Brillouin scattering, which can induce the change of the refractive index through electrostriction forces [157]. As a result, the soliton's velocity is affected by acoustic waves produced by the preceding solitons, leading to the variation of the separation between the solitons. The opto-acoustic interaction has been experimentally observed in fibers by Smith and Mollenauer in 1989, as well as in ultra-long communication transmission links and for cavity solitons over astronomically large distances[120,133].

In the theory, the acoustic vibrations for the components of displacement vector $U_i$ are governed by the equation for the material density perturbation $\rho$ [129,157]:

$$\rho_0 \partial^2 U_i / \partial^2 t - C_T^2 \partial^2 U_i / \partial x_k^2 + \left(C_L^2 - C_T^2\right) \partial^2 U_k / \partial x_k \partial x_i = \partial \sigma_{ik} / \partial x_k \tag{44}$$

where $\rho_0$ is the material density of the fiber, $C_L$= 5.99×10$^5$ cm/s is the velocity of longitudinal acoustic waves in silica glass, $C_T$ = 3.74×10$^5$ cm/s is the velocity of the shear acoustic wave, and $\sigma_{ik}$ = - $(2P_{11}\varepsilon^2 E_i E_k + P_{12}\varepsilon^2 E_l^2 \delta_{ik})/(8\pi)$ is the electrostrictional stress tensor, while $E_{i,k,l}$ are components of the electric field. Further, $P_{11}$ = 0.121 and $P_{12}$ = 0.27 are the strain-optic coefficients for fused quartz, $\varepsilon$ = 2.1 is the dielectric permittivity, and $\delta_{ik}$ is the Kronecker's delta. In a cylindrical coordinate system, two types of acoustic modes can be excited in the optical fiber: radial ones $R_{0m}$, featuring solely the radial component of the displacement vector, and mixed torsional-radial modes $TR_{mn}$, with the displacement vector varying as cos (2$\varphi$) [158]. Excitation of the $R_{0m}$ acoustic modes leads to an isotopic perturbation of the dielectric permittivity [157]:

$$\delta\varepsilon_{xx} = \delta\varepsilon_{yy} = \varepsilon^2 \left(P_{11} + P_{12}\right)\left(S_{rr} + S_{\varphi\varphi}\right). \tag{45}$$

$$\delta\varepsilon_{xy} = \delta\varepsilon_{yx} = 0 . \tag{46}$$

where $S_{rr} = \partial U_r / \partial r$ and $S_{\varphi\varphi} = U_r / r$ are the components of the strain tensor in cylindrical coordinates. Acoustic modes $TR_{mn}$ cause a birefringent perturbation in the fiber's dielectric permittivity (the principal axes of the tensor of this perturbation coincide with the axes of the polarization ellipse):

$$\delta\varepsilon_{xx} = \varepsilon^2 P_{11} / 4\left(\partial U_r^{(2)} / \partial r - \partial U_\varphi^{(2)} / \partial r + U_r^{(2)} / r - U_\varphi^{(2)} / r\right), \tag{47}$$

$$\delta\varepsilon_{yy} = -\delta\varepsilon_{xx}, \quad \delta\varepsilon_{xy} = \delta\varepsilon_{yx} = 0 , \tag{48}$$

where $U_r^{(2)}$ and $U_\varphi^{(2)}$ are the radial dependences of the displacement vector for the $TR_{mn}$ modes. Based on this theory, one can calculate the electrostrictional excitation of acoustic waves in single-mode fibers, and thus predict the interaction of the solitons with the acoustic wave. Small changes of the effective refractive index of the fiber, $\delta n$, can be calculated by means of the perturbation theory,

$$\delta n = (2n)^{-1} \int \delta\varepsilon F^2(r) ds / \int F^2(r) ds , \tag{49}$$

where $F(r)$ describes the intensity of the electric-field modal distribution, and the integration is performed over the fiber's cross section. The variation of the effective refractive index $\delta n_R(t)$ and $\delta n_{TR}(t)$, induced by the acoustic responses of $R_{0m}$ and $TR_{mn}$ modes with the linear polarization are shown in Figs. 16(a) and 16(b). The optoacoustic responses of the optical fiber have also been measured experimentally[159–161]. The response frequency is, generally, on the order of several hundreds of MHz, the lifetime of these waves in optical fibers is being ~100 ns[159,161].

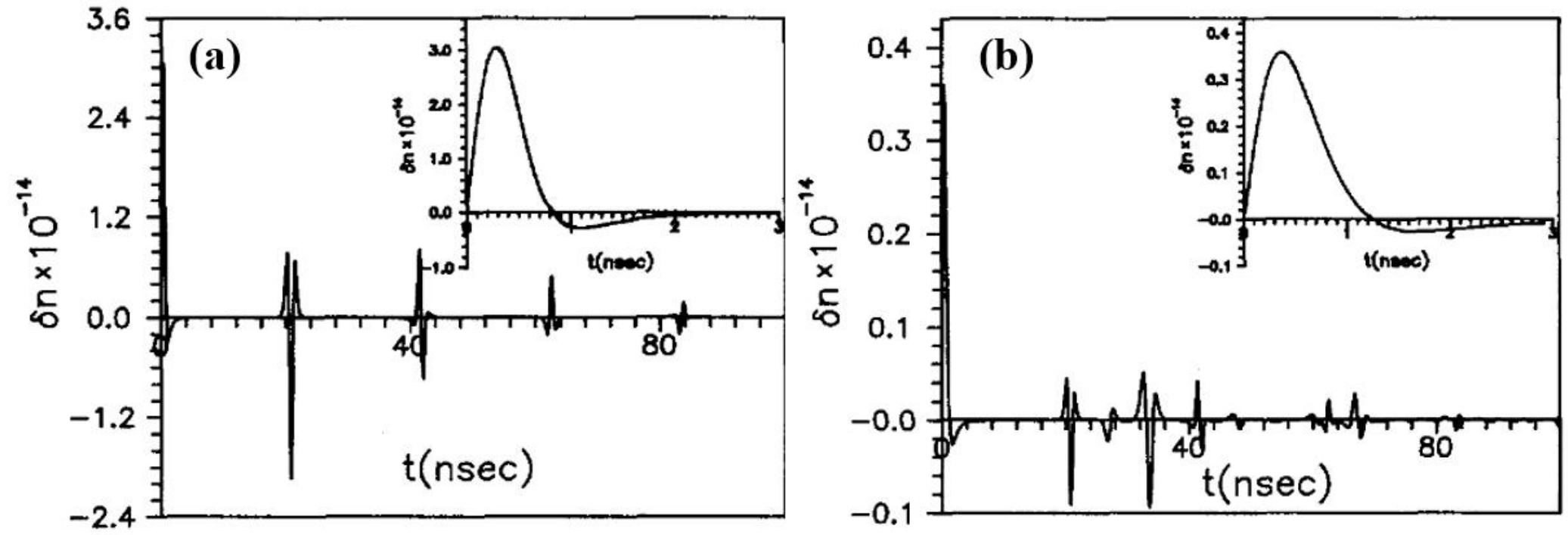

**Fig. 16** Acoustic responses to the soliton for **(a)** $R_{0m}$ modes and **(b)** $TR_{2m}$ ones, as per Ref. [157].

When two solitons propagate along an optical fiber, they increase their separation under the action of the optoacoustic driving force [157]. The acoustic interaction between pulses in passively mode-locked fiber ring lasers is similar to the interaction in long-distance transmission lines. However, the situation is different in the case of several intracavity pulses with a nearly equidistant separation. For simplicity, only temporal positions $t_i$ and the resulting frequency shift $\delta\Omega$ are used to describe a sequence of pulses in the laser cavity [162], leading to the evolution equations:

$$\frac{\mathrm{d}t_i}{\mathrm{d}\xi} = -\frac{(\delta\Omega_i)\lambda^2}{2\pi c}D, \tag{50}$$

$$\frac{\mathrm{d}(\delta\Omega_i)}{\mathrm{d}\xi} = -\frac{\omega}{c}\sum_l \left.\frac{\mathrm{d}(\delta n)}{\mathrm{d}t}\right|_{t_i - t_l} - \beta\delta\Omega_i, \tag{51}$$

$$\frac{\mathrm{d}(\delta\Omega_i)}{\mathrm{d}\xi} = -\frac{\omega}{c}\sum_l \left.\frac{\mathrm{d}(\delta n)}{\mathrm{d}t}\right|_{lT + \tau_i - \tau_l} - \beta\delta\Omega_i, \tag{52}$$

where $\tau_i$ is the deviation of the $i$-th pulse from its position in the unperturbed periodic train. The sum in Eqs. (51) and (52) includes the acoustic response of all the preceding pulses. Then, the pulse evolution is calculated by considering an $N$-pulses train in the laser cavity with period $T$. Analyzing the stability of the soliton train, it converges to the periodic pulse train with equal separations between some pulses, as shown in Fig. 17, which explains the formation condition for the harmonic-mode locking (HML) induced by the optoacoustic effect [159,162]. When the arrangement of the intracavity pulses is close to equidistant, the acoustic resonances are excited to provide gain for some specific multiple integer values of the pulse-repetition rate [130,163]. As a result, the periodic refractive-index modulation, induced by the enhanced acoustic resonance, can provide a trapping potential and stabilize the HML state, when its repetition rate matches the acoustic frequency. The HML fiber lasers were studied in many works, with the aim to generate high-repetition-rate pulses [130,162–165]. However, the dominating resonance frequency is generally below 1 GHz, that cannot meet the requirement of applications to optical communications, imaging and sensing. Then, it was found the photon crystal fiber (PCF) and micro/nano-fibers support higher acoustic resonance frequencies [166–169]. These frequencies can be flexibly controlled by designing fiber parameters. As result, HML fiber lasers operating at ~2 GHz have been reported, being locked by the $R_{01}$ acoustic mode [170].

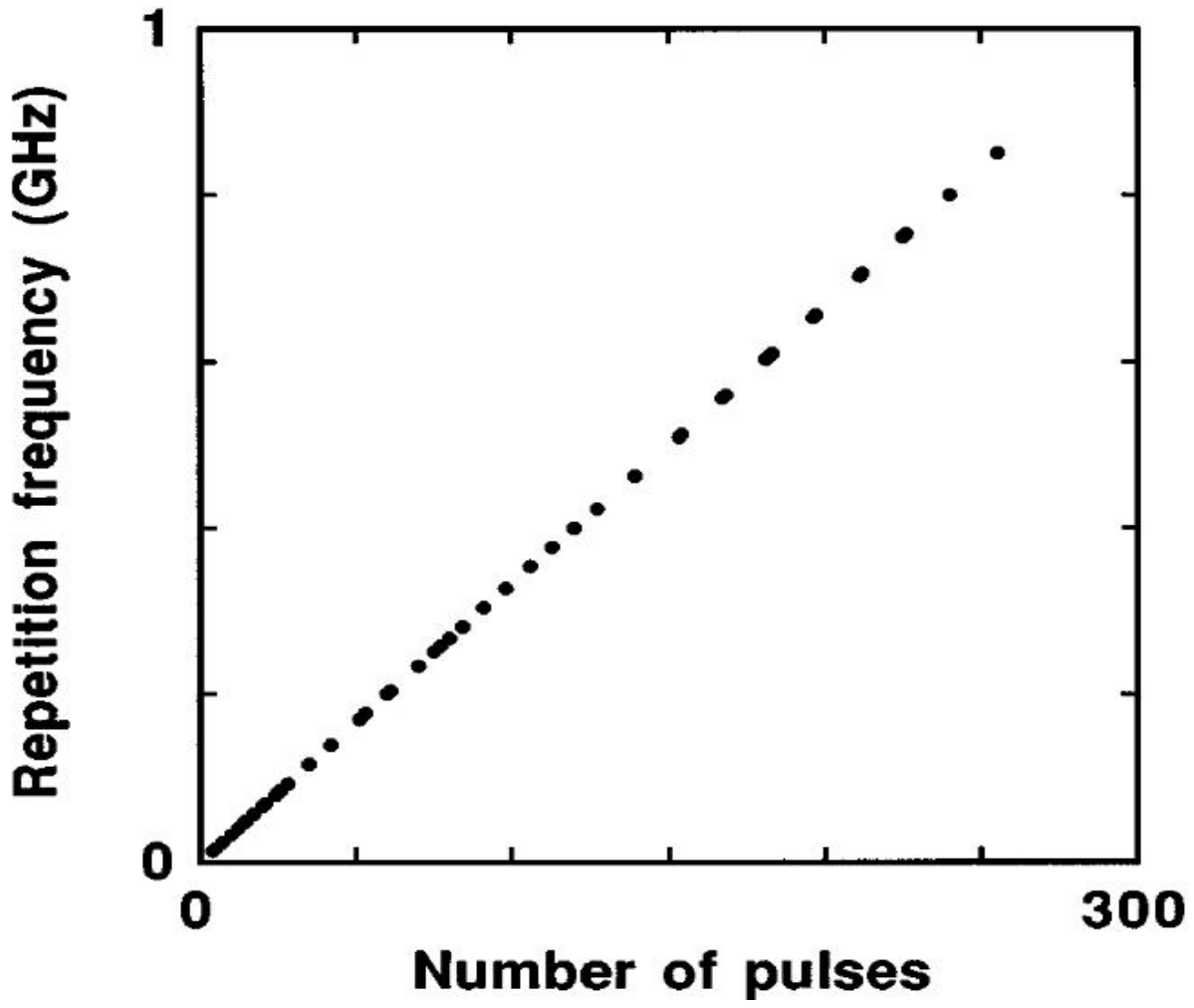

**Fig. 17** Repetition frequencies of stable HML operation as a function of the number of pulses ($N$), as per Ref. [162].

The gain effect in fiber lasers also changes the pulse separation, and even produces a dominant effect onto the pulse evolution dynamics in some cases of the long-range interaction [171,172]. Qualitatively, the population inversion in the gain medium is depleted through the energy transfer to a traversing pulse, as shown in Fig. 18(a) [171,173]. The leading edge of the pulse gets more gain than the trailing edge, producing a force driving the group-velocity drift towards the region of higher gain. Due to the recovery time needed to recover the population distribution of active ions, the next pulse experiences a time-dependent gain, see Fig. 18(b). Thus, the two pulses experience different temporal drifts owing to the different gain.

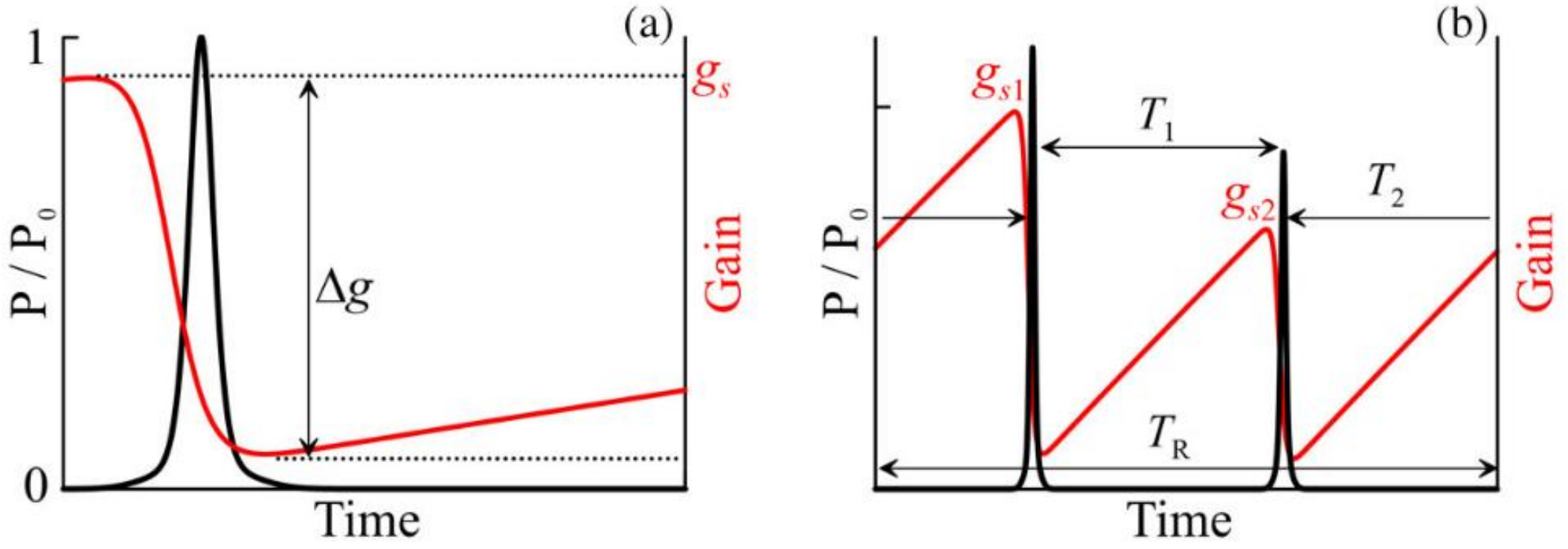

**Fig. 18 (a)** The schematic distribution of the time-dependent gain across the amplified soliton pulse. **(b)** The scheme of the interaction through gain depletion and recovery (GDR) in the cavity with a period $T_R = T_1 + T_2$ ($T_{1,2}$ are time intervals between the pulses). The results were reported in Ref. [173].

The effect of the gain depletion and recovery on the two solitons in Fig. 18(b) is represented by the evolution equations for the time drift[171]:

$$\frac{dT_1}{d\xi} = \frac{1}{2}\left(G_1 - G_2\right)(\xi), \tag{53a}$$

$$\frac{dT_2}{d\xi} = -\frac{1}{2}\left(G_1 - G_2\right)(\xi). \tag{53b}$$

Here, $T_1$ is the temporal separation between the two solitons in one roundtrip $T_R$, akin to the previously defined $\Delta\tau$, while $T_2= T_R- T_1$ is the separation between the second and first solitons in the next roundtrip. In Eqs. (53), $G_1$ and $G_2$ are the variations of the gain for the first and second pulses, respectively, induced by the depletion. By subtracting two equations, one obtains

$$\frac{d}{d\xi}\left(T_1 - T_2\right) = \left(G_1 - G_2\right)(\xi)\,. \tag{54}$$

This equation can describe the soliton- interaction dynamics affected by the change of the soliton separation, which is induced by gain effect. To model the dynamics of the gain recovery, it is assumed that the gain increases linearly when the pulse separation is much larger than the pulse duration $\tau$. The recursion relations between the instantaneous gain (prior to the passage of the pulses) for the two pulses, $g_{s1}$ and $g_{s2}$, can be expressed as

$$g_2(\xi) = g_1(\xi) - G_1(\xi) + \gamma T_1(\xi) \tag{55a}$$

$$g_1(\xi + \Delta\xi) = g_2(\xi) - G_2(\xi) + \gamma T_2(\xi) \tag{55b}$$

Here, $\gamma = T_0/\tau << 1$. The lasers always work near the steady-state condition, so that the energy difference of the pulses is very small. The gain variation can be simplified as $G_i(\xi)= g_i/\alpha$, $i$=1, 2., where $\alpha$ is a coefficient which depends on the pulse energy and magnitude of the total gain. This also implies that the gain is recovered in the course of the roundtrip. Thus, the following equation is obtained,

$$\frac{d}{d\xi}\left(T_1 - T_2\right) = -\frac{\gamma}{2\alpha - 1}\left(T_1 - T_2\right), \tag{56}$$

which gives rise to the solution

$$T_1 - T_2 = \left(T_1 - T_2\right)_0 \exp\left(-\frac{\gamma}{2\alpha - 1}\xi\right) \tag{57}$$

This suggests that the pulse separation is changed by the time-dependent gain, tending to a steady state with the equally spaced pulses, $T_1 = T_2$. The rule also works for the configurations with more pulses, which indicates the existence of the HML operation[171,174].

Actually, the gain-related interaction mechanism is affected by many factors. For example, soliton bunching with the unequal interpulse distances were found if the slow saturable absorption is considered[175], and the effect of dispersive waves, generated by the loss and periodic gain, gives rise to complex evolution of the interacting solitons. The gain effect also plays an important role in some other multi-pulse states, such as the soliton rain, soliton crystal, and soliton bunching with a quasi-periodic structure [127,172,176]. It is difficult to analytically explain the complex interaction mechanism with the help of above model, therefore the CGLE-based numerical simulations were broadly employed to study the gain effect on the soliton interaction [174]. In the simulations, the gain depletion and recovery are modeled by the time-dependent gain coefficient,

$$g\left(\xi, t_1\right) = g_0 \frac{1 - \exp\left[-T_{\mathrm{RT}} / T_{\mathrm{relax}}\right]}{1 - \exp\left[-T_{\mathrm{RT}} / T_{\mathrm{relax}} - Q(\xi) / E_{\mathrm{sat}}\right]}, \tag{58}$$

where $T_{\mathrm{relax}}$ is the population relaxation time of the dopants, $E_{\mathrm{sat}}$ is the saturation energy, $T_{\mathrm{RT}}$ is the roundtrip time, and $Q(\xi)$ is the pulse energy.

It is also found that the loosely-bound SM can be formed with orders-of-magnitude longer time separations, where the gain effect dominates the interaction mechanism[177]. The separation and phase of the loosely-bound SM displays periodic variations induced by the time-dependent gain effect [178]. Due to the different gain for two solitons, bound states of solitons with unequal pulse energy exist too, which can induce the sliding phase evolution [179]. Furthermore, the change of the saturable pump power directly changes the pulse separation. When the modulated pump is employed, the soliton is vibrating accordingly, and the resonant frequency can be found, indicating the intrinsic time scale of the energy exchange, which influences the gain effect [180,181].

XPM is another incoherent interaction mechanism which can connect solitons with different parameters (the frequency, polarization, and spatial mode), making the buildup of multidimensional solitons possible [91,182,183]. Solitons can be bound at the same time position via XPM to overcome the mismatch produced by the birefringence and chromatic and mode dispersion in the fiber, which is generally called the "soliton trapping" effect [2,184]. As the relative phase between mutually trapping solitons changes rapidly, the phase-related difference is averaged, exhibiting phase-independent results. The effect of the XPM-induced soliton-trapping mechanism is discussed above in Section 2.3. The experimental results about multidimensional solitons in fiber lasers are reviewed below in Sections 4 and 5.

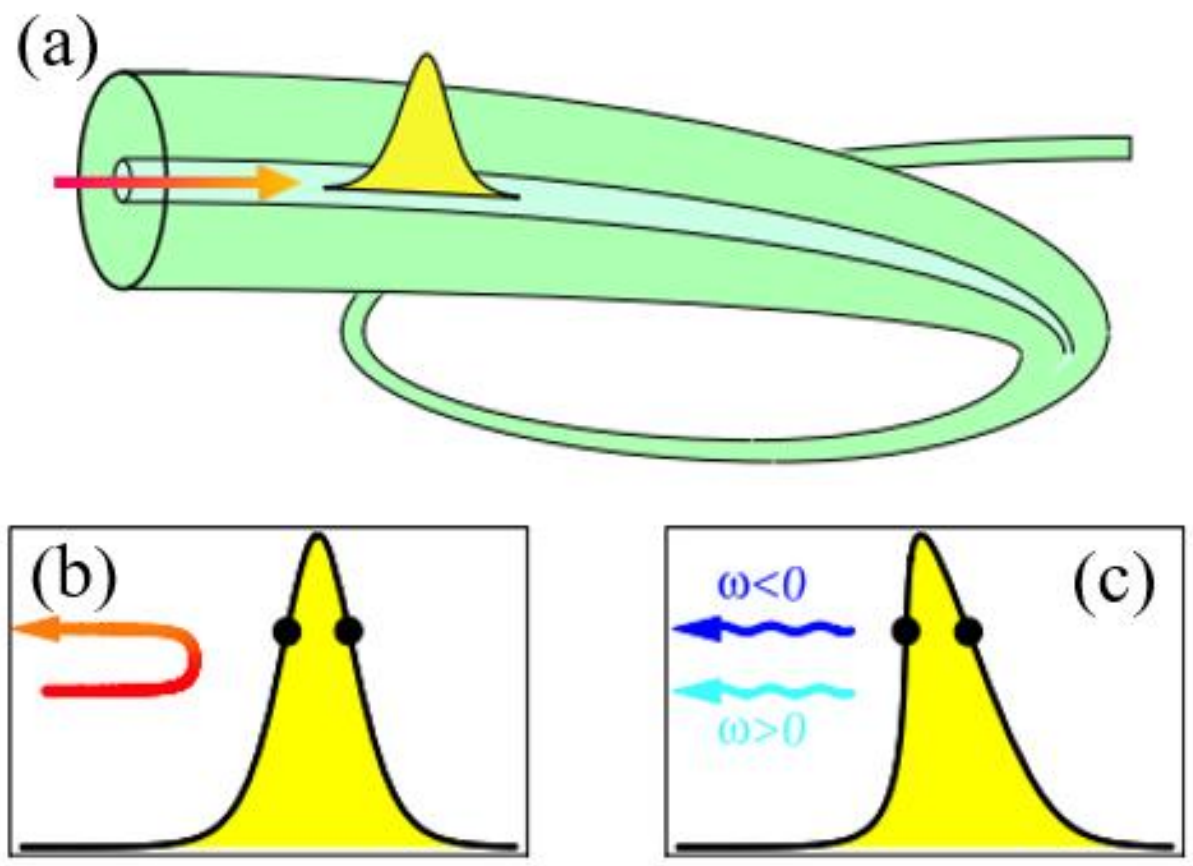

**Fig. 19** Fiber-optical horizons. **(a)** The light pulse in a fiber slows down the infrared probe light. The plots placed below pertain to the reference frame co-moving with the pulses. **(b)** Classical horizons: the probe is slowed down by the pulse until its group velocity matches the pulse speed at the points indicated by black dots. **(c)** Quantum pairs: even without the probe light, the horizon emits photon pairs, as per Ref. [185].

XPM acting between distinct polarization states in optical fibers has sparked considerable research interest. A notable phenomenon is the optical event horizon, which encapsulates the energy exchange between the two pulses. When the propagation speed of one pulse is altered due to the nonlinear effects and dispersion, and it momentarily matches the speed of the other pulse, an "event horizon" is established, *viz*., a white-hole horizon at the back of the pulse, and a black-hole horizon at its front, see Fig. 19[185]. The probe light is blue-shifted at the white hole until the optical dispersion releases it from the horizon. This condition significantly enhances the interaction and energy exchange between the pulses, leading to an irreversible frequency transfer or energy loss. A quintessential example of this phenomenon in fiber lasers is the Cherenkov radiation generated through the energy exchange between solitons and dispersive waves [186,187]. The phenomena can also be observed in the generation of supercontinuum spectra[188–191].

## 4 Bound states of solitons in the time domain

## 4.1 Characterization of two-soliton bound states in the time domain

Soliton bound states represent one of the multiple soliton operational modes in mode-locked fiber lasers. Their formation is intricately linked to multi-pulse operation and the effective interactions among solitons within fiber laser cavity[192]. In the highly nonlinear regime, mode-locked fiber lasers exhibit multi-pulse behavior[124]. The existence of numerous solitons within the cavity is also observed under the soliton energy quantization effect[193,194], which establishes a constraint on the peak power of individual solitons under specific conditions. Substantial advancements have been achieved in uncovering the mechanisms underlying multi-pulse formation, which is crucial for the optimization and design of fiber lasers[195–197].

The field of the two-soliton bound states, alias SM, with individual solitons $E_1(t)$ and $E_2(t)$, carried by a common frequency $\omega_0$, is written as[198,199]:

$$E(t) = \mathrm{Re}\{[E_1(t) + E_2(t)]\exp(i\omega_0 t)\}, \tag{59}$$

where $E_2(t)$ can be replaced by $E_1(t+\tau)\exp(i\phi)$ with temporal separation $\Delta\tau$ and relative phase difference $\Delta\phi$. Applying the Fourier transform to the time-domain electric-field envelope, the pulse separation is mapped into a modulation observed as an interferogram, $I(\omega) = |E(\omega)|^2$, in the frequency domain, and the phase of the fringe pattern at $\omega = \omega_0$ encodes the relative phase of the pulses in the two-solitons:

$$I(\omega) = \left|E(\omega)\right|^2 = 2\left|E_1(\omega - \omega_0)\right|^2 [1 + \cos(\omega\Delta\tau - \omega_0\Delta\tau + \Delta\phi)]. \tag{60}$$

Consequently, the temporal separation results in the phase factor $\exp(i\omega_0\Delta\tau)$, which modulates the spectrum $E(\omega)$ with a fringe period $\Delta v=1/\Delta\tau$. A modulation depth of 100% indicates that the interference produces either total constructive or total destructive interference at varying frequencies, dependent upon the phase difference and frequency offset. The intensity at the center of the spectrum (i.e., at $\omega = \omega_0$) is contingent upon the phase difference $\Delta\phi$, as seen in Fig. 20. When $\Delta\phi = 0$, the two solitons are in phase, leading to strongest constructive interference at the center of the spectrum. For $\Delta\phi = \pi$, the two solitons are out of phase, resulting in destructive interference and null intensity at the center. For $\Delta\phi = \pm\pi/2$, the interference produces a spectral profile with intermediate characteristics. The distinct reliance of the spectral intensity on the phase difference $\Delta\phi$ facilitates the identification of the relative phase between the solitons through the analysis of the optical spectrum. This approach offers an effective means for ascertaining the phase relationship between the bound solitons.

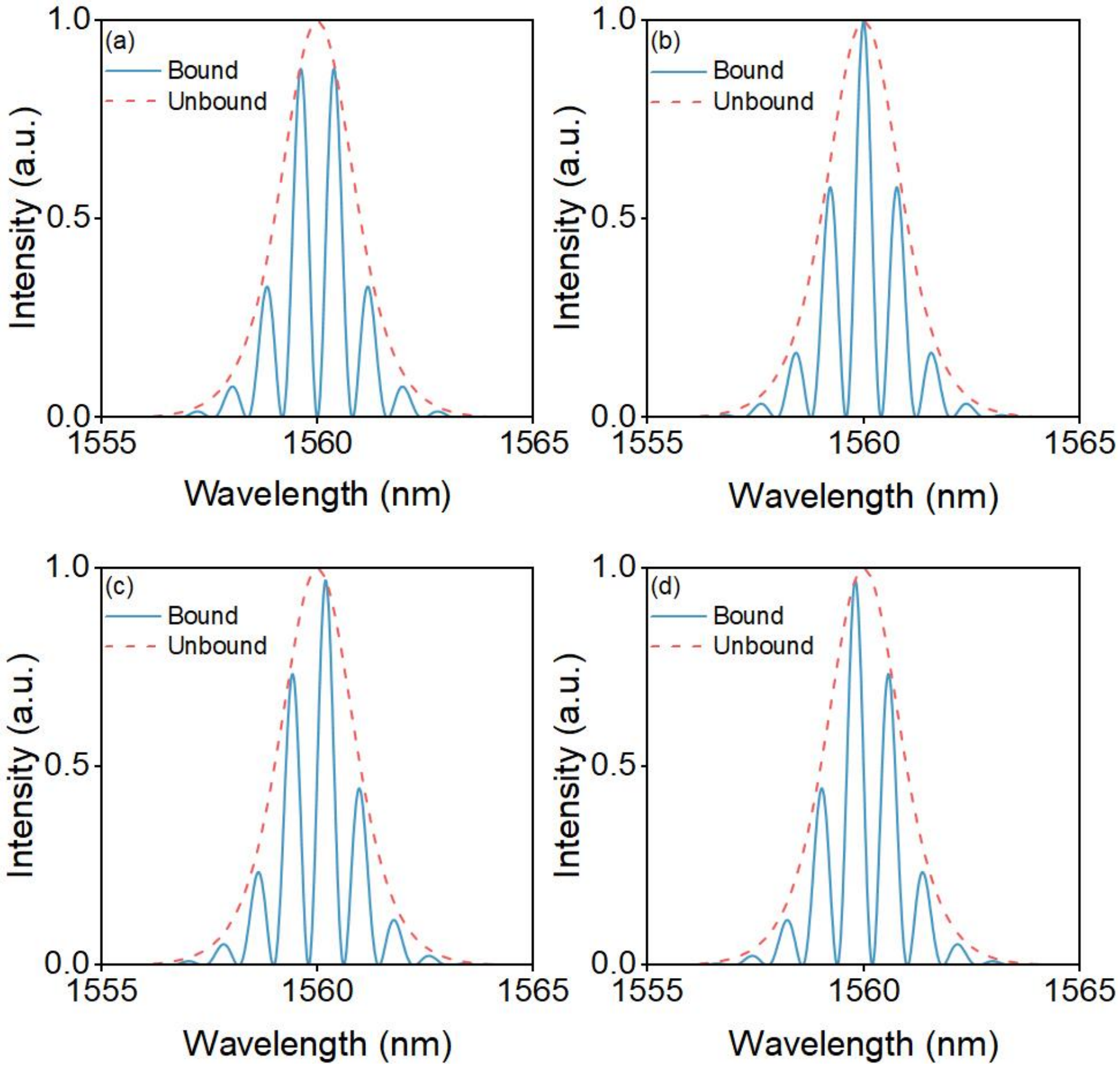

**Fig. 20** Simulated spectra of two-soliton bound states with **(a)** $\Delta\phi = \pi$; **(b)** $\Delta\phi = 0$; **(c)** $\Delta\phi = -\pi/2$; and **(d)** $\Delta\phi = \pi/2$. The blue and red dotted curves represent bound and unbound solitons, respectively. All figures utilize two 0.8-ps sech²-shaped pulses with temporal separation of 6 ps.

In Fig. 20 one can see that out-of-phase and in-phase bound states have symmetrical spectra with respect to the center wavelength, while the bound states with the $\pm\pi/2$ phase difference feature asymmetrical spectra. The asymmetry observed in the latter case results from a directed spectral shift caused by the phase difference. The spectral peaks of the bound states with the $\pm\pi/2$ phase difference approximately adhering a relationship for their intensities, applicable when the modulation period ($\Delta v$) is relatively small, which can be expressed as[123],

$$I_k \approx (I_{k-1} + I_{k+1})/2 \ (k = 2,3,...), \tag{61}$$

where $I_k$ denotes the intensity of the $k$-th maximum peak. This relationship provides insight into the energy distribution across the spectrum of these bound states. It indicates that the intensity ratio between consecutive peaks can be explained by considering the effects of relative phase, modulation depth, and the characteristics of the single-soliton spectrum. For small modulation periods, this behavior offers a means to characterize the spectral structure and identify the nature of the bound states.

To confirm the observation of the stationary bound states and retrieve details of their temporal shape in the experiment, the autocorrelation trace can be used. The autocorrelation function is defined as[200]

$$S_{AC}(t') = \int_{-\infty}^{+\infty} E(t)E(t-t')dt \, . \tag{62}$$

Physically, this formula represents a direct mapping of the experimental process to mathematical operations. The interaction of the two pulse replicas, $E(t)$ and $E(t - t')$, in a nonlinear crystal ($t'$ represents the relative time delay between the replicas) generates a signal proportional to their product, $E_{SHG}(t) \propto E(t)E(t - t')$. This step corresponds to an instantaneous nonlinear response that strictly obeys the causality principle. However, since the response of a standard photodetector is orders of magnitude slower than the pulse duration, the measurement physically performs an integration of this signal over the entire time domain. This is done by assuming an appropriate temporal profile (such as Gaussian or sech$^2$) and measuring the full width at half maximum (FWHM) of the autocorrelation trace. In the case of a "polymer" SM, composed of $n$ identical solitons, which differ solely by phases, the autocorrelation trace shows $(2n-1)$ peaks, according to the Wiener–Khintchine theorem. The intensity of these peaks starts from 1, rises to $n$, and then falls back to 1 in a stepwise manner. According to the spectral intensity of two-soliton bound states, the relation between the spectral intensity and field autocorrelation is

$$S(t') = 2\left|\int_{-\infty}^{+\infty} E_1(t)E_1^*(t-t')dt\right| + \left|\int_{-\infty}^{+\infty} E_1(t)E_1^*(t-\Delta\tau-t')dt\right| + \left|\int_{-\infty}^{+\infty} E_2(t)E_2^*(t+\Delta\tau-t')dt\right|. \quad (63)$$

Thus, the autocorrelation of two-soliton bound states has three peaks, reflecting the overlap and constructive interference of individual solitons. The width of each individual peak in the autocorrelation trace is proportional to the width of a single soliton. We define the three terms in Eq. (63) as

$$S(t') = G_{\text{cent}}(t') + G_{\text{left}}(t') + G_{\text{right}}(t'), \quad (64)$$

where the central-part term $G_{\text{cent}}(t')$ denotes the incoherent superposition of the optical intensity of the pulses, and the terms $G_{\text{left}}(t' = -\Delta\tau)$ and $G_{\text{right}}(t' = \Delta\tau)$ contain the phase information. One rewrites $G_{\text{right}}(t' = \Delta\tau) = P_0\exp(i\Delta\phi)$ and derives the phase $\Delta\phi$ from the imaginary part of $\ln[G_{\text{right}}(t' = \Delta\tau)]$[201]. The phase retrieval process may be symmetrized by employing both $G_{\text{left}}(t' = -\Delta\tau)$ and $G_{\text{right}}(t' = \Delta\tau)$. Figures 21(a) and 21(b) demonstrate that this relationship is valid for both single solitons and SMs, with the peak width reflecting the temporal duration of the individual solitons. The autocorrelation trace reveals the temporal separation between adjacent pulses, as well as their amplitude ratio. This information is related to the modulation period and depth in the frequency spectrum of the SMs. Nonetheless, the autocorrelation approach is incapable of ascertaining the phase relationship between neighboring solitons. The autocorrelation trace is solely responsive to pulse intensity, disregarding phase information. Consequently, although it offers significant temporal information, it lacks the comprehensiveness of optical spectrum recording, which can reveal the phase differences between the bounded solitons.

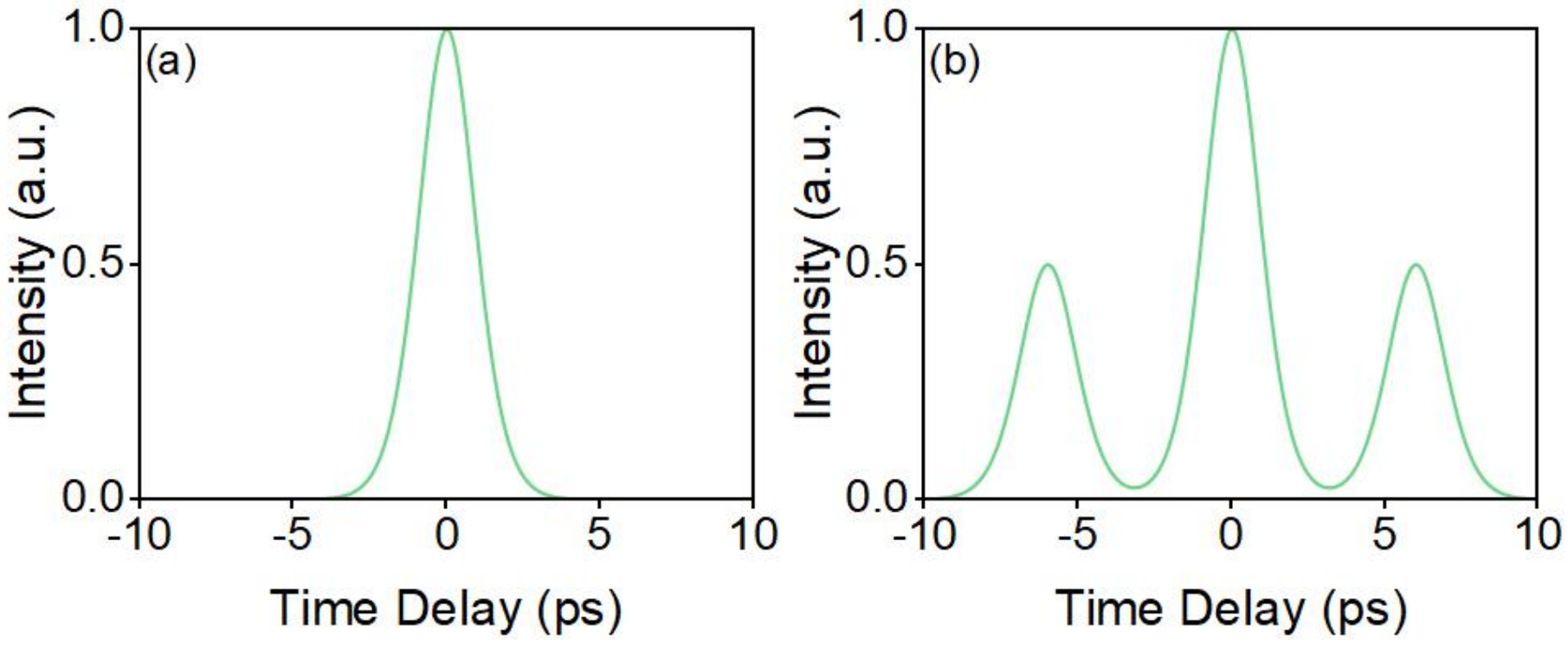

**Fig. 21** Simulated autocorrelation traces of **(a)** a single soliton; **(b)** a two-soliton bound state. Chirp-free

sech-shaped pulses with pulse width of 0.8 ps and temporal separation of 6 ps are used.

### 4.2 Generation of two-soliton bound states

To date, the generation of two-soliton bound states has been investigated in various fiber-laser cavities in simulations and experiments. Naturally, pairs of bound solitons are the most studied structures[123,202,203]. Beyond the simplest bound state, states including more constituents – in particular, soliton triplets [204], and even quartets, composed as pairs of two-soliton bound states, have also been reliably revealed in fiber lasers. Numerous experimental observations of soliton bound states have been documented in the fiber lasers employing the nonlinear polarization evolution (NPE)[205–208], nonlinear amplification loop mirrors (NALM)[209,210], and real saturable absorbers, such as semiconductor saturable absorbers (SESAM)[211], topological insulators[212,213], carbon nanotubes (CNT)[198,214–216], graphene[213,216–218], hybrid no-core fibers, gradient-index multimode fiber structures[219], VSe2/GO nanocomposites[220], and MXenes[221–225] ($MoS_2$, InSb, chromium sulfide, SnTe, etc.). As these saturable absorbers may be polarization-independent, bound states of vector solitons were also reported in fiber lasers[114,226–229]. The presence of both positive- and negative- dispersion fibers in the laser results in diverse properties of bound states, which are contingent upon GVD. This section focuses on the generation and related properties of bound states across various dispersion regions.

#### 4.2.1 The anomalous-GVD regime

Fiber-laser cavities operating in the anomalous-GVD regime facilitate the formation of conventional sech-profiled solitons, which are stable solutions resulting from the interplay between nonlinearity and dispersion[208,230]. This regime has been widely studied due to its potential applications to fiber-optic telecommunications, security, and other fields. Stable multi-soliton states can be established in the cavity when sufficient pump power is provided. The bound states in the case of anomalous GVD were first observed in a passively mode-locked fiber laser in 2001 by Tang et al.[12]. The observed bound states exhibited remarkable stability, characterized by discrete and fixed temporal separations among the solitons. The energy-quantization principle for solitons indicates that higher pump powers allow for the formation of more solitons in the cavity[194]. Through precise adjustments of the polarization states via the rotation of the polarization controller (PC) or wave plate, it is possible to observe bound states of solitons with varying phase differences[198,231,232]. Bound states exhibit a constant phase relationship, leading to high-contrast interference patterns. Figure 22 illustrates an example of a modulated spectrum for a pair of pulses separated by the temporal distance of 20.8 ps, which takes a few minutes to be recorded[232]. Consequently, the two pulses must be phase-locked with precision to preserve the interference pattern's contrast, as any timing jitter at the femtosecond scale would degrade the fringe contrast. The interferometric autocorrelation record, illustrated in Fig. 22(b), further substantiates the evidence of phase locking. The side peaks represent the interferometric cross-correlation of the two pulses. The insets of Fig. 22(b) illustrate that the intensity of side peaks exhibits fringes ranging from 0.5 to 4.5, while the central peak displays fringes extending from 0 to 8.

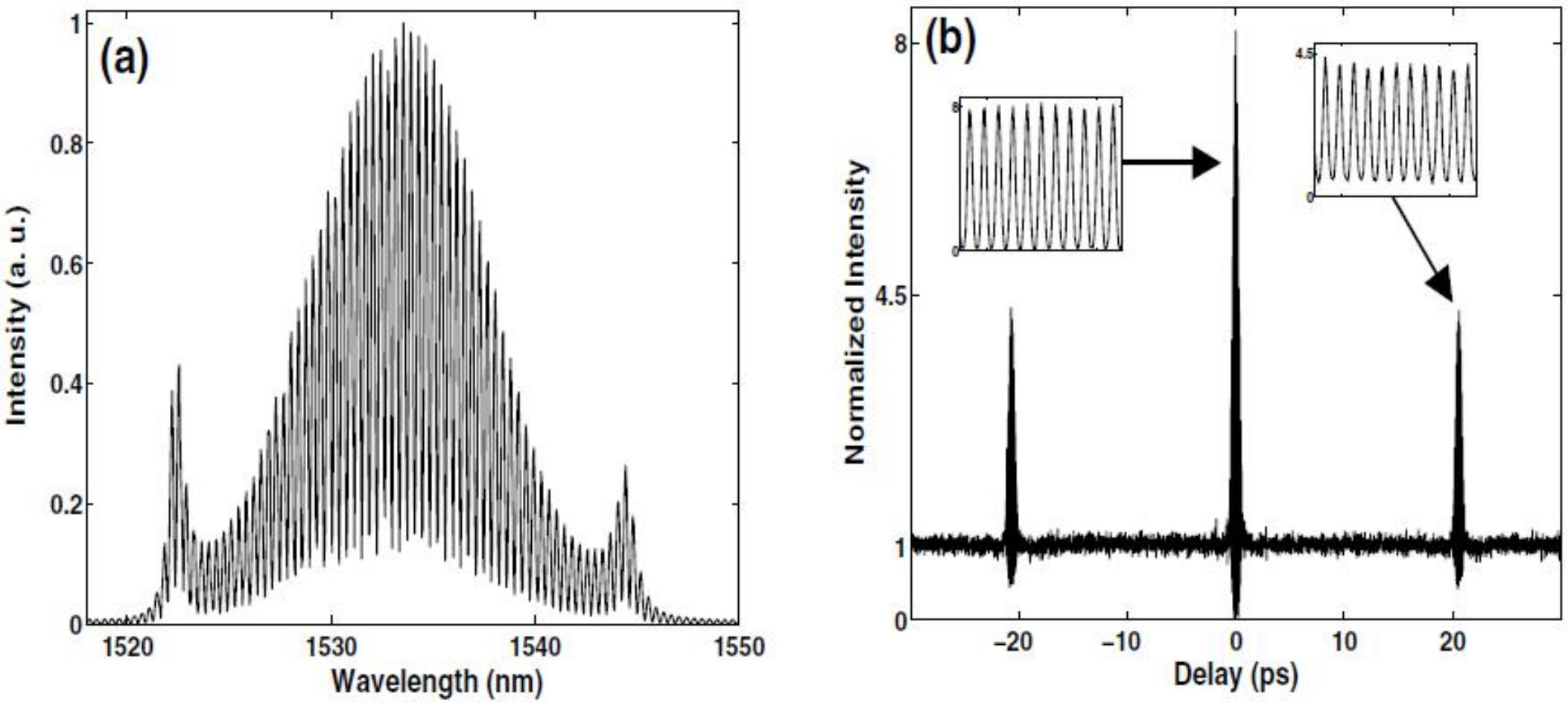

**Fig. 22** Measured result of two-soliton bound states featuring a fixed phase relation, as per Ref. [232]. **(a)** The modulated optical spectrum, and **(b)** the corresponding interferometric autocorrelation trace. The insets in (b) provide enlarged views of the central peak and a lateral one, illustrating interference patterns.

Under suitable pump power and PC orientation, experimental observations of triple-soliton bound states can be achieved, arising from the equilibrium between repulsive and attractive forces within the cavity[233]. With the further increase of the pump power, soliton quartet and quintuplet states were also observed[234]. These observations demonstrate the existence of tightly-bound SM in fiber lasers[235], as evidenced by the fully modulated spectral fringes indicating a fixed phase relationship. These structures are generated when the ratio of pulse separation to pulse width falls within a range of several to several tens. This range is determined by the optical Kerr and Raman effects, which are predominant when the two pulses exhibit partial overlap. Weak and long-range interactions mediated by the CW component can result in loosely bound SM that are more widely spaced[177,236]. The temporal separations between such SMs may reach up to 100 times the pulse width. Standard silica optical fibers demonstrate significant anomalous GVD at approximately 2 μm, resulting in passively mode-locked thulium fiber lasers generally operating within a regime of a large-net-anomalous-GVD. Combined with intense pumping, this regime also facilitates the generation of diverse multi-pulse states, including noise-like ones, pulse bunches, harmonic mode locking[237], and SMs[238]. The comprehension and regulation of these bound states are crucial for the advancement of optical systems and may have significant applications in high-capacity optical communication.

#### 4.2.2 The near-zero dispersion regime

Near the zero-dispersion point, soliton bound states exhibit high sensitivity to external disturbances, such as ambient temperature changes and pulse noise. The dispersion effect may be diminished; however, the nonlinear effect continues to enhance the attraction between solitons, resulting in a unique bound-state structure. Alternating the positive and negative group velocity dispersion segments of the fiber results in the formation of DM solitons, characterized by a more uniform energy distribution and enhanced resistance to disturbances. The bound state in this case is generally more stable, thereby minimizing pulse distortion in long-distance communications. In 1999, a model of antisymmetric DM bound state was proposed, consisting of a pair of solitons with opposite phases. This model was characterized by an odd Hermite-Gauss function[239,240]. Maruta et al. conducted numerical studies on in-phase DM bound states [241]. This research was extended by means of higher-order multi-scale asymptotic analysis[242]. Similar results were obtained for solitons carried by adjacent wavelength-divided channels[240], and for fibers equipped with Bragg gratings used for the dispersion compensation[43]. The first experimental demonstration occurred in 2005[13], followed by the detailed amplitude and phase characterization[243] and explanation of the binding mechanism[244].

In the near-zero dispersion region, pulses do not experience rapid broadening as in the anomalous-GVD regime, nor do they compress as quickly as in the normal-GVD region. Here, nonlinear effects, such as self-interaction and four-wave mixing, dominate the pulse dynamics. Multiple solitons can remain stably bound together through nonlinear interactions, forming SMs. Recently, considerable progress has been made in the study of DM bound states. In 2015, Lin et al. experimentally investigated bound states of multiple DM solitons in a passively mode-locked ytterbium-doped fiber laser with the net positive GVD, utilizing the nonlinear polarization rotation mechanism and precise control provided by waveplates at a specific pump power. They successfully observed multiple bound solitons and complex soliton bound states, whose intensity autocorrelation traces resemble those of single and double bound soliton states, offering new insights into the SM dynamics[245]. In 2017, Wang et al. revealed quantized temporal separations in phase-locked soliton pairs in a DM thulium-doped fiber laser at the 2 μm wavelength. These soliton pairs were formed due to the nonlinear interaction between solitons and dispersive waves, particularly the interaction driven by the two lowest-order Kelly sidebands, which the separation between the pulses in the soliton pairs fixed to a discrete set of values, deepening the understanding of the SM stability[246]. Luo et al. captured a broad range of structures, from tightly-bound DM soliton pairs to loosely-bound ones, and from three-pulse to eleven-pulse SM, as shown in Fig. 23[247]. They also observed DM bound states with unequal intervals between the bound pulses, including (2+1)-type and (2+3+1)-type SMs, greatly enhancing the diversity and complexity of DM bound states. In 2020, Xia et al. studied the dynamics of simple oscillating soliton pairs, which exhibited quasi-periodic and chaotic phase oscillations. They also observed various regimes of oscillatory motion and sliding-phase evolution, introducing oscillation frequencies to validate the complexity of the SM dynamics[248]. More recently, Hu et al. observed the formation of novel "polyatomic" SMs in a single-mode fiber laser with near-zero GVD[125]. These SMs consist of different types of solitons, including structures made of a scalar dark and bright solitons, two scalar dark solitons and one scalar bright one, and a vector dark soliton coupled to a vector bright soliton (referred to as ordinary dark-bright solitons). These findings not only expand the list of SM species, but also open new avenues for studying the dynamics and potential applications of SMs.

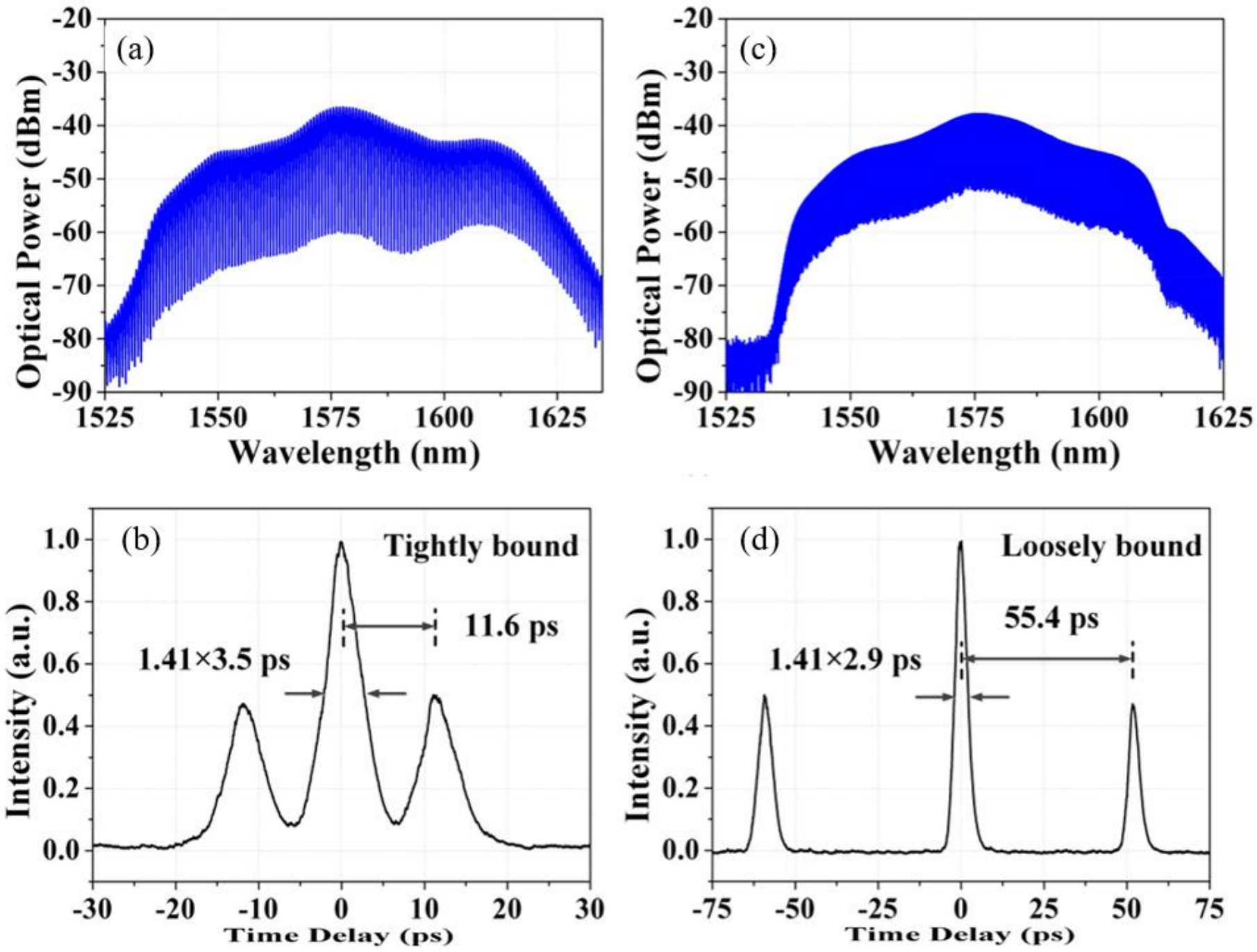

**Fig. 23** The characterization of two different DM soliton pairs, as per Ref. [247]. **(a)** the optical spectrum and **(b)** the autocorrelation trace of the tightly bound DM soliton pairs; **(c)** the optical spectrum and **(d)** the autocorrelation trace of the loosely bound DM soliton pairs.

#### 4.2.3 The normal-GVD regime

In the normal-GVD region, the dispersion effect causes pulse broadening, necessitating a stronger nonlinearity, such as the Kerr effect, to facilitate the attraction between solitons. Stabilizing the bound states in this region is often challenging, necessitating the use of periodic modulation or supplementary components, such as fiber Bragg gratings. The interaction between two dissipative temporal solitons can lead to the establishment of a stable stationary bound state, characterized by a phase difference of $\pm\pi/2$ between the components. Stable soliton pairs were first theoretically predicted in 1997 using the CGLE model[72]. Grelu experimentally confirmed the prediction in 2002[231]. In 2009, Zavyalov et al. numerically demonstrated the existence of discrete robust bound-state solutions in fiber lasers with instantaneous saturable absorbers, operating in the normal-GVD regime. They also discovered the bistability of robust dissipative SMs with independently evolving phases in normal-GVD mode-locked fiber lasers[249]. In 2010, Liu et al. confirmed, in numerical and experimental studies, that dissipative SMs exhibit spectral modulation patterns with a modulation depth of $\approx$ 3 dB (measured as an average). They also observed that these SMs showed vibrating separations and independently evolving phases[250]. In the same year, Ortaç et al. experimentally generated two bounded SMs in an all-polarization-maintaining ytterbium-doped fiber laser with normal GVD. These SMs display independently evolving phases and regular spectral modulation patterns with an average modulation depth of 80%. Numerical modeling confirmed that the limited spectral modulation depth arises from the evolution of phase differences between the pulses[251]. Also in 2019, Wang et al. explored SMs with variable compositions in L-band mode-locked fiber lasers under normal GVD. They extended the SM concept to include pulsating mechanisms and demonstrated, for the first time, that repulsive interactions, caused by energy-exchange-induced intensity distortions, can drive the dissociation of pulsating SMs[252]. Zhou et al. investigated the formation and dissociation dynamics of dissipative SMs in normal-GVD mode-locked fiber lasers. Their theoretical models show that, under varying saturable-absorber transmission characteristics, the background noise can directly form a pair of solitons. Those solitons either evolved into SMs via strong repulsive interactions or dissociated into single solitons, featuring transient annihilation and energy transfer[253].

Adjusting the dispersion conditions in fiber lasers facilitates the generation of gain-guided solitary pulses[254,255], which are characterized by a nonlinear frequency chirp, distinguishing them from self-similar pulses. Zhao et al. observed that gain-guided solitary pulses can form bound states, characterized by a periodic modulation of the optical spectrum, as illustrated in Fig. 24[136]. The observed ratio of pulse separation to width indicates significant interactions that exceed standard soliton interactions. Numerical simulations indicate that these bound solitons maintain stability within the cavity despite significant stretching and compression, demonstrating a complex interaction of pulse dynamics. The findings indicate that the formation of solitons is a fundamental characteristic of mode-locked fiber lasers, demonstrating independence from particular cavity parameters, including dispersion regime, spectral window, and saturable absorbers. The persistence of these bound states for a limited range of experimental parameters suggests that SMs are quantified and sensitive to system characteristics, highlighting the delicate balance in the dynamics of these dissipative systems. The findings highlight the complex relationships among DM, pulse dynamics, and the development of intricate structures, such as SMs in fiber lasers. Manipulating and understanding these interactions facilitates the advancement of laser technologies and their applications in telecommunications, medical devices, and scientific instruments.

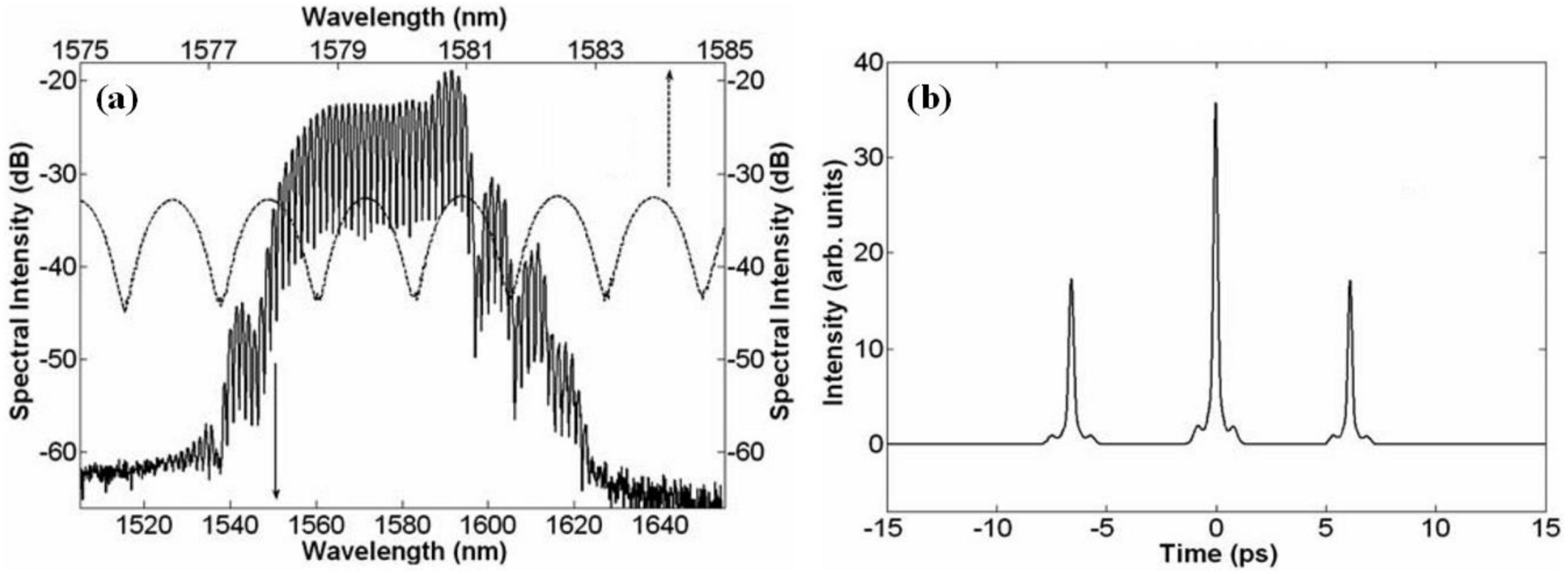

**Fig. 24** Measured result of a bound state of gain-guided solitons, as per Ref. [136]. (**a**) The optical spectrum and **(b)** the autocorrelation trace.

### 4.3 Simulations of two-soliton bound states

The study of bound states originated from the numerical simulations of nonlinear models, particularly NLSE [11] and CGLE [72]. In parallel, various analytical techniques, such as the perturbation theory [11] and the inverse-scattering transform [256], have provided simplified frameworks for studying soliton interactions and the formation of bound states. It is common practice [257] to model a laser as a distributed system, assuming that the pulse shape changes only slightly during each round trip. However, in a real laser, different effects occur at various points in the cavity. As a result, the solution typically experiences substantial changes during each round trip, which may make the distributed models inadequate [258]. Even if this is not the case, such models fail to account for phenomena arising from the periodicity of effects that occur during each round-trip in the laser cavity. The periodicity was incorporated into the model in a simplified form, using the CGLE with periodically varying parameters [258–260].

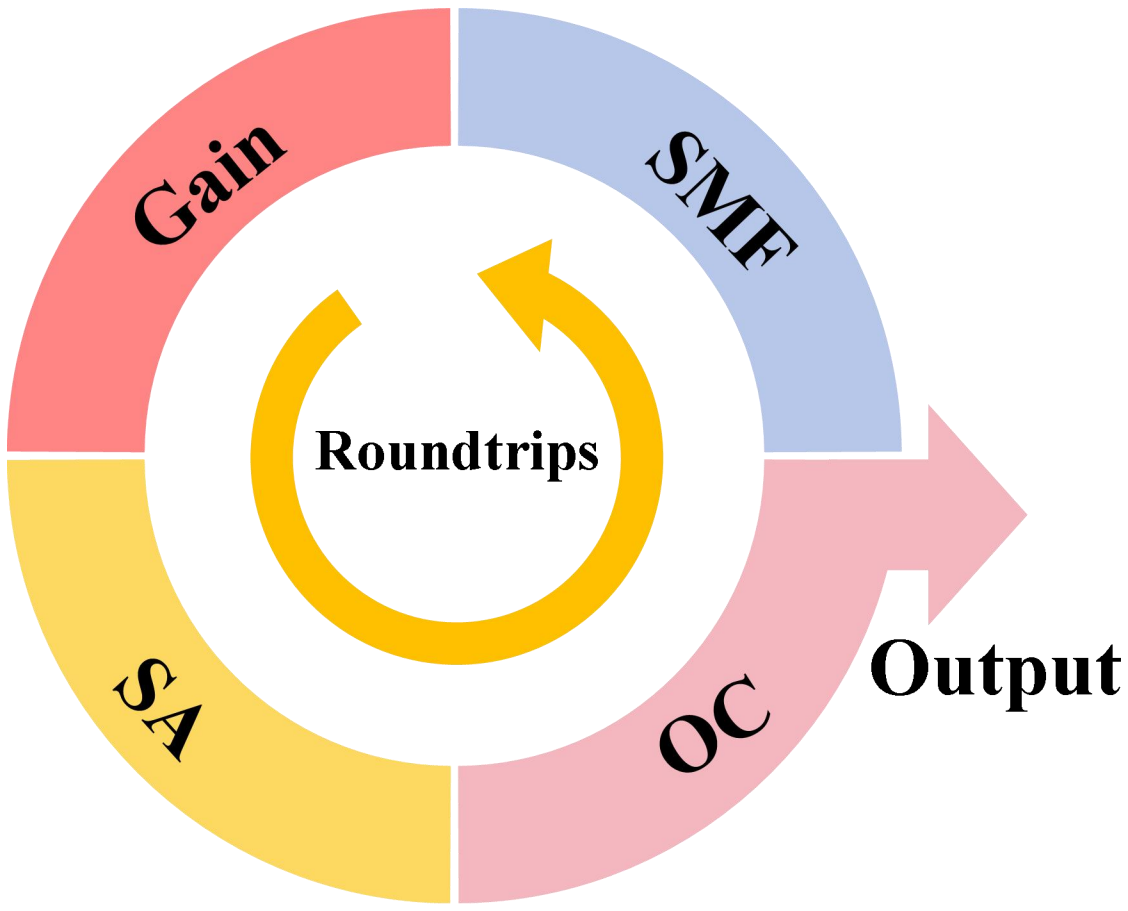

**Fig. 25** The scheme of the numerical model.

For a fiber ring laser, in which experimental results for multi-soliton complexes have been observed, the simulation model is shown in Fig. 25. The system consists of a gain fiber, a saturable absorber (SA), a birefringent single-mode fiber (SMF), and an optical coupler (OC). The wave propagation in the gain fiber is described by the extended NLSE with an additional gain term that includes saturation and limited bandwidth [124,208]:

$$\frac{\partial u}{\partial z}+\frac{i\beta_2}{2}\frac{\partial^2 u}{\partial t^2}=\frac{g-\alpha}{2}u+\frac{g}{2\Omega_g^2}\frac{\partial^2 u}{\partial t^2}+i\gamma\,|u|^2\,u\,, \tag{65}$$

where $t$ and $z$ are the time and propagation distance, $\beta_2$ and $\gamma$ represent the second-order GVD coefficient and cubic nonlinearity of the fiber, respectively, $g$ is the gain coefficient, $\Omega_g$ is the gain spectral bandwidth, and $\alpha$ represents the fiber loss. The gain function $g$ can be modeled as

$$g=g_0(z)\exp\left(-\frac{\int_{-\infty}^{t}|u|^2\,dt}{E_{\mathrm{sat}}}\right), \tag{66}$$

where $g_0$ is the small-signal gain and $E_{\mathrm{sat}}$ the saturation energy. The small-signal gain generally depends on position $z$ in the fiber, because of the pump depletion. However, when the fiber length and pump power are appropriately chosen, this dependence is weak and may be neglected. It was verified that using the first-order approximation, such as

$$g_0(z)=g_{0i}+(g_{0f}-g_{0i})z/L_{\mathrm{gain}}\,, \tag{67}$$

where $g_{0i}$ and $g_{0f}$ are the initial and final small-signal gain, respectively. In Eq. (67), $L_{\mathrm{gain}}$ is the length of the gain fiber, which does not significantly affect the main physical results. For simplicity, one may assume the linear polarization fixed in the gain fiber. The nonlinear birefringence, which is necessary for the pulse formation in this laser, occurs during the propagation in the passive SMF, which is governed by the following system of coupled NLS equations:

$$\begin{aligned}\frac{\partial u_x}{\partial z} &= i\Delta\beta u_x - \delta\frac{\partial u_x}{\partial t} - i\frac{\beta_2}{2}\frac{\partial^2 u_x}{\partial t^2} + i\gamma\left(\left|u_x\right|^2 + \frac{2}{3}\left|u_y\right|^2\right)u_x + \frac{i\gamma}{3}u_y^2 u_x^* + \frac{g-\alpha}{2}u_x + \frac{g}{2\Omega_g^2}\frac{\partial^2 u_x}{\partial t^2},\\ \frac{\partial u_y}{\partial z} &= -i\Delta\beta u_y + \delta\frac{\partial u_y}{\partial t} - i\frac{\beta_2}{2}\frac{\partial^2 u_y}{\partial t^2} + i\gamma\left(\left|u_y\right|^2 + \frac{2}{3}\left|u_x\right|^2\right)u_y + \frac{i\gamma}{3}u_x^2 u_y^* + \frac{g-\alpha}{2}u_y + \frac{g}{2\Omega_g^2}\frac{\partial^2 u_y}{\partial t^2},\end{aligned} \tag{68}$$

where $u_x$ and $u_y$ denote the normalized slowly varying envelopes of two orthogonally polarized pulses in weakly birefringent fibers. Further, $2\Delta\beta=2\pi\Delta n/\lambda$, $2\delta=2\Delta\beta\lambda/2\pi c$, and $\Delta n$ are the wavenumber difference, inverse group- velocity difference, and refractive-index difference between the two polarization modes, respectively. Observing the pulse at a certain point of the cavity in each round-trip, we can see its convergence to some stationary profile.

The laser generates single pulses over a wide range of parameter values. As the gain increases, two or more pulses emerge. The interaction between them, which leads to the formation of multi-soliton complexes, strongly depends on whether the path-averaged GVD is anomalous or normal. Equations (65) and (68) are solved with an arbitrary localized input as the initial condition. When a two-pulse solution appears, its relative phase difference and separation may converge to fixed values. The phase difference for these states is close, but not equal, to $\pi/2$, and gets closer to $\pi/2$ as the separation increases. It is important to note that, by symmetry, the same number of attractors are found around the phase difference of $-\pi/2$. These values of the phase difference result in an asymmetric solution, where only one side of the spectrum is attached as the sideband. Taking this spectral asymmetry into account, the time separation $\rho_N$ between the centers of the two solitons can be calculated as

$$\rho_N = \frac{2N-1}{4\delta\nu_1} \quad , \tag{69}$$

where $\delta\nu_1$ is the frequency displacement of the lowest-order Kelly side-band that may be taken as[150]

$$\delta\nu_1 = \pm\frac{1}{2\pi\tau_p}\sqrt{1+8\frac{Z_0}{L}} \, . \tag{70}$$

where $\tau_p$ is the pulse FWHM, $Z_0=0.5\pi t_p^2/|\beta_2|$ is the actual soliton period, and $L$ is the length of laser cavity. This frequency displacement can also be directly taken from experimental data. For small separations, the stronger interaction between the two solitons causes the phase difference to depart significantly from the $\pi/2$ value.

To achieve the bound state in the simulation, the SA plays a predominant role. As a result, for the generation of tightly and loosely bound SMs, the transmission functions of the SA may differ [261]. In the case of tightly bound states, the transmission function may be considered as[261]

$$T = R_0 + \Delta R \sin^2(0.5\pi \times P / P_0) \, . \tag{71}$$

where $R_0$ is the unsaturable reflectance, $\Delta R$ is the saturable reflectance, $P$ is the pulse instantaneous power, and $P_0$ is the saturation power. While generating a loosely bound state, the transfer function can be modified as[261]

$$T = R_0 + \Delta R\left(1 - \frac{1}{1 + P / P_0}\right), \tag{72}$$

Figure 26(a) presents the simulation results regarding the formation of a tightly bound SM. The figure illustrates the observation of mode locking and soliton splitting. The interaction between the two solitons and their vibrations is observable prior to the establishment of a stationary SM. The simulation results indicate that the final separation between the solitons in the tightly bound soliton molecule is 2 ps. Figure 26(b) presents the simulation results for the formation of a loosely bound SM. Initially, the two solitons attract each other to a minimum separation, subsequently repelling before ultimately forming the final SM. The enlarged depiction of the dashed rectangle in Fig. 26(b) illustrates that the two solitons do not merge; rather, they display a notable intensity difference, as indicated by the grey arrow.

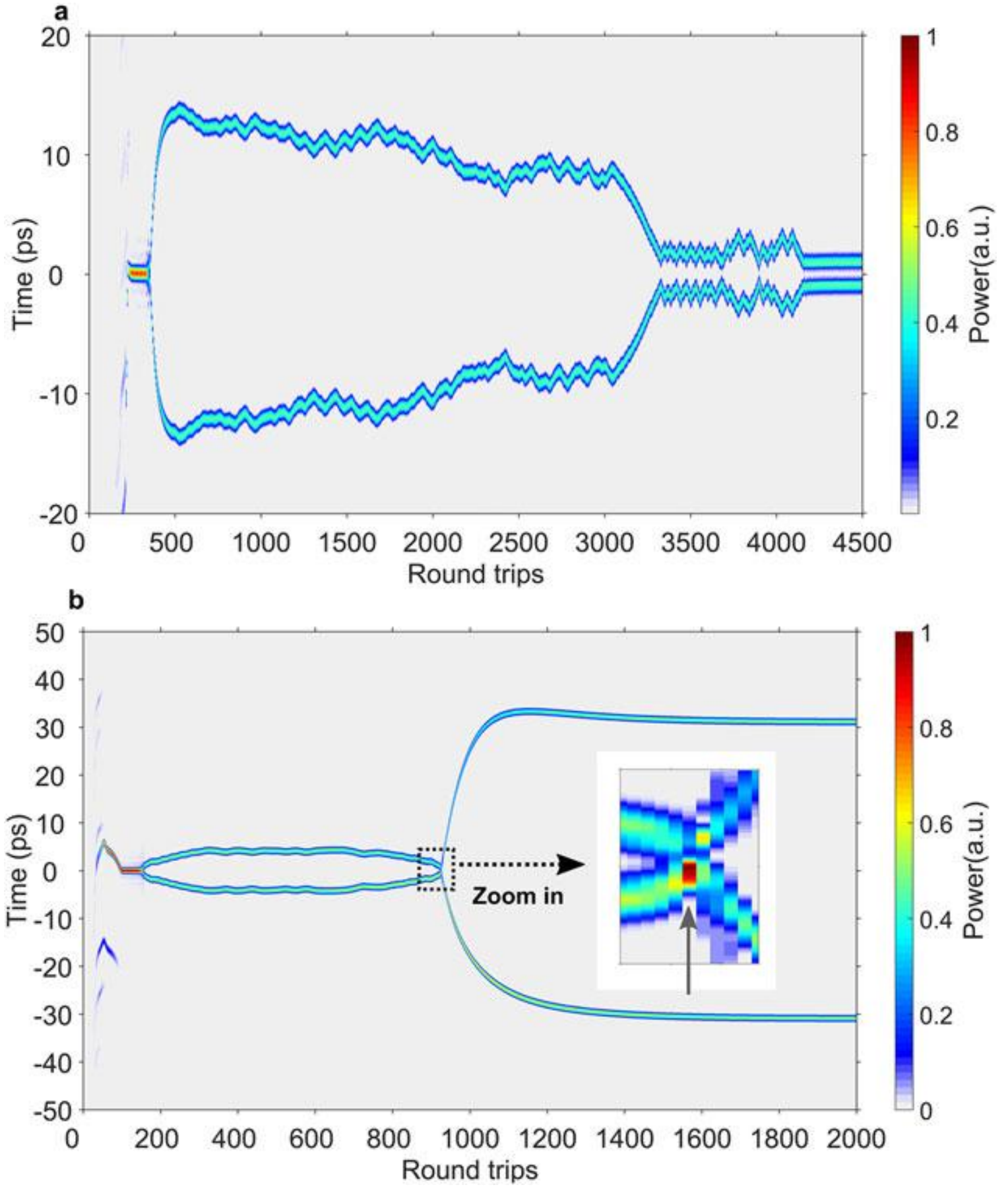

**Fig. 26** Simulation results for the build-up phase of the **(a)** tightly- and **(b)** loosely bound states, as per Ref. [261].

The mechanism of long-range interaction has been observed in mode-locked lasers, resulting from gain depletion effects in the presence of noisy quasi-CW light[127,262]. This interaction arises from the reduction of optical fluctuations caused by gain depletion after a pulse traverses the gain medium. The suppression of fluctuations reduces the temporal jitter of the pulses, thereby influencing the jitter and establishing the pulse drift motion. The noise-mediated interaction (NMI) mechanism exhibits notable parallels with the Casimir effect in quantum electrodynamics, wherein macroscopic objects undergo an effective attraction resulting from the suppression of electromagnetic field fluctuations. In cases where solitons are sufficiently separated, the NMI typically exhibits attractive characteristics[127]. When solitons are sufficiently close, resulting in overlapping pedestals, the interaction may become repulsive. Thus, the NMI may result in the formation of bound states of solitons, with their steady-state separation comparable to the width of the pedestal. The temporal dynamics of dissipative solitons in fiber lasers can be effectively represented by the Haus master equation applicable to mode-locked lasers[263]. To illustrate that the dynamical transition results from the NMI mechanism, a generalized master equation encompassing all essential physical components for observing both the NMI and coherent nonlinear interactions is presented by[264]:

$$\frac{\partial u}{\partial z}=-\frac{\alpha}{2}u-\frac{i\beta_2}{2}\frac{\partial^2 u}{\partial t^2}+\frac{g(u)+\kappa(z,t)}{2}(u+\frac{1}{2\Omega_g^2}\frac{\partial^2 u}{\partial t^2})+(i\gamma+\delta-\sigma|u|^2)|u|^2 u+\eta(z,t), \quad (73)$$

In this context, $\kappa$ represents the gain coefficient, while $\delta$ and $\sigma$ denote the fast-saturable-absorption constants that characterize the bleaching and saturation of the absorber, respectively. Additionally, $\eta(z, t)$ is defined as a complex Gaussian white-noise term, satisfying the relation $\langle\eta(z, t)^* \eta(z', t')\rangle = \zeta\delta(z-z')\, \delta(t-t')$, where $\zeta$ indicates the noise strength. The noise term produces an inhomogeneous quasi-CW noise floor in the waveform, acting as the medium for the NMI mechanism. The amplifier gain is depleted, and the slow recovery of gain results in a nonuniform depleted gain profile $g(u)+\kappa(z,t)$, which indicates that the quasi-CW intensity is also nonuniform. The inhomogeneous gain is described by the equation

$$\kappa(z,t)=-g_d\int_{-t_R/2}^{t}|u(z,t)|^2\,dt+\left(t+\frac{t_R}{2}\right)g_d P_{av}(u), \quad (74)$$

where $g_d$ represents the gain-depletion coefficient. To achieve a zero temporal meaning for $\kappa(z,t)$, one must subtract its meaning following the application of Eq. (74). In Equation (74), the recovery rate is approximated as a constant due to the recovery time of the erbium-doped fiber gain medium being significantly longer than the round-trip time[263].

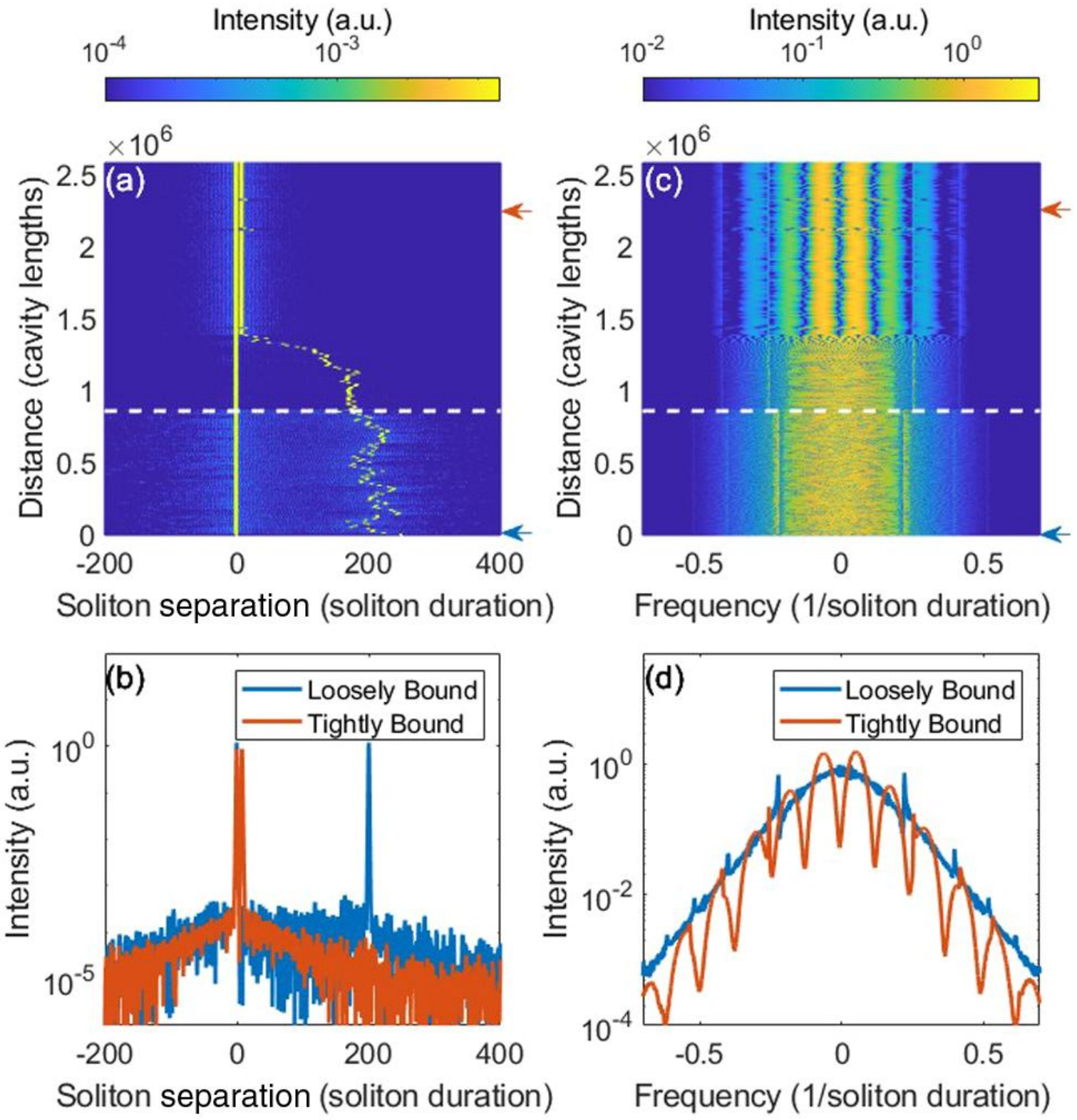

**Fig. 27** The simulated transition dynamics between tightly and loosely bound SMs, see Ref. [264]. **(a)** The simulated temporal dynamics. **(b)** Temporal cross sections of tightly and loosely bound SMs before (the blue arrow) and after (the red arrow) the gain was changed. **(c)** The simulated spectral dynamics. **(d)** Cross sections of the spectrum of tightly and loosely bound SMs.

The Haus model averages the effects of laser elements, including nonlinear propagation, dispersion, gain, and saturable absorption, over one round trip in the laser cavity, resulting in a continuous-variable equation[264]. Figures 27(a) and 27(c) illustrate the simulated temporal and spectral transition dynamics between tightly and loosely bound SMs, respectively. Reducing the unsaturated gain alters the energy carried by the pedestals surrounding each soliton. The pedestal energy decreases by 75% following the sudden reduction of $g_0$ [Fig. 27(b)]. The observed decrease aligns with experimental changes in the Gordon–Kelly sidebands subsequent to a reduction in the pump current. The difference between phase-locked and unlocked states can be identified by comparing the averages of successive spectral traces presented in Fig. 26(d). Stable spectral fringes are observed in a steady state for the low unsaturated gain value, indicating that the solitons are tightly bound (the red curve). In contrast, high values of unsaturated gain result in the absence of stable interference fringes (the blue curve), suggesting that the pulse phases lack coherent locking. In this condition, the solitons are loosely bound, maintaining a constant mean separation.

The interplay between solitons evolution within an active cavity and Raman excitations in the crystal can also be analyzed using a master-equation model based on an adapted complex CGLE[143]. The governing equation is expressed as:

$$\frac{\partial u}{\partial z}=(t_f^2+\frac{i\beta_2}{2})\frac{\partial^2 u}{\partial t^2}+\left[g(t)-\alpha-\frac{\Gamma_{sat}}{1+\sigma_{sat}|u|^2}+i\gamma_0|u|^2\right]u-s\frac{\partial}{\partial t}(|u|^2u)+iu(R_p+iR_a)\int_{-\infty}^{t}d\tau h_R(t-\tau)|u|^2. \quad (75)$$

In this model, $g(t)$ represents the laser gain, $\alpha$ denotes the output coupling loss, $\gamma_0$ corresponds to SPM, $s$ accounts for the self-steepening effect caused by higher-order nonlinear SPM. The effective saturable absorption, due to electronic Kerr-lensing, is included via the modulation depth $\Gamma_{sat}$ and the saturation parameter $\sigma_{sat}$. The spectrally confined laser gain is characterized by the inverse fluorescence bandwidth $t_f$. Additionally, transient laser gain depletion, in the absence of pump recovery on short timescales, is approximated through cumulative gain depletion, given as:

$$g(t) = g_0\left(1-\kappa\int_{-\infty}^{t} d\tau \left|u\right|^2\right), \tag{76}$$

where $\kappa$ is the depletion coefficient and $g_0$ is the linear gain. To model the refractive index perturbations caused by the Raman response, the corresponding temporal waveform is computed by convolving the field intensity $|u|^2$ with the linear response function:

$$h_R(t) = R_1 \exp(-\frac{t}{\tau_1})\sin(2\pi\Omega_1 t) + R_2 \exp(-\frac{t}{\tau_2})\sin(2\pi\Omega_2 t). \tag{77}$$

This function describes the phonon excitation in the impulsive limit. The effective Raman response of the gain medium is governed by two dominant phonon modes at $\Omega_1$ = 16 THz and $\Omega_2$ = 13 THz, with lifetimes $\tau_1$= 2 ps and $\tau_2$= 6 ps, respectively. Relative amplitudes of the phonon modes are $R_1$ = 0.5 and $R_2$ = 1. These phonon modes contribute to the Raman-induced self-phase modulation ($R_p$) and self-amplitude modulation ($R_a$) of the system. The gain depletion effect introduces a notable dynamic: a trailing soliton experiences slightly lower gain compared to a leading soliton. However, due to the similar group velocities of the solitons (as shown in Fig. 28), a constant inter-soliton separation can be maintained (blue curve). On the other hand, self-steepening introduces intensity-dependent group velocity variations, where the leading soliton exhibits a lower group velocity. This causes the solitons to gradually approach each other (orange curve). The full model captures metastable steps in the allowable inter-soliton separations, as shown by the gray curve.

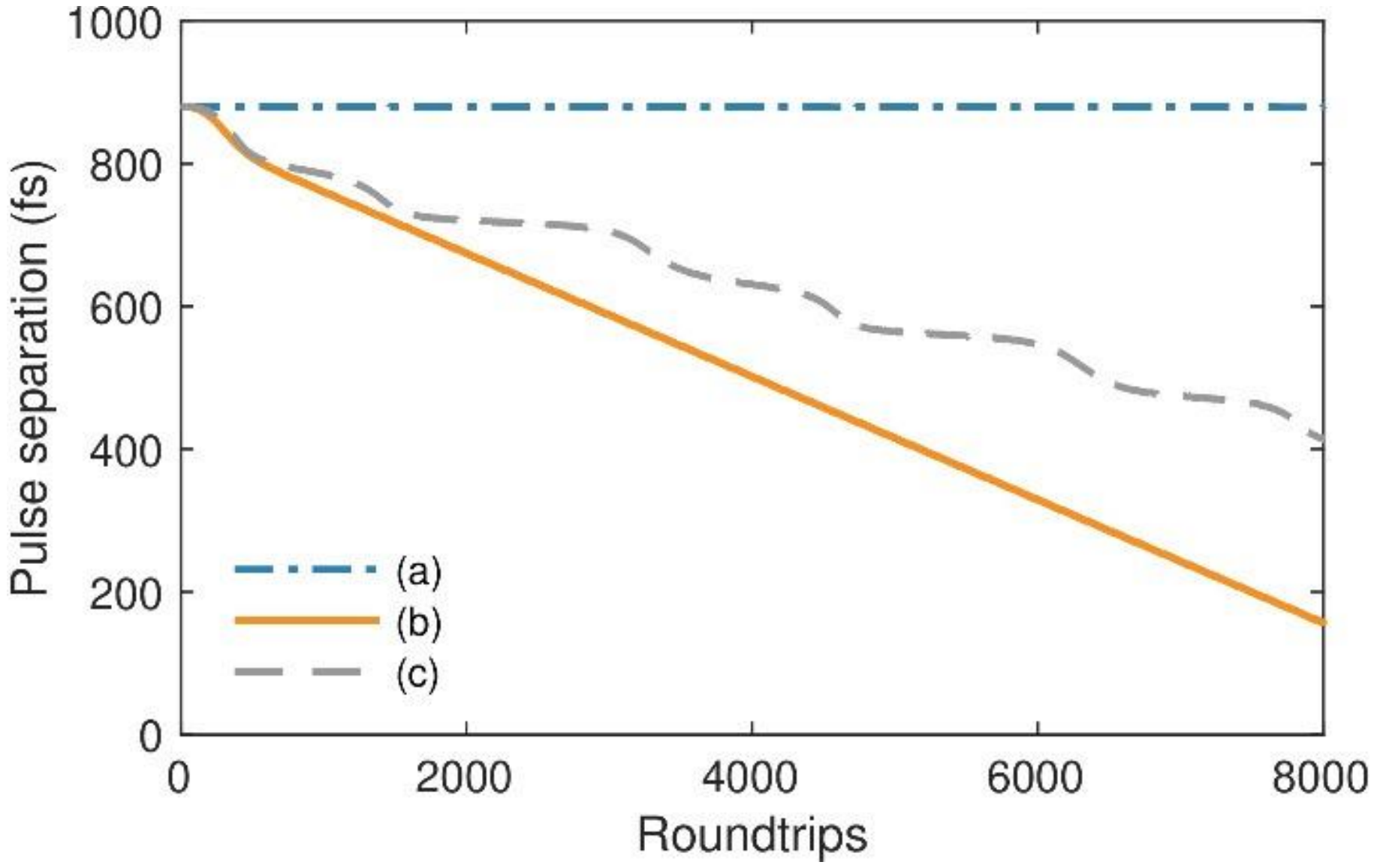

**Fig. 28** The simulated internal dynamics between two solitons when **(a)** $R$ = 0, $s$ = 0, **(b)** $R$ = 0, and **(c)** the full model. The result was reported in Ref. [143].

### 4.4 The generation of multi-soliton bound states

The dynamics of dissipative solitons in fiber cavities are influenced by various interactions and governed by multiple parameters. The spatial distribution of multi-soliton states in the cavity allows for a classification into SMs, soliton bunching, soliton rain, and HML regimes[265]. the temporal separation between bound solitons is typically small, comparable to the pulse width, which renders short-range interactions between the solitons significant[62]. States with solitons positioned at greater distances are termed soliton bunching[266]. If all pulses in a single round trip are uniformly spaced, this configuration is referred to as HML, with the pulse repetition rate being a multiple of the fundamental repetition rate[267]. In both soliton bunching and HML, long-range interactions among the pulses are predominant in the formation of multi-soliton complexes. A notable phenomenon frequently observed in multi-soliton states is the quantization of output parameters [193,194,196,266]. Multi-soliton systems exhibit the notable characteristic that the final output states within a specific cavity can vary when the laser is turned off and subsequently restarted. This indicates that the steady-state configuration of the multi-soliton system is contingent upon the initial conditions of the laser[268], suggesting the presence of multiple attractors within the phase space of the mode-locked fiber laser. The final attractors and evolution paths may vary based on the initial conditions. Alongside steady attractors, there exist attractors characterized by periodic variations or unstable properties. In these instances, multi-soliton states may evolve periodically, such as in the phenomenon of soliton rain, or display instability, resulting in noise-like pulses or rogue waves. Mode-locked lasers can support the coexistence of pulses with varying wavelengths, polarizations, or propagation directions within a single cavity [110,269,270]. The propagation velocities of these pulse trains may vary due to differences in group velocity [270,271], or they may synchronize through internal interactions, resulting in identical propagation velocities[272,273].

### 4.4.1 Bound soliton complexes

Once two-soliton molecules may behave as units in the ultrafast laser cavity, these molecules may also interact with each other. The long-range repulsive interaction resulting from gain-relaxation dynamics promotes the establishment of HML between the two-soliton molecules, a phenomenon that has been confirmed by real-time spectrally resolved data obtained in experiments. It is also interesting to explore the formation of SM complexes, in which multiple two-soliton molecules bind together by intermediate-range attractive forces. Back in 2000, Akhmediev et al. highlighted the significance of multi-soliton complexes [72]. In simple terms, such a complex refers to a self-localized state formed by the nonlinear superposition of multiple fundamental solitons. In optics, they may manifest themselves as a single beam of light or pulse created by the nonlinear superposition of solitons, all traveling at the same velocity. The nonlinear superposition may be either coherent or incoherent, meaning that the phases of the solitons in the bundle are correlated or independent. These solitons may bind together or remain close due to their identical initial velocities. The dynamical behavior of SMs in fiber lasers is highly complex, involving their formation, stabilization, and potential dissociation under specific conditions. The dynamics are influenced not only by interactions between the solitons but also by other factors present in the laser cavity, such as the gain saturation, loss, and dispersion [152]. This has been recently investigated in the case of 2 + 2 optical soliton molecular complexes (SMC) [201]. In this study, two identical soliton-pair molecules, each formed through strong pulse-tail interactions, were linked by a weaker dispersive wave-mediated interaction. Figure 29(a) illustrates the sliding-phase spectral evolution of a 2+2 SMC, revealing two distinct spectral fringe sets. The long-period set (with a period of 1.1 nm) remains stationary, while the short-period one shifts to higher frequencies as the number of round trips increases. This observation highlights the differences between the two types of bonds, which exhibit different dynamical behavior. Two main dynamical regimes can be identified for the SMCs, *viz*., the sliding-phase and oscillating-phase ones. They were observed through the spectral analysis of successive laser outputs and confirmed by numerical simulations. These findings open the way to investigating complex dynamics of intricate multiple-pulse patterns, though the involved multiple time and frequency scales pose new challenges for the experimental characterization. Similarly, soliton *supramolecules* involve the aggregation of multiple soliton molecules to form more complex, larger-scale structures[274].

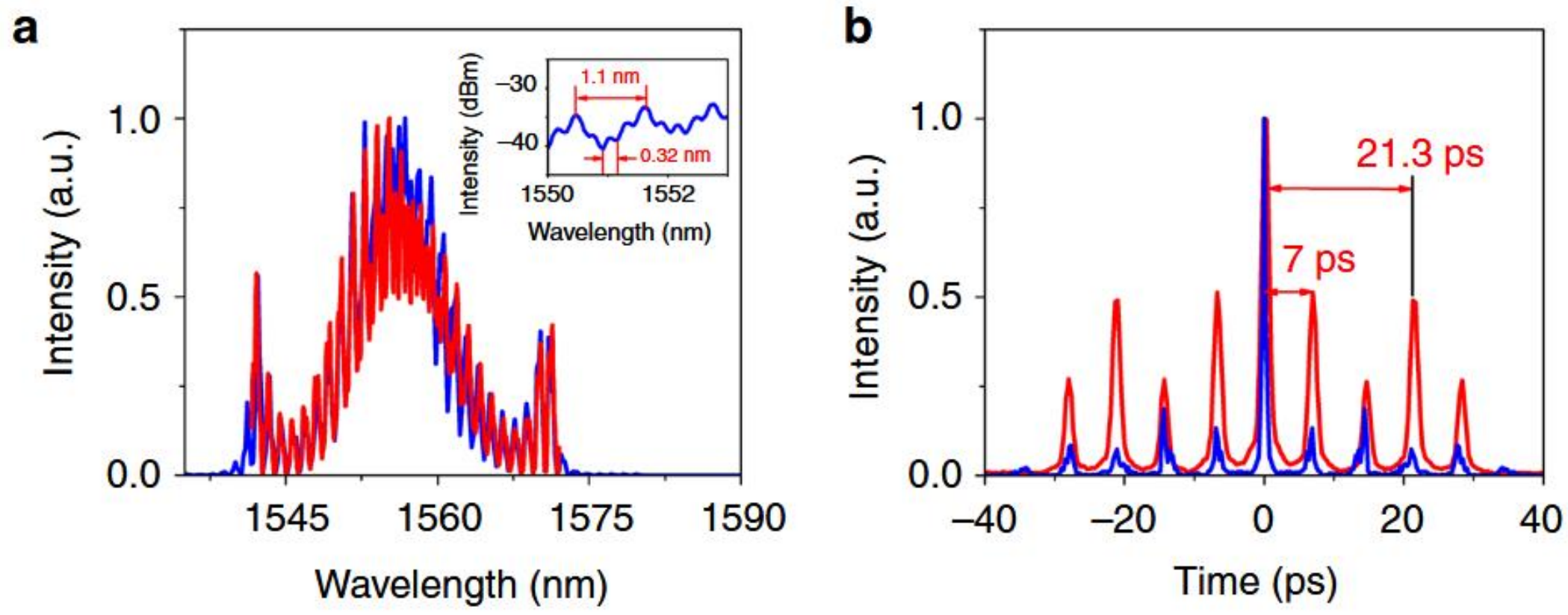

**Fig. 29** Characterization of soliton molecular complexes. **(a)** The optical spectrum, as directly captured by the optical spectrum analyzer (blue curves), and as the mean of 4800 successive single-shot spectra (red curves). The inset displays an enlarged view of the spectral periods. **(b)** The second-order multi-shot-averaged autocorrelation trace (depicted by the red curve) and an example of a single-shot first-order autocorrelation trace (represented by the blue curve) generated using the Fourier transform of the single-shot spectrum. The results were reported in Ref. [201].

### 4.4.2 HML (harmonic mode-locking)

Various dynamical regimes were observed in the early experiments that reported multiple pulsing in fiber lasers. The intra-cavity pulses may exhibit erratic relative motions, stabilize themselves at more or less arbitrary relative positions, or distribute themselves routinely along the cavity. The latter regime, known as HML, appeared to be highly appealing for the purpose of attaining a high output repetition rate of the fiber laser, specifically for telecommunications applications. The weak interaction force's action, which is a result of the gain depletion and recuperation time between consecutive solitons, is the explanation for the trend toward the spontaneous establishment of HML with regular soliton spacing[171]. In the laser medium, this force is dissipative. When a soliton passes through the gain section, it extracts energy and reduces population inversion. The next soliton requires the gain medium to recover the inversion in order to produce the same gain level. The result is a faint repulsive force that exists between the successive vibrations. The force's vulnerability is a result of the significant disparity in the ratio between the cavity transit time (100 ns, for instance) and the excited state longevity of the erbium ion (10 ms). Laser-cavity dynamics may be susceptible to minute effects that do not influence the single-transmission experiments. This is due to the fact that the significant pulse-pattern evolution may occur on a time scale of approximately one second, which is equivalent to the propagation distance of approximately 200,000 km.

In fact, the lasers that operate within the HLM regime do not completely satisfy the requirements of telecom sources. In particular, the pulse-to-pulse separation is subject to a significant amount of temporal jitter, and the feeble interaction previously mentioned is unable to mitigate it. As a result, numerous endeavors have been undertaken since 1994 to achieve HML through the use of an intra-cavity modulator that is externally operated by a reference clock. Active mode-locking is a practical method for the development of telecom sources that are functional. Nevertheless, it eliminates numerous intriguing aspects of cavity dynamics, which are the subject of the current chapter. The cavity repetition rate is restricted by the pace of the driving electronics as a result of the active mode-locking. Nevertheless, encouraging alternatives have been recently suggested. One potential approach is to create a short-cavity passively-mode-locked fiber laser that requires fewer pulses to accomplish the telecom repetition rates[275]. The "passive-active mode-locking" is an alternative approach. It essentially entails the stabilization of pulse rate by utilizing the optical output to produce an electrical signal that is then applied to an internal modulator that functions as a feedback element[276,277]. In the context of active mode-locking, it is also feasible to leverage the higher harmonics of phase-modulated light to increase the pulse repetition rate beyond the electronics' maximum speed [277].

### 4.4.3 Other multi-soliton bound states

Pulse-bunching is another dynamic regime of intra-cavity soliton pulses that is readily observed[278]. Initially, it did not attract much attention, as it was overshadowed by the prospects of HML. Pulse-bunching refers to the ability of several pulses to group into a stable and tightly packed cluster, with a duration much shorter than the cavity round-trip time. In many cases, resolving the bunched structure requires recording an optical autocorrelation trace, as the delay between consecutive solitons may be on the order of a few picoseconds—smaller than the response time of the photodiode. Pulse bunches can comprise up to several tens of equal soliton-like units [279], which may be regularly distributed to form a high-repetition-rate pulse train. The origin of this phenomenon initially appeared mysterious. Several mechanisms have been proposed, including direct soliton-soliton interaction, modulational instability[280], birefringence-induced modulated spectrum formation[30], or optoacoustic interaction. The latter arises from the electrostrictive force created by light pulses[129], which causes a slight radial distortion of the fiber core. This distortion changes the refractive index in the wake of the pulse, exerting a weak force on subsequent pulses. Depending on the time delays between pulses, this force can be either attractive or repulsive. It plays a significant role in the bunching of solitons separated by temporal distances greater than 100 ps, where direct soliton tail interaction becomes negligible[162,267]. Self-stabilization is another aspect of soliton bunch formation when multiple solitons coexist in the cavity. Pulse bunching produced highly stable stationary states, enabling the laser to remain mode-locked with consistent multi-pulse characteristics for several hours without external stabilization of the cavity length or temperature[281]. To add more solitons into the bunch, increased pump power and slight adjustments to the cavity were applied. Fig. 30 shows that the separation between consecutive solitons remained constant at $20.7 \pm 0.1$ ps[281]. As expected for a bunch of identical solitons, the overall shape of the autocorrelation function is triangular.

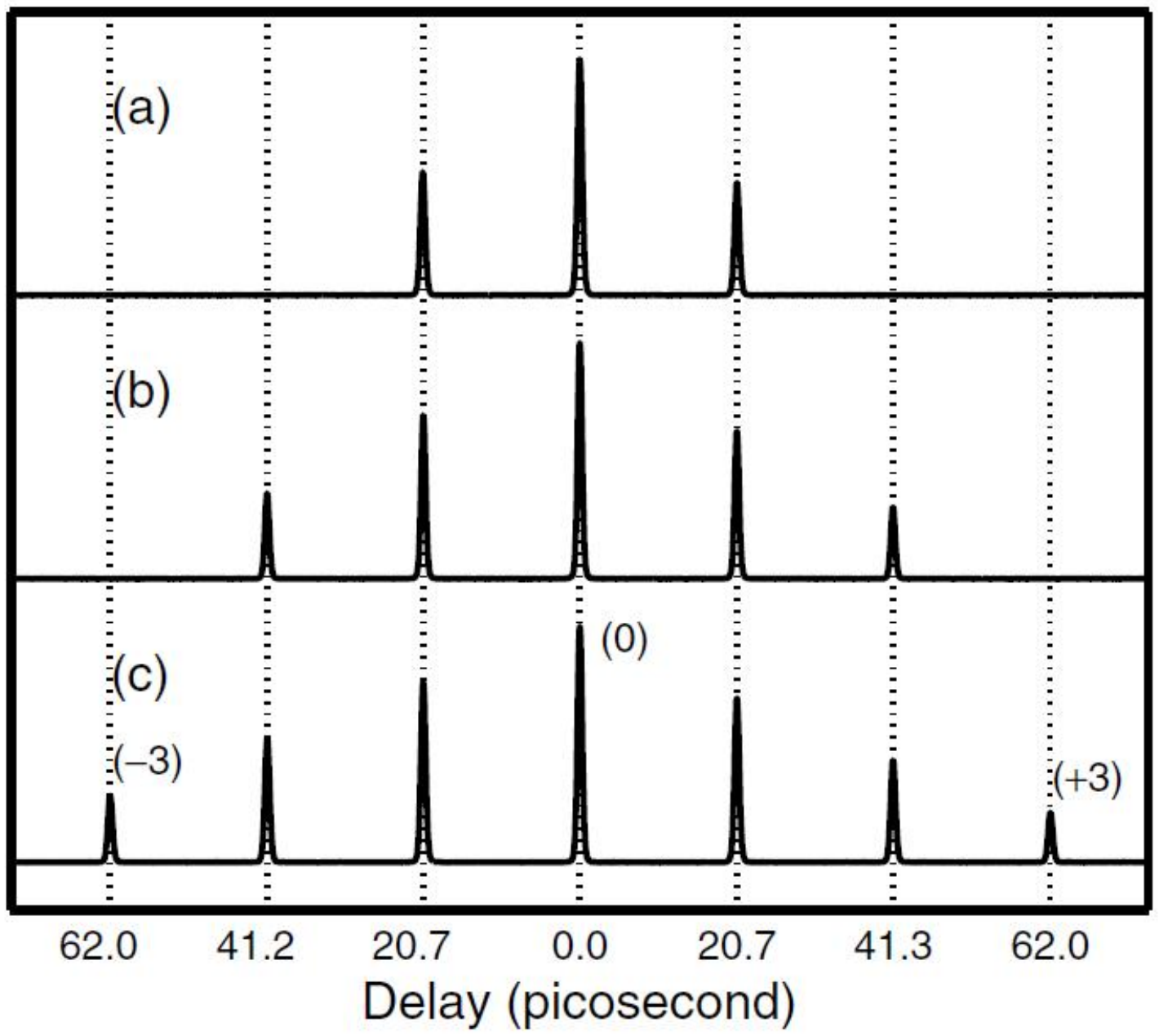

**Fig. 30** Auto-correlation traces showing, as per Ref. [281], the formation of a sequence of uniformly spaced identical solitons, consisting of **(a)** two, **(b)** three, and **(c)** four solitons.

Soliton showers were first observed in an anomalous-GVD Er-doped fiber laser[282,283]. The term "rain" derives from the resemblance to the evaporative cycle of water[176]. Figure 31 illustrates that in a soliton rain, numerous closely spaced dissipative solitons create a condensed phase, while new weak dissipative solitons, referred to as drifting dissipative solitons, spontaneously emerge from the noisy background and drift towards the condensed phase, where they collide and annihilate. The development of many solitons (several tens) and a disparity in the center frequencies between the condensed phase and the floating solitons are two essential prerequisites for the formation of soliton showers. The prerequisites may be readily satisfied in the anomalous-GVD domain; hence, the "rains" were initially seen in anomalous-GVD fiber lasers. Conversely, sustaining multiple-soliton functioning in the normal-GVD domain at moderate pump power is more challenging, since pulses traveling through the normal-GVD medium endure more nonlinear phase changes [284]. Soliton showers were also recorded in a normal-GVD fiber laser by the implementation of a dual-filter configuration[285]. The dual filter comprises a narrowband Gaussian filter and a broadband birefringent filter with multiple transmission peaks. The narrowband filter facilitates the division of the mode-locked pulse into numerous pulses[194,285], whereas the multiple-transmission-peak filter induces the emission of dissipative solitons in the condensed phase and drifting dissipative solitons at varying center wavelengths. They facilitate relative mobility owing to the disparity in group velocity.

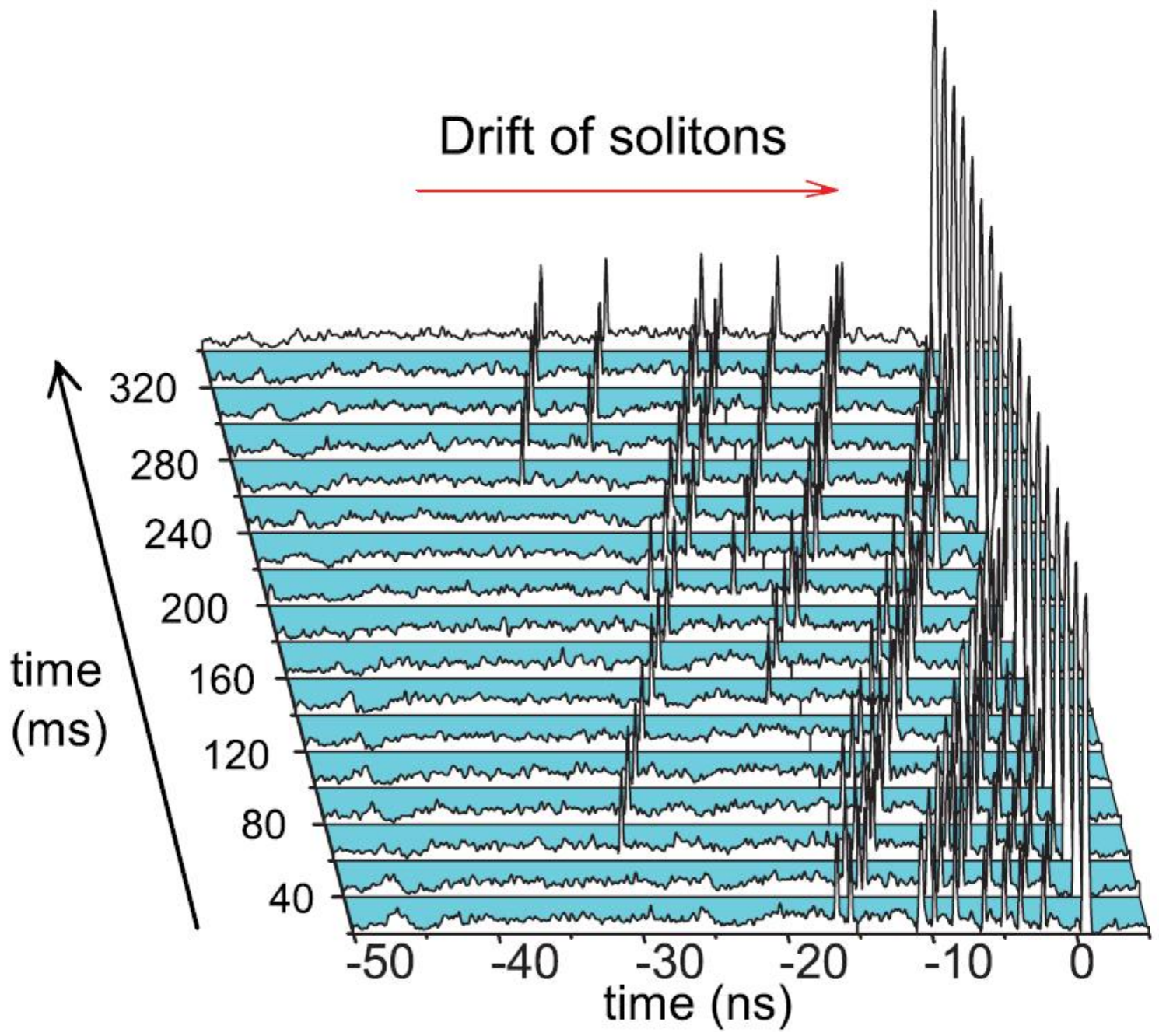

**Fig. 31** Sequential oscilloscope recordings of the output laser intensity illustrate stroboscopically the temporal dynamics of soliton rain. The results were reported in Ref. [176].

### 4.5 The transient dynamics of soliton bound states

#### 4.5.1 The buildup dynamics

In 2005, while studying soliton bound states, Stratmann first coined the term SMs ("soliton molecules") to describe the structure of bound solitons and experimentally observed stable SMs in a passively mode-locked soliton fiber ring laser[13] (note, however, that the same term was earlier (in 2003) applied to theoretically investigated bound spatiotemporal solitons in optics[14]). These SMs were observed in the time domain, revealing their stability in long-distance propagation and suggesting their potential as a new information encoding method to enhance the transmission rate in fiber-optic telecommunication. Subsequently, Grelu elaborated the SM concept, explaining that SMs are stable structures formed by the interaction of multiple solitons in a nonlinear medium[286]. Until 2017, the SM dynamics had not been fully explored. Constrained by the time-domain scanning speed and spectral range of detection devices, research into complex transient phenomena relied predominantly on high-speed oscilloscopes for time-domain monitoring and lacked experimental exploration of the real-time spectral evolution. Then, with the rapid advancement of real-time measurement technologies, Herink et al. employed the time-stretch dispersive Fourier transform (TS-DFT) technique, which maps pulse spectra into temporal waveforms, using time-stretching methods to acquire real-time spectral information[199], as shown in Fig. 32. That work was the first to investigate the evolution of femtosecond SMs in a few cycles of a mode-locked laser cavity, tracking dual-soliton and tri-soliton bound states over hundreds of thousands of consecutive cavity roundtrips, identifying fixed points and periodic as well as aperiodic SM orbits, and revealing various complex transient dynamics mechanisms in mode-locked ultrafast lasers. The TS-DFT technique enabled the experimental observation and detailed documentation of the complex transient dynamics in solitons created in mode-locked fiber lasers, including the soliton accumulation, SM formation, and soliton explosions[202,287,288]. In contrast to conventional spectrometers, the TS-DFT technique offers the additional benefit allowing the dispersive medium to function as an optical amplifier, optimizing the balance between the sensitivity, speed, and resolution through the distributed amplification of optical pulses in dispersive fibers[288–290]. Subsequently, the TS-DFT technique has been extensively employed for observing and analysing complex soliton dynamics, paving the way for the real-time detection of interactions in complex nonlinear systems.

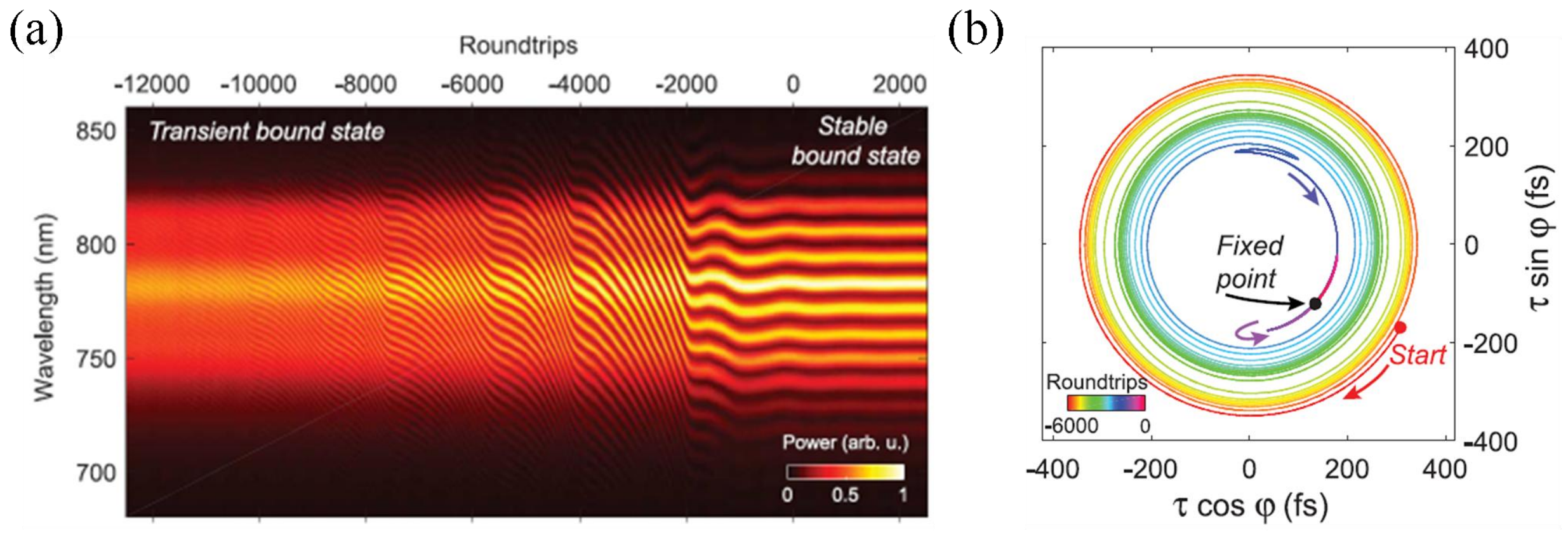

**Fig. 32** Formation Dynamics of a stable SM, as per Ref. [199]. **(a)** Measured real-time interferograms during the formation of a stable soliton bound state. **(b)** Interaction plane of the soliton bound state. The radius and angle represent the momentary pulse separation and relative phase, respectively.

The exploration of the SM dynamics, becoming one of central topics in nonlinear optics, has sparked widespread research interest. In 2022, Du et al. conducted a comprehensive study on the binding mechanism and internal dynamics of asymmetric SMs composed of solitons with unequal short-range interactions[291]. They found that these solitons achieve the balanced separation and energy distribution through periodic energy exchange mediated by overlapping tails. In parallel, Zhang et al. integrated a time-lens system with the TS-DFT technology to elucidate the spectral-temporal characteristics of dissipative SMs in a passively mode-locked erbium-doped fiber laser, revealing various dynamical phenomena, such as the attraction, crawling, period-doubling breathing, and collisions[292]. In 2023, Hu et al. further advanced the study of SMs by demonstrating the formation of a new type of SM, composed of bright solitons in a fiber laser with nearly-zero GVD[125]. This discovery represented a significant advancement in the exploration of the soliton interactions. In 2024, Sun et al., employing a linearly chirped fiber Bragg grating as a dispersive element in the TS-DFT measurements, achieved direct real-time observation of soliton explosions and pulses in a mode-locked fiber laser operating at the wavelength of 2 μm [293]. By adjusting the cavity parameters, they successfully observed a dual-wavelength soliton rain. These research achievements reflect the ongoing development of soliton research in recent years.

With advancements in the real-time detection technology, the complete process of establishing long-lived SMs, akin to single solitons, has been elucidated. By minimizing external disturbances, optimizing the laser system, and employing CNTs as mode-lockers, Liu et al. successfully tracked the formation and evolution of SMs[287]. They demonstrated that the formation dynamics of stable SMs proceed through five distinct stages: the elevated relaxation oscillation (RO) stage, the bouncing dynamics stage, the transient single-pulse stage, the transient bound state, and finally, the establishment of the stable bound state. During the elevated RO stage, it was observed that the pulse evolution follows a pattern in which only the strongest pulse ultimately survives.

Additionally, during the same RO stage, pulses periodically reappear at the same time positions of all laser peaks (referred to as memory capability), though this capability diminishes between different RO stages. The RO stage is a characteristic feature of the transient behavior in fiber lasers. Experimental results indicate that, during the RO stage, multiple sub-nanosecond pulses emerge, with only one dominant pulse evolving into a steady-state mode-locked soliton. Furthermore, Peng et al. documented the formation of tightly bound and loosely spaced solitons[261]. In both cases, the formation phase involves three nonlinear stages: mode-locking, soliton splitting, and soliton interaction. For tightly spaced bound solitons, interactions exhibit a wide range of behaviors in repeated measurements, including attraction, repulsion, collisions, vibrations, and annihilation. For loosely spaced bound solitons, repulsive interactions predominantly prevail. Subsequently, in 2020, Zhou et al. reaffirmed the process of the SM formation: under normal GVD and varying the transmission characteristics of the saturable absorption, a single soliton initially forms from noise in the time domain and subsequently splits ino two solitons, which then interact with each other[253]. Ultimately, stable sub-molecules with well-defined pulse separations are formed.

### 4.5.2 The vibration dynamics

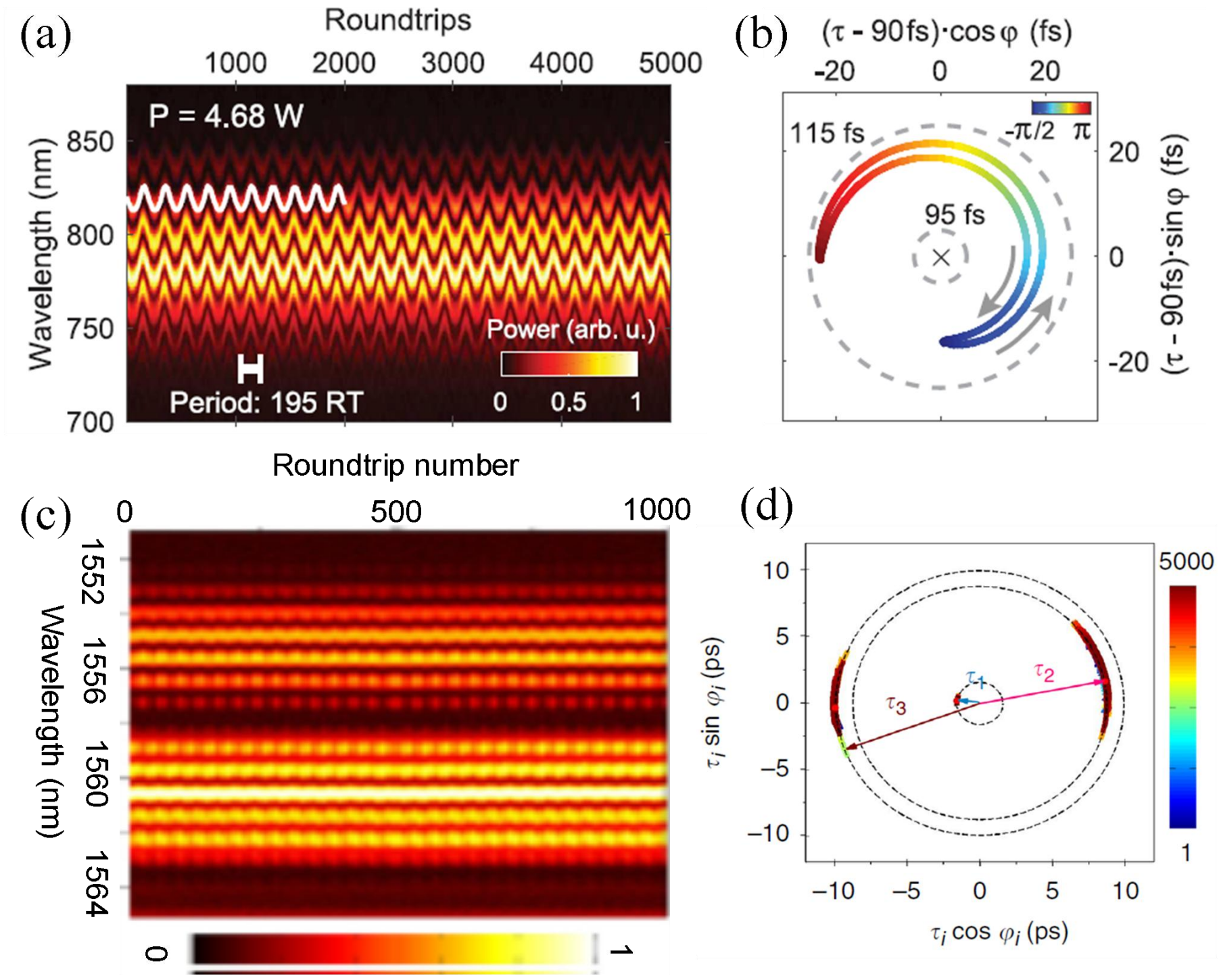

**Fig. 33 (a)** Vibration dynamics of two-soliton bound states, as per Refs. [199]. **(b)** The corresponding interaction plane. **(c)** Oscillating-phase dynamics of 2 + 2 soliton molecular complex, as per Refs. [201]. **(d)** The corresponding interaction plane, with the inter-pulse separation $\tau_i$=1,2,3 and relative phase $\varphi_i$=1,2,3.

SMs show diverse dynamical behaviors, including vibrations of an oscillatory pulse structure[294–296]. This is important for understanding complex wave interactions in nonlinear media. The vibration dynamics refer to the oscillations that arise during the propagation, being caused by external disturbances or internal nonlinear interactions. The vibrational dynamics of SMs in ultrafast lasers were predicted as early as in 2006[77], but real-time experimental validation was initially hindered by measurement limitations. Recently, with the advancement in real-time measurement techniques, the TS-DFT has become a standard tool for observing the soliton dynamics in real time, advancing the study of SM vibrations. The key features of the SM dynamics are the relative temporal separation and phase shift between the bound solitons[297].

In 2017, Herink et al. used TS-DFT to study the internal dynamics of SMs in real time[199], observing both periodic and non-periodic oscillations caused by nonlinear phase delays in the gain medium. These delays caused dynamical changes in the solitons' relative phase and separation. They also suggested that the vibrational behavior might be linked to topological protection, in the framework of which some excitations exhibit topologically protected characteristics. Figures 33(a) and 33(b) illustrate the typical spectral evolution of SMs oscillations, dominated by the oscillations of the relative phase. Meanwhile, Krupa et al. identified two types of the vibrations: one where both the relative phase and temporal distance between the pulses oscillate together (similar to diatomic-molecule vibrations), and another one with only phase slipping[202]. In 2018, Peng et al. introduced a conceptually distinct type of SM, known as intermittent vibrating SMs that transition between vibrating and static states[261]. Wang et al. [201] used TS-DFT to show that SM vibrations synchronize with the periodic appearance of the Kelly sidebands, i.e., emitted dispersion waves[150]. It shows that two soliton pairs can combine to form stable SMC of the (2+2) type, highlighting differences between intramolecular and intermolecular bonds. Figures 33(c) and 33(d) illustrate the spectral evolution of the complex of the 2+2 type, revealing two distinct spectral fringe patterns and emphasizing the differences in the bond dynamics. In 2019, Chen et al. found that SMs can pulsate simultaneously in energy, separation, and phase. In double SMs, the separation and phase pulsate synchronously with each round-trip, showing either regular or irregular behavior[298]. Additionally, they also observed similar dynamics in triple SMs. Regular pulsations occur when the phase and separation remain consistent in each cycle, while irregular pulsations happen when the synchronism condition does not hold. This indicates that the locking or periodic evolution of the pulse separation and phase difference is crucial for achieving regular pulsations in SMs. Besides, symmetric or asymmetric distortions in their profiles and energy exchange processes can also be observed[297]. Peng et al. experimentally confirmed the existence of breathing SMs in normal-GVD mode-locked fiber lasers, overcoming the prior limitation of their confinement to micro-resonator platforms[299]. The following year, more complex dynamics were discovered, In particular, Xia et al. discovered, beyond the simple oscillatory dynamics, two types of oscillating soliton pairs exhibiting quasiperiodic and chaotic phase oscillations, respectively[248]. These complex internal dynamics are governed by subtle energy flows between the components, influenced by the gain dynamics and soliton interactions.

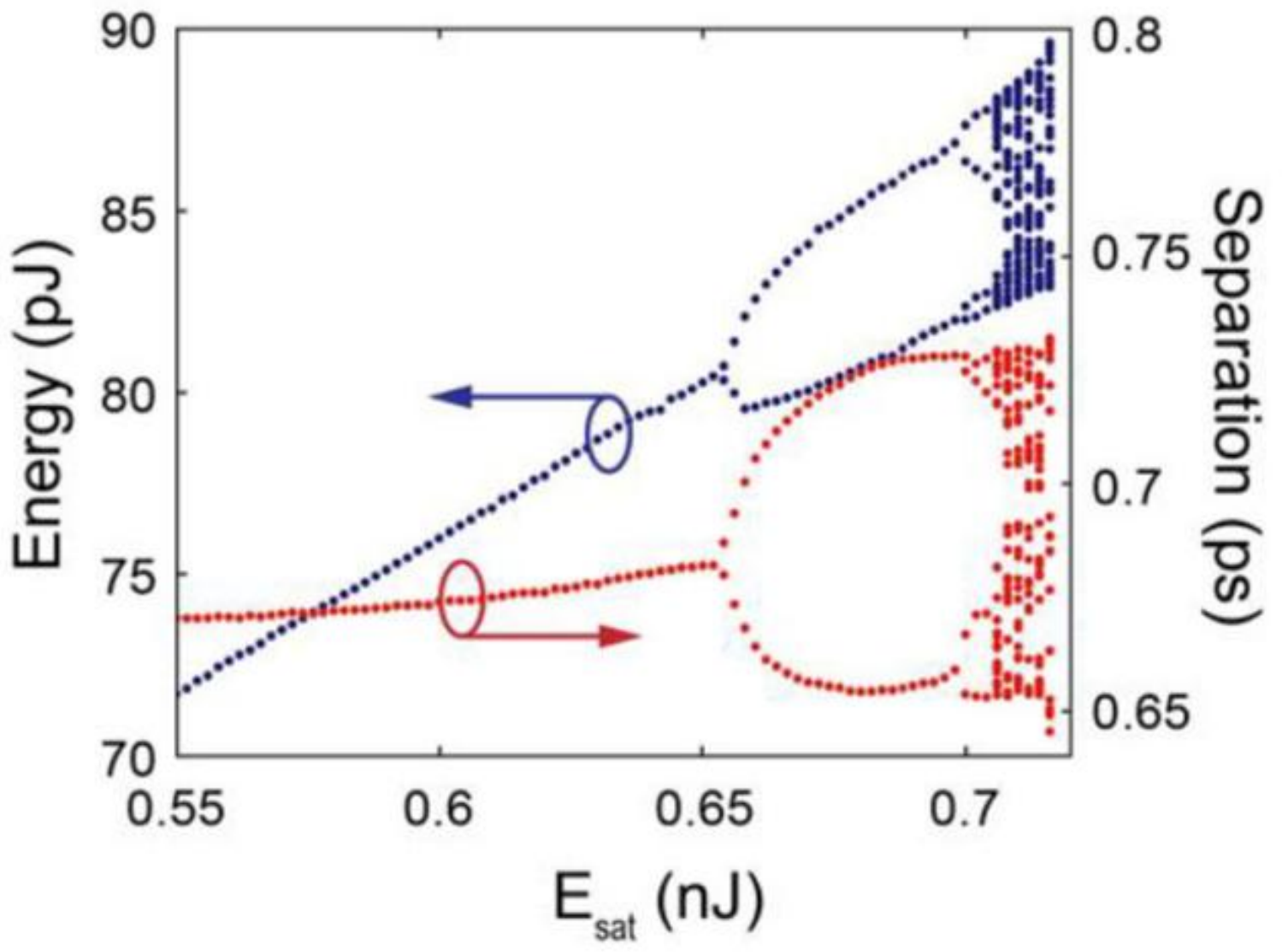

**Fig. 34** The bifurcation diagram of the SM energy and pulse separation, as per Ref. [304].

In 2022, Hamdi et al. successfully applied the super-localization technique to observe the external motion of SMs in a fiber ring-cavity laser in the time domain, showing that oscillating SMs, separated by several nanoseconds, can be synchronized[300]. Additionally, Zou et al. experimentally confirmed the synchronization of internal vibrations of self-excited oscillating SMs by injecting modulation signals into the laser cavity[301]. They scanned for rational multiples of the SM's free vibrational frequency and used high-precision balanced optical cross-correlation to monitor real-time responses. They discovered sequences of locked states where vibrational frequency was controlled by the injected signal, following the Arnold's-tongue model. This demonstrated efficient synchronization across subharmonics, fundamental harmonics, and superharmonics. In 2023, Wu et al. further investigated synchronization in breathing soliton molecules[302], showing that they can be synchronized with the laser cavity's second harmonic frequency. During desynchronization, hysteretic effects were observed between the internal components of the SM and an intermediate state, characterized by subharmonic breathing and anharmonic sidebands. Song et al. in 2023, also observed chaotic SMs in an ultrafast fiber laser using the balanced-optical-cross-correlation method to measure the temporal separation between the bound solitons with sub-femtosecond precision in real time[303]. By injecting modulated optical signals, they demonstrated a possibility to switch between ordered and chaotic states, revealing a bifurcation diagram based on monitoring the local maxima, as shown in Fig. 34, which enables optical control of SMs and paves the way for soliton-based logic gates and chaotic communications. Zou et al. later expanded this research by observing quasi-periodic behavior and transitions to chaos in SMs, showing intrinsic frequency locking and its independence from the laser repetition rates[304]. By means of the simultaneous time-frequency analysis, phase diagram analysis, and calculation of Lyapunov exponents, they were able to demonstrate a connection between the quasi-periodic scenario and chaotic dynamics. These studies offer new insights into the complex dynamics of optical SMs, providing a basis for further research and control.

### 4.6 Manipulations of the bound states in the time domain

To further understand the formation dynamics and explore the potential applications of SMs, it is necessary to study methods allowing precise control of soliton bound state, which may enable more efficient data-transmission schemes and more accurate optical measurements [13,305].

Changing the pump power is the direct and effective technical means to produce and change the bound states. The formation of these states necessarily needs the increase of the pump power with respect to the single- soliton state. The gain effect can effectively affect various characteristics of the bound states, including but not limited to the soliton amplitude, width, and frequency, as well as the evolution of the separation between the bound solitons and phase shift between them[179,180]. However, soliton bound states are influenced by many other factors, making it challenging to produce desired bound states. Kurtz et al. found that a rapid drop of the pump power can trigger a switch of the bound state from one value of the separation to another [180]. The multi-state switching was achieved through a non-perturbative approach, leading to promising applications in ultrafast spectroscopy, information encoding, and logic operations. Meanwhile, the resonant vibration behavior of bound states was detected when applying a modulation to the pump power, identifying the resonance frequency of the energy exchange of solitons. The dynamical response was probed for the harmonic modulation of the pump power, $P(t) = P_0(1 + M_f\cos(2\pi ft))$, where $P_0$ is the average power, $M_f$ the relative modulation, and the frequency $f$ is adiabatically swept within 2 ms from 100 kHz to 1 MHz. The SM primarily responds by oscillations in the relative phase $\Delta\varphi(t)$, as is shown in Fig. 35 by the periodically shifting fringes. Further, a universal formation mechanism was proposed to produce bound states with an arbitrary separation between the solitons. This mechanism makes use of a possibility to use an echo of the leading soliton at some interfaces, to produce nonlinear cross-absorption with the other soliton in SA [143]. The amplitude of the SM pulsations and evolution of the phase shift between the coupled solitons are also strongly affected by the gain effect [306]. It is found that the phase of the bound state shows distinct sliding evolution and velocity at other values of the pump power, which is attributed to the gain depletion and recovery process [171,175].

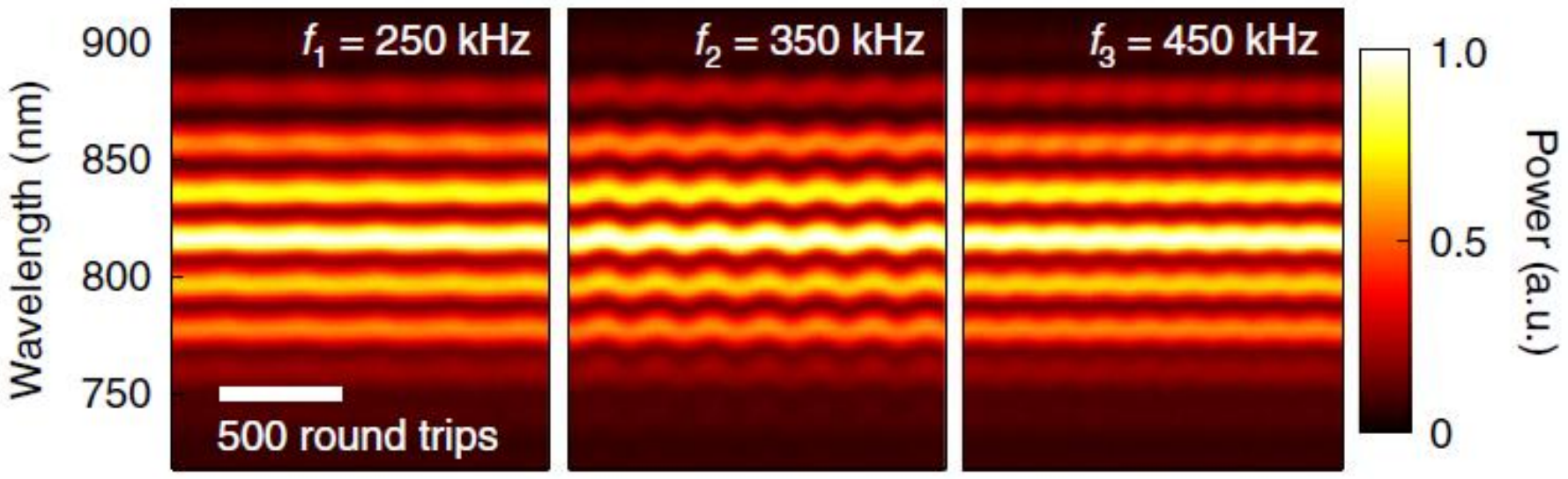

**Fig. 35** Three segments of experimental real-time spectra during the frequency-swept excitation, displaying the resonant behavior at frequency $f_2$ = 350 kHz, as per Ref. [180].

Controlling the bound states by means of an external optical injection is an effective method, as the added laser can not only induce the coherent interaction in the bound state, but also precisely change the soliton energy via the gain-competition effect. Hu et al. demonstrated directional traction and locking of the vibration frequency in dynamical SM states by applying the sinusoidal modulation to an injected continuous optical signal[301]. Thus, they observed and controlled subtle interaction modes inside SMs, further achieving synchronized control of the SM internal dynamics. He et al. employed CW lasers integrated with electro-optic modulators, synchronizing them to the laser cavity's temporal lattice[307]. By precisely timing the address pulses to interact with the solitons in target time slots, it was possible to reversibly and selectively edit each soliton element without disturbing other solitons. Chang et al. used the CW optical injection to perturb the gain dynamics of soliton bound states in the fiber-cavity laser and observed spontaneous collapse and revival of the states[308].

With the above methods, the generation and switching among different soliton bound states have been demonstrated, while the parameters of the bound states cannot be designed on demand. Liu et al. reported the generation of soliton bound state with predetermined parameters in a passively mode-locked laser by controlling the second-order GVD and dispersion losses[309]. In this connection, it is relevant to consider the CGLE-based model,

$$\frac{\partial u}{\partial z} = i\sum_k \frac{\tilde{\beta}_k}{k!}\left(i\frac{\partial}{\partial t}\right)^k u + \frac{g-\alpha}{2}u + i\gamma|u|^2 u + \frac{g}{2\Omega_g^2}\frac{\partial^2 u}{\partial t^2} + \text{H.O.E.}, \tag{78}$$

Here H.O.E stands for terms representing higher-order nonlinear effects, which includes the saturation of the nonlinear gain, saturation of the nonlinear refractive index, etc. The higher-order terms are essential for the normal operation of the system. The dispersion can be written as $\widetilde{\beta_k} = \text{Re}(\widetilde{\beta_k}) + i\text{Im}(\widetilde{\beta_k})$, where the imaginary part is the frequency-dependent loss. For the description of the experiment, we define the dimensionless complex coefficient $\tilde{n}$ to encode the hologram by means of its real and imaginary parts:

$$\text{Im}\left(\tilde{\beta}_{2,\text{holo}}\right) = kI \bullet \text{Im}\left(\tilde{n}\right) \quad , \tag{79}$$

Figure 36(a) shows the average temporal separation extracted from the TS-DFT spectra, where the blue line is the fitting result, produced by the linear model in the following form: TS(ps) = 0.112× Im( $\tilde{n}$ ) + 2.342, 6 ≤ Im( $\tilde{n}$ ) ≤ 28. Close to Im( $\tilde{n}$ ) = 29, a large jump in the TS occurs, and then it does not change conspicuously with the further increase in Im( $\tilde{n}$ ). Here, only the range of Im( $\tilde{n}$ ) between 6 and 28 is considered, where the TS can be tailored at will from 3.014 ps to 5.478 ps, and the SMs can be switched gently without a hysteresis. Such features are essential and useful to perform quaternary encoding and optical switching in SMs. The simulations were performed to verify the physical mechanism. The Fourier transform of these final output spectra produces the autocorrelation function, which is displayed in Fig. 36(b).

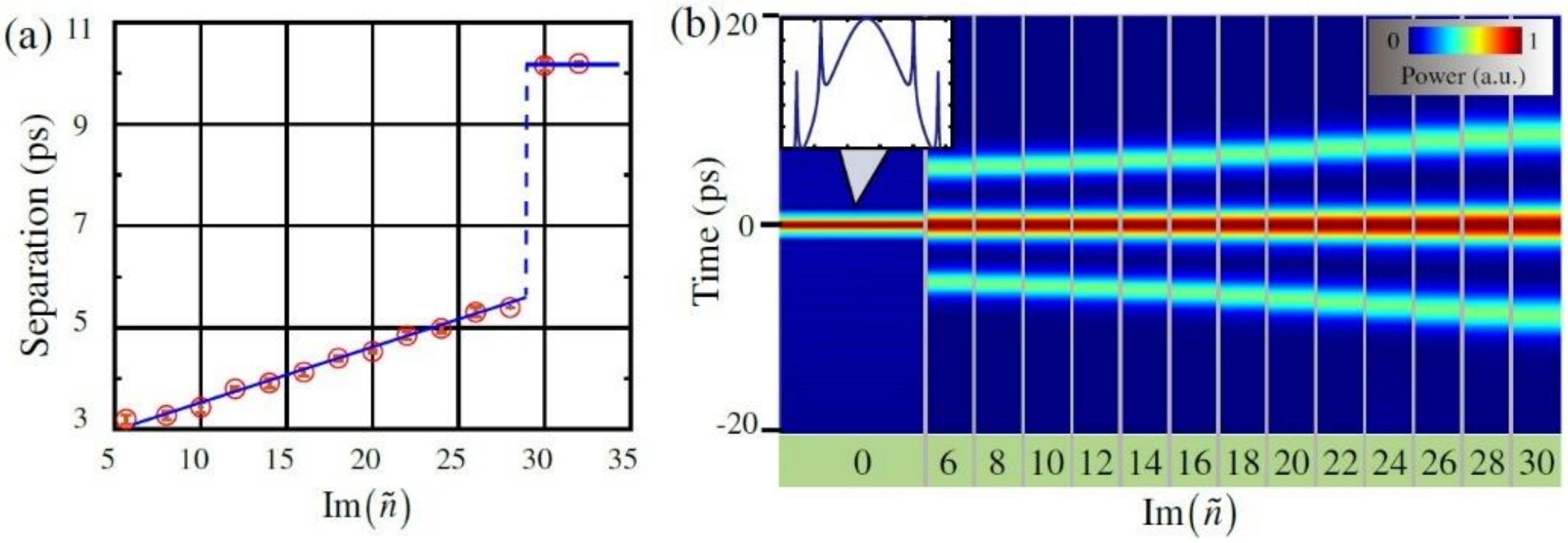

**Fig. 36** (a) Temporal separation between the bound solitons extracted from the spectrum recorded by TS-DFT. (b) Results produced by simulations of the theoretical model based on the CGLE and pulse-shaping equations. The inset in the top left corner is the spectral intensity produced by the simulations in the single-soliton regime, where the bandwidth is 4.37 nm. The results were reported in Ref. [309].

Manipulations of bound states based on an SLM pulse shaper has also been implemented with the intelligence algorithm to produce the soliton bound states with a controllable temporal separation[310]. This is a kind of black-box regulation method in the framework of which the soliton state is achieved by searching in the solution space with the help of a control device, the intelligence algorithms contributing to the fast convergence of the solutions [311] . This technique has been widely utilized to achieve self-start mode-locking for the nonlinear polarization-rotation technique, using electric polarization controllers (EPCs)[312,313]. The adjustment of the controller can optimize the birefringence and change the parameters of saturable absorption and intracavity loss, so that the parameter space of the system can be explored fully to obtain any theoretically designed soliton state. Liu et al. investigated the temporal arrangement and internal dynamics of soliton triplets by means of the EPC-assisted control[306]. They analysed how the polarization adjustment affects the internal dynamical characteristics of the triplets, uncovering the polarization control mechanism for the three-pulse bound states. Furthermore, by combining the DFT technique with machine learning algorithms for controlling EPC, breathing SMs can also be intelligently controlled [314].

## 5. Bound states of solitons in the frequency domain

### 5.1 Characterization of bound states in the frequency domain

As single-mode fibers exhibit slight birefringence resulting from the strain and bend, the inclusion of polarization-sensitive components in fiber lasers can give rise to Lyot filters. Moreover, due to the effect of inhomogeneous broadening, the gain coefficient of an active fiber changes with the wavelength, resulting in an uneven gain spectrum[315]. By leveraging the birefringence-induced filtering and the uneven gain spectrum, multi-wavelength mode-locked solitons have been demonstrated in ytterbium-, erbium-, and thulium-doped fiber lasers[316–319]. In the regime of anomalous GVD, solitons at different wavelengths exhibit characteristic spectral sidebands and demonstrate the chirp-free property. Affected by GVD, such multi-wavelength solitons have different group velocities, and thus collide periodically in the laser cavity[320]. However, due to the particle-like nature of solitons, the pulses quickly recover to the initial states after collisions. In the normal-GVD regime, the dissipative soliton is strongly chirped because of the SPM effect, giving rise to the pulse duration ranging from several to tens of picoseconds[321]. Although the interaction time is much longer than that of the chirp-free solitons, the dissipative solitons can also recover into the initial state after collisions[322,323]. The bound states of solitons in the frequency domain are composed of pulses with distinct wavelengths that are temporally overlapped. The formation of this type of bound-state solitons relies on three factors: first, the implementation of multi-wavelength mode-locking is crucial, typically achieved through the utilization of a spectral filter or an uneven gain spectrum. Second, the group-delay difference among pulses at different wavelengths should be minimized to achieve synchronization between these pulses. Furthermore, it is imperative to ensure the precise overlapping of multi-wavelength pulses with the aid of saturable absorption and cross-phase modulation effects.

To study the propagation dynamics of two-color bound sates, one can consider a modified NLSE of the form[324]

$$\frac{\partial u}{\partial z} = \frac{i\beta_2}{2}\frac{\partial^2 u}{\partial t^2} + \frac{i\beta_4}{24}\frac{\partial^4 u}{\partial t^4} + \gamma |u|^2 u\,. \tag{80}$$

The linear part of Eq. (80) includes the higher-orders dispersion, with $\beta_2 > 0$ being positive GVD coefficient, and $\beta_4 < 0$ a negativeed fourth-order dispersion coefficient. Considering the discrete set of angular frequency detunings $\Omega$ and using the identity of the spectral derivative[325], the frequency-domain representation of the propagation constant is given by the polynomial expression

$$\beta(\Omega) = \frac{\beta_2}{2}\Omega^2 + \frac{\beta_4}{24}\Omega^4 . \tag{81}$$

The frequency-dependent inverse group-velocity of a mode at detuning Ω is

$$\beta_1(\Omega) \equiv \frac{\partial \beta(\Omega)}{\partial \Omega} = \beta_2 \Omega + \frac{\beta_4}{6}\Omega^3 \tag{82}$$

with group-velocity (GV) $v_g(\Omega) = 1/\beta_1(\Omega)$, and the GVD is

$$\beta_2(\Omega) \equiv \frac{\partial^2 \beta(\Omega)}{\partial \Omega^2} = \beta_2 + \frac{\beta_4}{2}\Omega^2 . \tag{83}$$

Subsequently, using parameter values $\beta_2 = 1$ fs²/μm, and $\beta_4 = -1$ fs⁴/μm, resulting in the model dispersion characteristics shown in Fig. 37 [324]. As seen in Fig. 37(c), the GVD profile given by Eq. (83) has a concave downward shape with two zero-dispersion points at $\Omega_{Z1}$ and $\Omega_{Z2}$. It exhibits anomalous dispersion at $\Omega < \Omega_{Z1}$ and $\Omega > \Omega_{Z2}$. Inspecting the inverse group velocity shown in Fig. 37(b), it is seen that two frequencies are GV matched to $\Omega = 0$. Due to the symmetry of the propagation constant, these are given by the pair $\Omega_1 = -\Omega_2$, uniquely characterized by $\beta(\Omega_1) = \beta(\Omega_2)$ and indicated by the open and filled circles in Fig. 37. In fact, for the considered propagation constant, GV matching of three distinct modes can be realized as long as the frequency loci in AD1 and AD2 lie within the range of frequencies shaded in red in Fig. 37(b). Note that the GV matching for two optical pulses at vastly different frequencies, supported by the propagation constant (81), is methodologically different from the GV matching that supports quasi-copropagation of different modes with similar frequencies. Nevertheless, both methods allow for quasi co-propagation of optical pulses under different circumstances, supporting similar XPM-induced propagation effects.

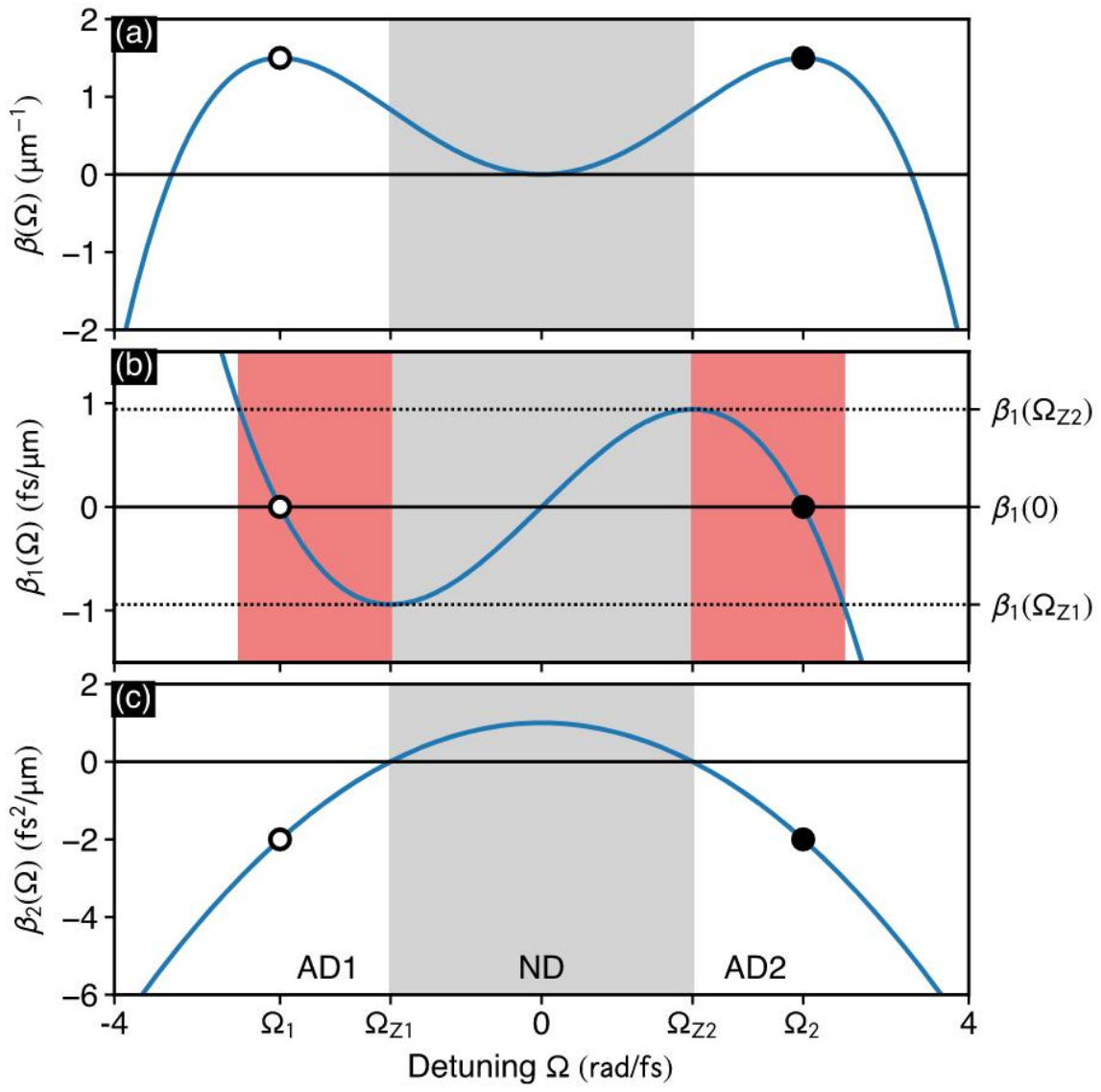

**Fig. 37** Details of the frequency-dependent propagation constant supporting two-color bound state, as per Ref. [324]. **(a)** The propagation constant. **(b)** The inverse group velocity. **(c)** The group-velocity dispersion.

**5.2 Generation of bound states in the frequency domain**

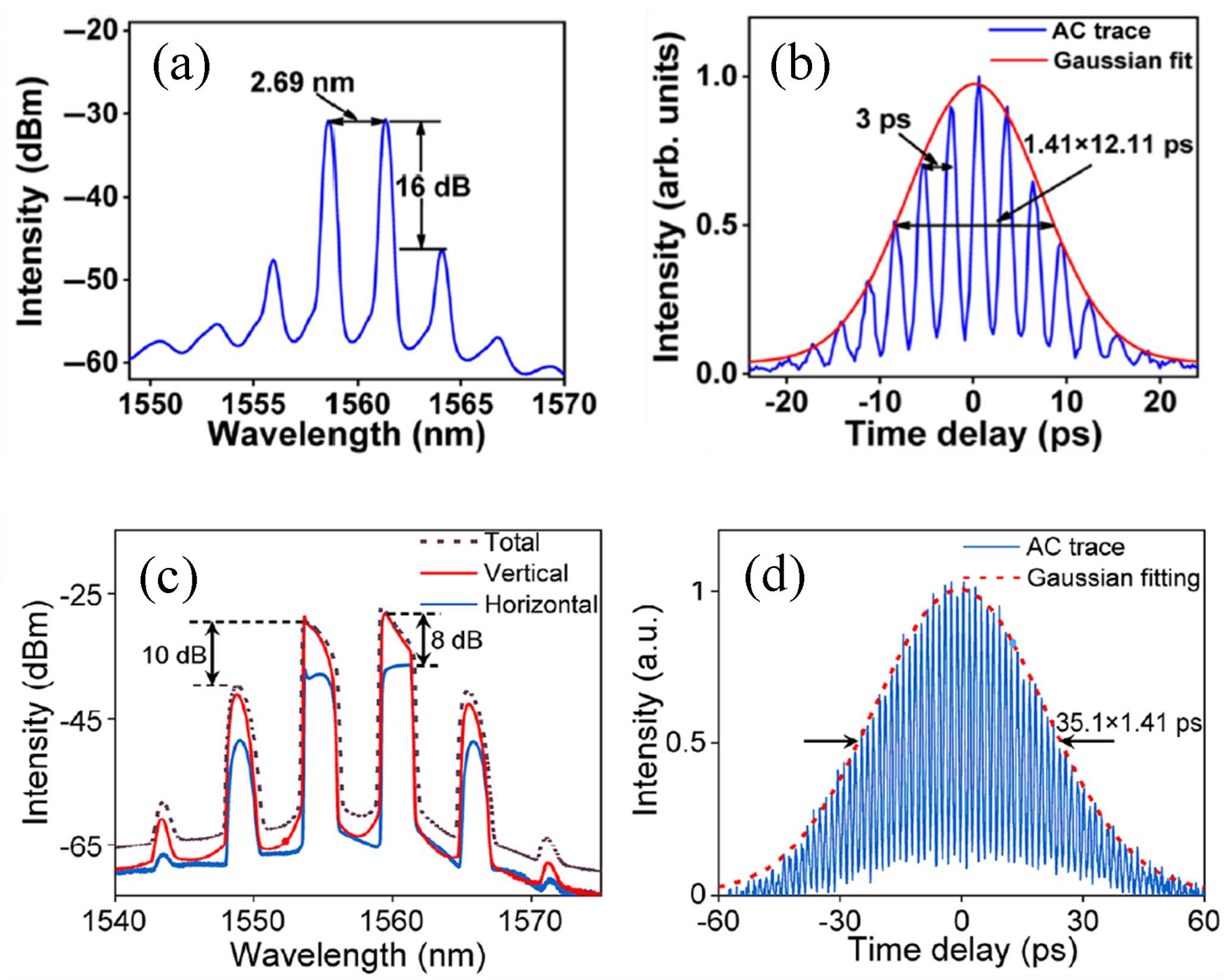

**Fig. 38** Dual-color bound-state solitons in the nearly-zero GVD regime with the mode-interference-induced filtering effect, as per Refs. [327]: **(a)** the optical spectrum and **(b)** the autocorrelation trace. Dual-color bound-state solitons in the slight-normal-GVD regime, as per Refs. [328]: **(c)** optical spectra of the bound-state soliton and its two orthogonally polarized components; **(d)** the corresponding autocorrelation trace.

In the nearly-zero-GVD regime, the group- delay differences between different spectra approach zero. Thus, bound-state solitons can be easily obtained by modulating the gain profile and introducing the spectral filtering into the cavity[326]. By leveraging the mode-interference-induced filtering effect, Mao et al. successfully achieved the generation of dual-color bound-state solitons in a nearly-zero-GVD fiber laser comprising single-mode and two-mode fibers[327]. As shown in Figs. 38(a) and 38(b), the spacing between the two main spectra is about 2.69 nm, resulting in a multi-peak-structured pattern in the time domain with a modulation period of 3 ps, which corresponds to the group-delay difference of two modes in the fiber. Further, two sub-spectrums occur due to the four-wave mixing effect. In the slight-normal-GVD regime, giant-chirp dual-color bound-state solitons can be produced by incorporating a section of a polarization-maintaining fiber into the fiber laser[328]. In this case, the filtering effect is enabled by the cascaded mode-coupling between two orthogonally polarized components and nonlinear effects. As shown in Figs. 38(c) and 38(d), the wavelength spacing between the neighboring spectra (5.54 nm) is governed by the mode coupling related to the birefringence of the polarization- maintaining fiber. Both the autocorrelation trace and retrieved pulse of the bound-state soliton display the stable periodic modulated structure, confirming the synchronization and interference of pulses at these discrete wavelengths.

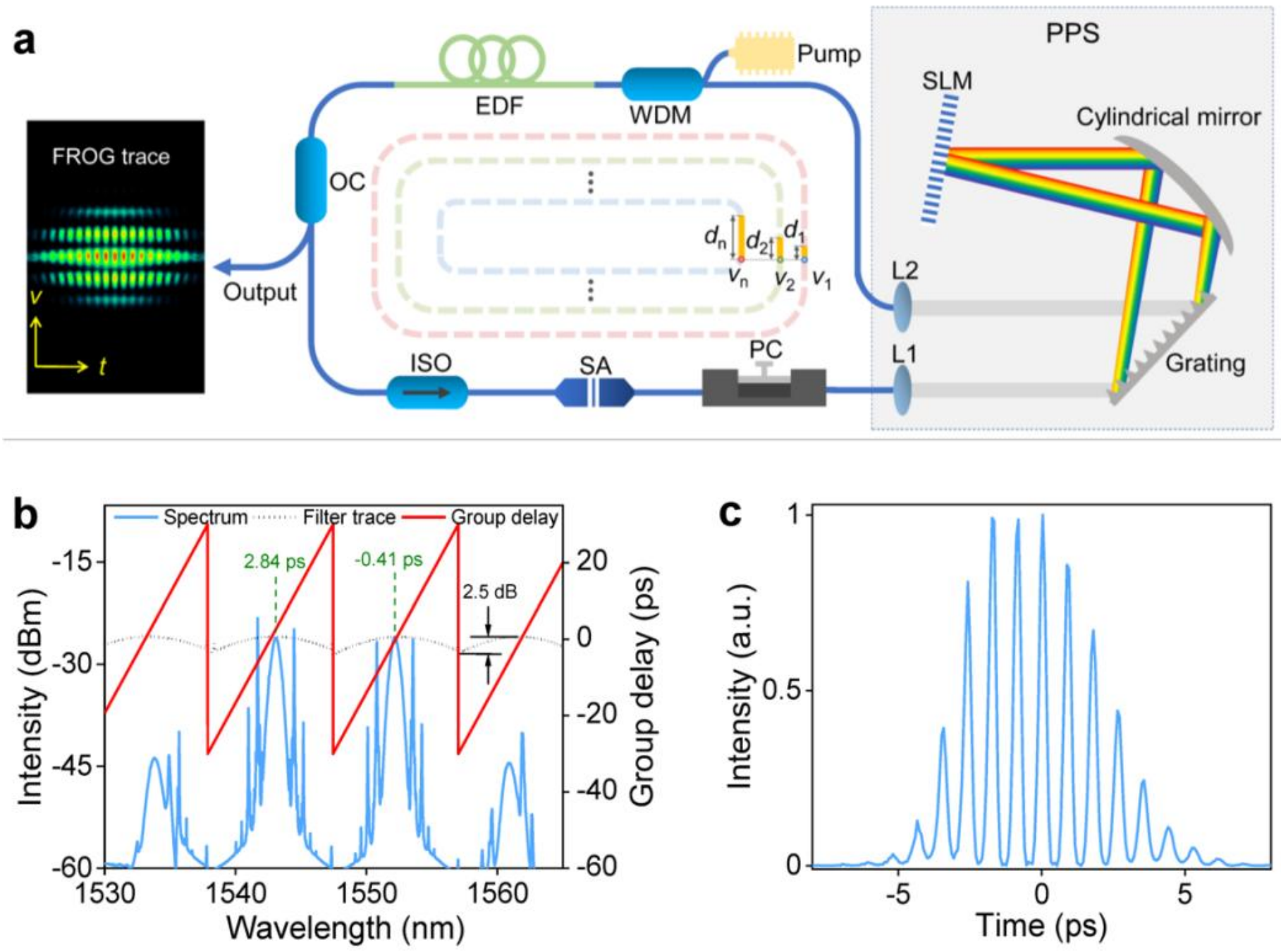

**Fig. 39** The bound-state soliton based on the group-delay compensation, as per Ref. [329]. (a) The experimental setup. (b) The spectrum, filter trace and group delay. (c) The pulse profile retrieved from the FROG spectrogram.

As discussed above, the formation of bound-state solitons in nearly-zero and slightly-normal-GVD regimes primarily depends on the filtering effect, while the pulse properties are determined by the cavity dispersion. However, the situation becomes much more complex in the non-zero-GVD regime as the dispersion-induced walk-off effect is pronounced, thus compensation of the group-delay difference is necessary to form bound-state solitons, using a pulse shaper [329–332]. It consists of a diffraction grating, a cylindrical mirror, and a liquid crystal on a silicon-based spatial light modulator, as shown in Fig. 39(a)[329]. By manipulating the voltage on each pixel of the spatial light modulator, both the phase and amplitude of diffracted spectral components can be manipulated. Subsequently, these components are directed back to the same grating by the cylindrical mirror and focused into the output port of the pulse shaper. Consequently, the group delay and amplitude can be adjusted for each spectral component. Such a fiber laser enables the direct generation of dual- to quintuple-color bound-state solitons with sub-pulses repetition rates ranging from 0.4 THz to 1.26 THz. Taking the dual-color bound-state solitons as an example, the spectra are centered at 1543.1 nm and 1552.2 nm with an intrinsic group-delay difference of 3.68 ps in the single circulation, as shown in Fig. 39(b). The bound-state solitons exhibit a modulated temporal profile, resulting in higher intensity when compared to single-wavelength pulses of equal energy (see Fig. 39c). Given that the constituent pulses in bound-state solitons are generated in the same dispersion regime, they exhibit similar properties, such as the spectral morphology and pulse chirp. Based on the Taylor-series expansion of the spectral phase, the parabolic convex-concave frequency phase corresponds to a linear group delay with opposite slopes and an anomalous-normal-GVD regime. Therefore, the opposite-GVD regimes can be achieved in the same laser resonator by introducing the convex-concave phase[330]. Besides, with the aid of the Kerr nonlinearity, the group-delay difference induced by the cavity dispersion is compensated by the modal dispersion of the multimode fiber, thereby facilitating the formation of bound-state solitons. Meanwhile, the birefringence-induced filtering effect and wide gain bandwidth endow the bound-state soliton with a large wavelength separation. Triple-color bound-state solitons can also be achieved by simply adjusting the orientation of the polarization controller. Since the frequency spacing between the adjacent spectra falls within the terahertz range, thus rendering such a pulse suitable as a pump source for generating terahertz waves based on the difference-frequency scheme[333,334].

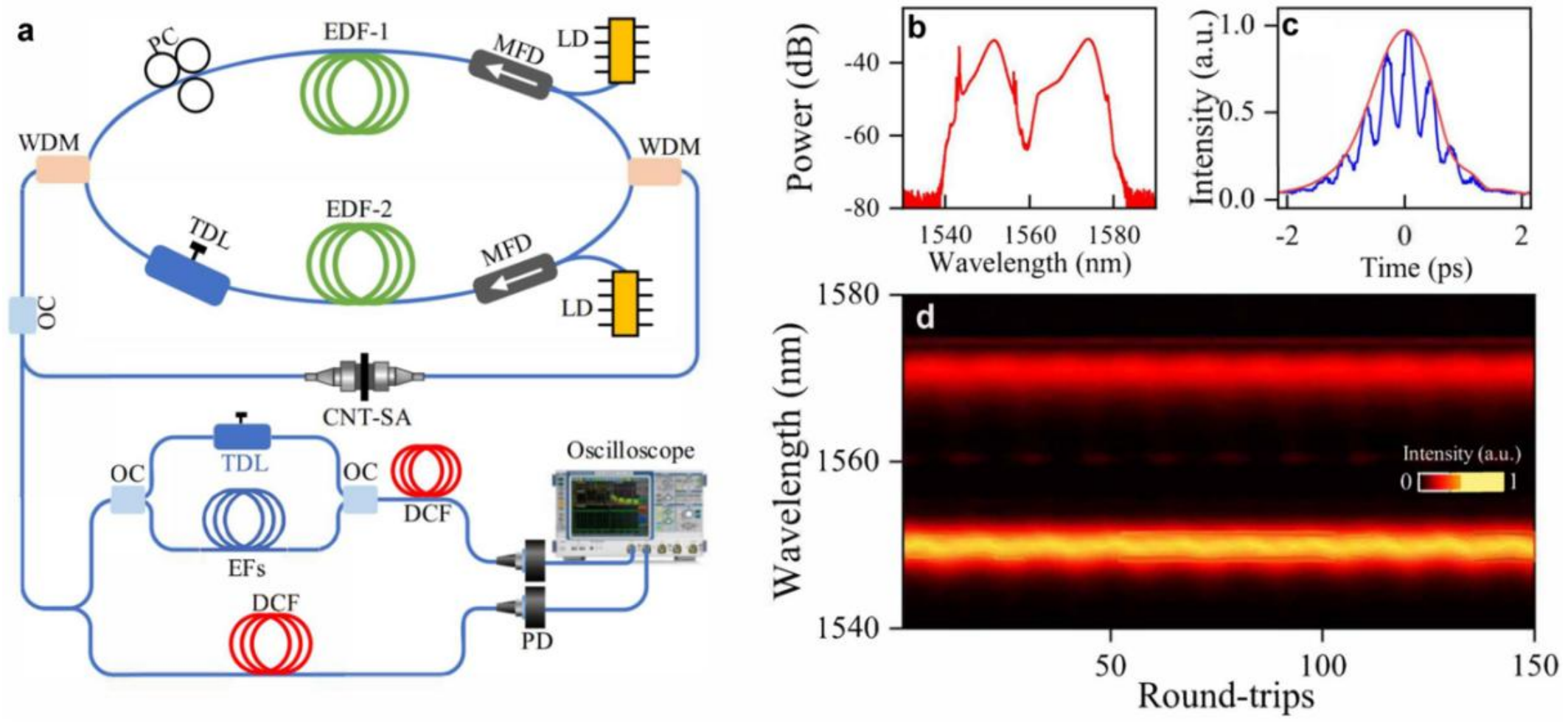

**Fig. 40** The bound-state soliton based on the XPM effect in the common saturable absorber, as per Ref. [338]. (a) The experimental setup. (b) The spectrum and (c) the autocorrelation trace. (d) The real-time spectral evolution.

Passive synchronization based on the XPM effect in a common saturable absorber is an alternative method that can effectively support the synchronization of two wavelengths[335–337]. As illustrated in Fig. 40(a), two wavelength-division multiplexers generate two distinct paths for optical signals carried by wavelengths of 1550 and 1570 nm[338]. Each path consists of a segment of erbium-doped fiber, a polarization controller, and a multifunctional device that integrates the pump with the signal while maintaining the unidirectional operation. A time-delay line adjusts the path difference for compensating the delay induced by GVD. Both paths share the same saturable absorber and output coupler. In this case, the pulse in each branch is a soliton with distinct spectra including the central wavelength and intensity (see Figs. 40b and 40c). The resultant bound-state soliton exhibits robust out-of-phase vibrations between two constituent pulses, with the vibration amplitude changing with the pump power (see Fig. 40d) [338]. Such evolution behavior is attributed to the Q-switching instability and represents a characteristic intrinsic mode of bound-state solitons. By adjusting the gain of each branch, it is possible to create a dichromatic soliton-molecule compound in the proposed fiber laser, which represents a hybrid state consisting of multiple bound-state solitons carried by two different wavelengths[339]. The dichromatic soliton-molecule compound is maintained by two different binding mechanisms, namely, the SPM interaction between solitons at the same wavelength, mediated by their tailing fields, and the cross-phase modulation interaction between solitons at different wavelengths, enabled by their direct overlapping. The available wavelength is contingent upon the specific branch of the wavelength-division multiplexer, which becomes significantly intricate in the case of multi-color bound-state solitons.

## 6. Bound states of vector solitons

### 6.1 Vector solitons

When the vectorial nature of light is used, the framework of vector solitons extends beyond their scalar counterparts. Governed by the coupled CGLSs, the vector soliton is essentially a multi-pulse complex with different polarization components [101]. The two orthogonally polarized components can trap each other and propagate as a non-dispersive unit in the laser cavity by oppositely shifting their central frequencies with respect to the net birefringence [340,341]. The vector dissipation solitons (VDSs) not only possess much more plentiful behaviors and richer dynamics than their scalar counterparts, but also pave a promising way for numerous applications, from nano-optics to high-capacity fiber optic communications. Since the first report on vector solitons in 1997 [342], varieties of these states with nontrivial polarization dynamics, such as polarization-locked vector solitons [110], polarization-rotating solitons [343], higher-order vector solitons [273], and dark-bright VDSs [344], have been observed in polarization-non-discriminating mode-locked fiber lasers. Besides, group-velocity-locked vector solitons (GVLVS) were demonstrated [184,188]. In Fig. 41(a), it is seen that the two orthogonally polarized pulses are carried by different wavelengths, with the horizontal and vertical components having bandwidths of 3.9 nm and 4.1 nm, respectively. In connection to the development of DM schemes and the goal of the creation of ultrashort pulses with high energy, the studies of the mode-locking operation gradually shifted into the normal-GVD regime, where the spectral filtering gets involved in the composite balance[345]. DSs formed in the normal-GVD regime are characterized by steep spectral edges and a large frequency chirp, as well as possessing much larger pulse energy and broader spectral bandwidth. Therefore, the studies of VDSs in the normal-GVD regime are twofold, motivated by the search for novel soliton dynamics and improved performance for applications.

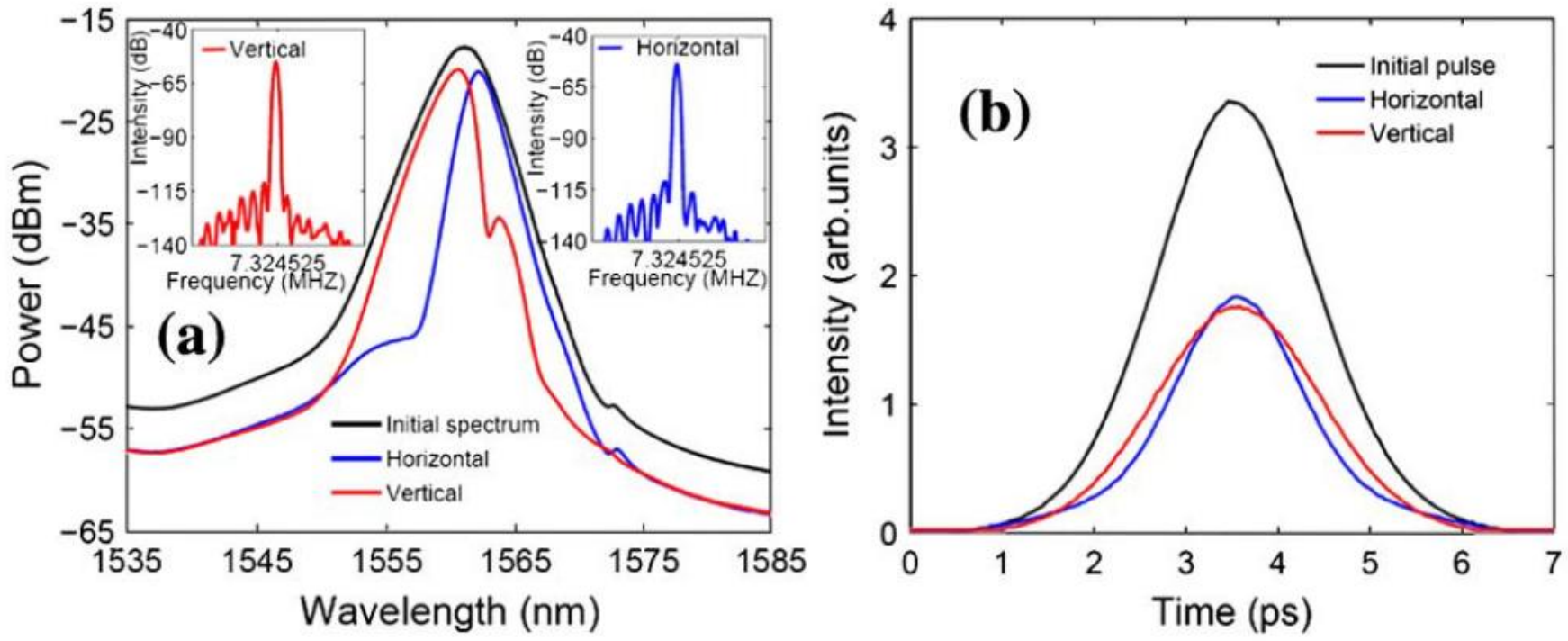

**Fig. 41** Characterization of group-velocity-locked vector solitons, as per Ref. [188]. **(a)** Optical spectra. The insets are radio-frequency spectra of the two components. **(b)** Corresponding autocorrelation traces.

Indeed, the vectorial nature of VDSs is universal with respect to different self-assembled forms of pulses. In particular, it suggests the consideration of vector-SM compounds, with their time-domain and polarization structure. Apart from the operation regime of regular pulses, passively mode-locked fiber lasers can also deliver the so-called noise-like pulse, which is, essentially, a pulse envelopes consisting of a bunch of randomly evolving femtosecond ultrashort pulses[346,347]. It is of great interest to investigate the dynamics of noise-like pulses beyond the scalar model.

More recently, TS-DFT techniques have been developed to trace the buildup of the mode-locking in ultrafast lasers. Benefiting from the polarization-resolved real-time spectrum, the formation mechanism of vector asymmetric solitons with the spectral period doubled under birefringent conditions can be recovered, showing that the XPM effect plays a major role for orthogonally polarized components[348,349]. In addition, the vector dynamics of incoherent short pulses can be elucidated through the utilization of a real-time, polarization-resolved spectro-temporal technique[350]. Furthermore, it has been demonstrated that a probabilistic distribution of the state of polarization during the dissipative soliton-buildup process can be retrieved, following an inverse- exponential pattern[351]. The utilization of the division-of-amplitude technique in the course of far-field transformation enables the analysis of the dynamics in the wavelength-dependent state of polarization[352].

### 6.2 Polarization-time bound states

The emergence of SMs and vectorial nature of the light suggests to consider VDS bound states formed in both time-domain and polarization directions[53-55]. It is found that VDSs can attract or repel each other, forming VDS bound states, providing a possibility for the two-fold increase of the optical-communications capacity. Zhao et al. reported the first experimental observation of a GVLVS bound states, as shown in Fig. 42[114]. The birefringence-enhanced fiber laser facilitates the generation of GVLVSs, where two orthogonally polarized components are coupled to form a multi-soliton complex. In addition to GVD and nonlinearity, the fiber's birefringence significantly affects properties of mode-locked pulses. For instance, low-birefringence single-mode fiber lasers enable the formation of group-velocity-locked[353] and polarization-locked VDSs[110,111]. Such vector soliton can be regarded as a unique bound state formed by the interaction of two orthogonally polarized components. Considering the interaction between multiple pulses, bound-state vector solitons can also be created in single-mode-fiber lasers. In the anomalous-GVD regime, Sergeyev et al. reported tightly and loosely bound vector solitons with locked of the polarization in a carbon nanotube mode-locked fiber laser[229]. Polarization-locked tightly-bound state VDSs were created when pump power is high, showing a fixed phase difference in this "molecule" and locked polarization. Adjustment of the polarization controller gives rise to the bistable operation between the loosely-bound state and a twin pulse. In the case of a large pump current, the bound-state vector solitons and oscillating polarization dynamics were realized, accompanied by the harmonic mode-locking operation, highlighting the complexity of the nonlinear laser system. L. M. Zhao et al. demonstrated bound states of VDSs in a mode-locked erbium-doped fiber laser with normal GVD[354]. Both the coherently and incoherently coupled vector dissipative solitons are group-velocity-locked without polarization evolution. Two types of bound-state VDSs have a fixed temporal separation that remains invariant with respect to different laser-operation conditions.

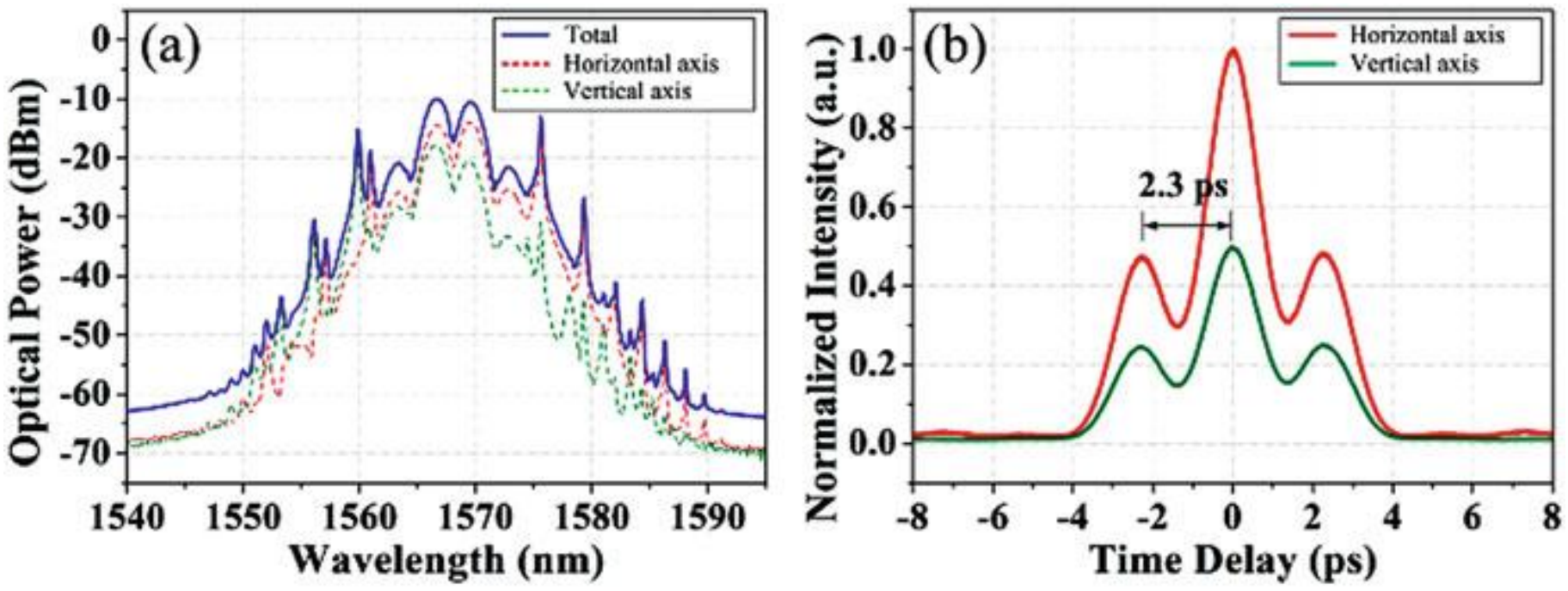

**Fig. 42** Polarization-resolved measurements of GVLVS bound states, as per Ref. [114]. (a) Polarization-resolved spectra of the GVLVS bound states; (b) Autocorrelation traces.

For hybrid-structure lasers comprising low-birefringent single-mode fiber and high-birefringent polarization-maintaining fibers, the polarization orientation and cavity effect should be considered. In the normal-GVD hybrid-structure fiber lasers, Mao et al. demonstrated near-chirp-free solitons with distinct spectral sidebands. The two orthogonally polarized components propagate in an asymmetric "X" configuration in the polarization-maintaining fiber and realize the self-consistent evolution due to the coupling between the two fibers[355]. They also demonstrated oscillating bound-state solitons in the birefringence-managed fiber laser with nearly-zero GVD. Due to the unequal coupling between the orthogonally polarized vector modes, the constituent pulses of bound-state solitons possess different intensities and result in a dual-roundtrip and large-period oscillations. Numerical simulations reproduce the experimentally observed phenomena and show that two vector modes display diverse behavior in the polarization maintaining fiber[356]. In the birefringence-managed fiber laser with normal GVD, asymmetric bound-state vector solitons, composed of two orthogonally-polarized components with mirrored profiles, can be formed through the mode coupling. Due to the dynamical balance between the fiber birefringence and GVD, the temporal dislocation between the two components changes along the cavity while reaching a self-consistent state per round trip. The asymmetric nature of sub-pulses for each component is primarily influenced by the unequal coupling between orthogonally polarized modes at the input interface of the polarization-maintaining fiber. The observation of these bound-state solitons at 1.03 μm and 1.55 μm wavelengths demonstrates the generality of the vector asymmetric structures[357].

## 7 Bound states of spatiotemporal solitons

### 7.1 Spatiotemporal solitons

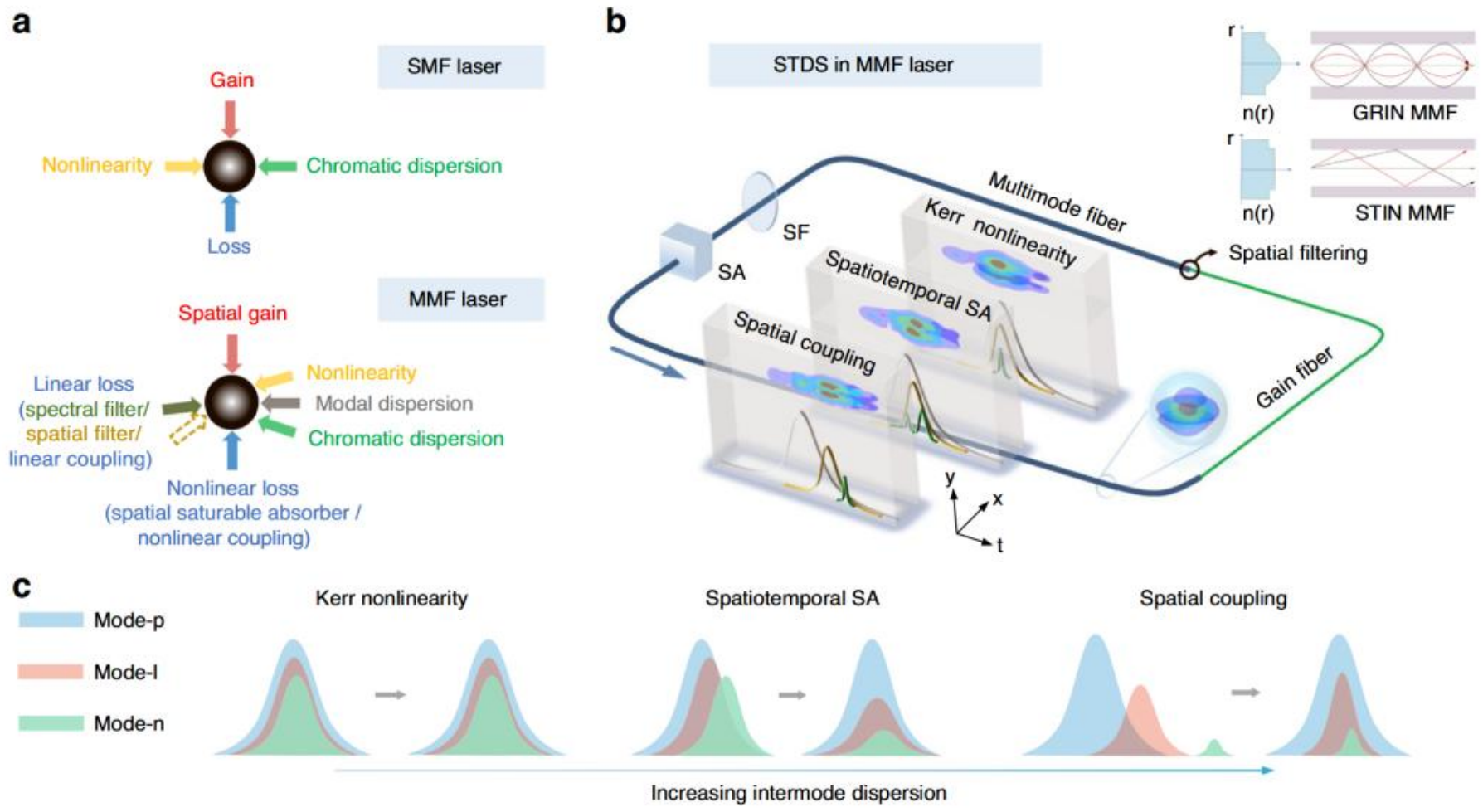

**Fig. 43** Concept of STML and STDSs in multimode fiber (MMF) lasers, as per Ref. [359]. **(a)** Schematic diagram of dissipative soliton formation in SMF lasers and MMF lasers. **(b)** An illustration of a typical STML MMF laser layout. **(c)** Schematic diagrams of pulse reactions for three mechanisms as intermodal dispersion increases.

For temporal solitons, the mode-locking in time relies on the nonlinear effects and dispersion management within the cavity to achieve pulse compression and stabilization, as shown in Fig. 43(a). Recently, the spatiotemporal mode-locking (STML) operations in multimode fiber lasers have been demonstrated through the counteraction of modal dispersion, along with strong spatial- and spectral-filtering effects, generating optical fields with durations ranging from picoseconds to femtoseconds and composed of multiple transverse modes[98]. The generated pulse manifests as spatiotemporal dissipative solitons (STDS) [358]. In this context, dispersion encompasses not only chromatic dispersion (intramodal dispersion) but also intermodal dispersion. Intermodal dispersion refers to the phenomenon where different transverse modes propagate at different group velocities in multimode fibers, causing the transverse modes to gradually separate and diverge within the fiber. How to balance dispersion in multimode fiber lasers, including intramodal dispersion and intermodal dispersion, especially intermodal dispersion, is an important issue for the realization of STDS[359], as shown in Fig. 43(a).

Figure 43(c) illustrates the different mechanisms, such as Kerr nonlinearity, spatiotemporal saturable absorbers, and spatial coupling/filters, and their varying speeds in compensating for intermodal dispersion. This information is crucial for understanding the nonlinear dynamics within spatiotemporal mode-locked lasers and optimizing the design of the lasers. At low intermodal dispersion, pulses of different spatial modes can be combined together through Kerr nonlinearity in graded-index (GRIN) MMFs[360]. When the modal walk-off is large in step-index (STIN) MMF lasers, the Kerr effect cannot bind the modal pulses together and they can walk off strongly. Under moderate intermodal dispersion, the pulses will slightly shift, and the spatiotemporal SA can reset the shift by saturable absorbing the weaker modes and redistributing the energy to other modes. At high intermodal dispersion, the pulses will significantly deviate, and spatial coupling can weaken the pulses of the weaker modes, redistributing the energy from the parent modes to the child modes, and aligning the timing of the mode pulses[361]. Kerr nonlinearity, spatiotemporal saturable absorption, and spatial coupling or filtering effects play key roles in mode locking, working together to balance intermodal dispersion and achieve stable mode locking. In 2017, Wright and colleagues used GRIN multimode fibers with low intermodal dispersion for their experiments[98]. By coupling light into different core diameter fibers and utilizing spectral filtering, they achieved spatial and spectral filtering, respectively, and thus balanced the dispersion (including normal intramodal and intermodal dispersion) with other intracavity effects (such as Kerr nonlinearity, spatial filtering, spectral filtering, and saturable absorbers). In 2022, Wu and colleagues conducted the first detailed study on the problem of laser spatiotemporal mode-locking with anomalous dispersion[362]. In their research, they balanced the dispersion through the intracavity Kerr nonlinear effects, spatial filtering, and saturable absorbers, ultimately forming stable STDS. They altered the misalignment of the fiber fusion splicing, resulting in output beams with different modal distributions in the beam profile and optical spectrum. By adjusting the coupling between the wave plate and the fiber, they observed various multimode mode-locking states but found it challenging to observe higher-order multimode states. These results provide an in-depth understanding of the pulse formation and evolution process within the laser cavity, especially under multimode conditions, how the pulse distributes energy among different modes, and their temporal and spectral characteristics.

### 7.2 Spatiotemporal bound states

Actually, the multiple transverse modes in the multimode fiber lasers operate in the bound state supported by the nonlinear interaction[357,363,364]. Manipulating the waveplates in the cavity, Xiao et al. not only achieved stable single-pulse spatiotemporal mode-locking, but also observed bound-state solitons in the same fiber laser [364]. They have produced various bound-state solitons, including soliton pairs, triplets, and quartets with pulse separations spanning from several picoseconds to tens of picoseconds. For certain combinations of the pump power and waveplate settings, a single soliton is emitted prior to the bound-state soliton, as the pump power increases. However, in other settings, the bound-state soliton appears before the single soliton, which is different from the case in single-mode fiber lasers, where the single pulse typically emerges first. Such unique phenomena result from the more complex interaction of light in the multimode fiber laser. The multi-soliton complexes discussed above were investigated in fiber lasers running on a single transverse (spatial) mode. There is great interest to nonlinear spatiotemporal dynamics in ultrafast lasers. The spatiotemporal mode locking (or mode locking in the multimode setting), in which both the longitudinal and transverse modes are locked simultaneously, was recently demonstrated in lasers composed of multimode fibers [98]. Using graded-index multimode fibers with small modal dispersion, the walk off of different transverse modes can be balanced by an intracavity spatial filter. The mechanism of the spatiotemporal mode-locking was analyzed, and several distinct forms of spatiotemporal dynamics, which have no analogues in single-mode lasers, were predicted[360]. In a hybrid cavity with both single-mode and multimode components, the formation of spatiotemporal SM involves the interplay between sections of single-mode and multimode fibers. As illustrated in Fig. 44, Guo et al. have constructed a novel all-fiber hybrid ring-cavity structure, which operates in the spatiotemporal SM mode, integrating single-mode and multimode fiber segments[365]. In the hybrid cavity, longitudinal and transverse mode locking is achieved by adjusting the polarization controller and multimode fiber segment, respectively. Adjusting the polarization controller, loosely and tightly bound soliton pairs were generated at the carrier wavelength 1560 nm. Replacing the quasi-single-mode active fiber with a multimode step-index active fiber, it was recently found that spatiotemporal mode-locking can also be supported in these lasers [361], making it possible to observe multi-soliton states. Using single-shot multispeckle spectral-temporal measurement technology, which leverages optical time-division multiplexing and TS-DFT, Yang et al. enabled the simultaneous observation of multiple speckle grains, providing insights into the long-term evolutionary dynamics of STDSs [366]. Under the spatiotemporal coupling effect, spatiotemporal SMs can be formed. The researchers further demonstrated the speckle-resolved spectral-temporal dynamics of spatiotemporal SMs in real time [367]. For the first time, they captured diverse real-time dynamics of spatiotemporal SMs, including their birth, spatiotemporal interactions, and internal vibrations. The investigation of nonlinear spatiotemporal dynamics in multimode fiber lasers is still at its early age, and we expect that new manifestations of nonlinear multisoliton dynamics will be revealed in multimode settings.

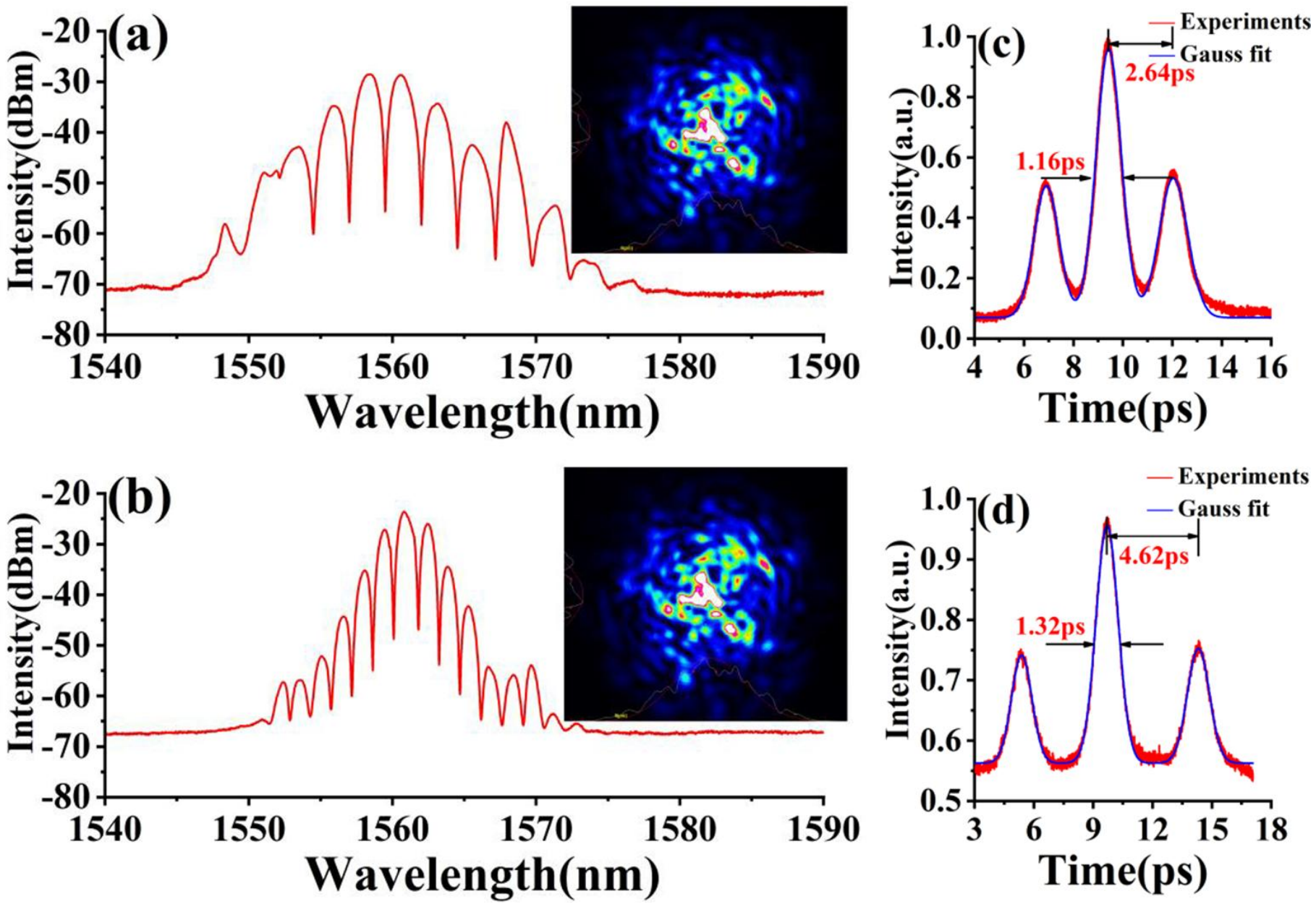

**Fig. 44** The output characteristics of the tightly bound spatiotemporal SMs, as per Ref. [365]. **(a)-(b)** The optical spectra. Inset: the output beam profile. **(c)-(d)** The corresponding autocorrelation trace.

## 8 Conclusions

This article presents a systematic review of results for diverse types of bound states of dissipative solitons (also called "soliton molecules") that were studied experimentally in fiber lasers and theoretically in the underlying models, which are based on complex Ginzburg-Landau equations and coupled systems of such equations. Both the experimental and theoretical findings are summarized here, including very recent ones.

The study of SMs has evolved from early theoretical predictions to experimental observation, and now, to deterministic control[368]. Initially, theoretical models suggested that stable bound states could exist, a prediction later confirmed in passively mode-locked fiber lasers. A turning point occurred with the adoption of TS-DFT techniques, which allowed researchers to visualize internal molecular motion and formation dynamics in real-time[369]. This capability clarified a critical distinction: simply observing SM phenomena is not the same as utilizing them as reliable, functional tools.

Recent work has thus shifted to actively engineering the internal parameters of the interacting soliton pairs, $\Delta\tau$ and $\Delta\varphi$. This control mechanism entails modulating physical quantities—such as the gain, loss, refractive index, and acoustic fields— with the aim to reshape the effective interaction potential. Deepening the potential wells enhances stability ("stabilization"), while shifting their positions allows for deterministic state selection ("programming"). Thus, key physical control modalities include: the pump/gain control, where direct modulation steers the gain dynamics to switch bound states [179,180], the frequency-domain control, employing the intracavity spectral filtering to reshape the dissipation structure for the "on-demand" selection of the separation between the bound solitons [309,370], the time-domain control, utilizing long-range optoacoustic or optomechanical interactions to create addressable temporal potential wells [274], and the spatial-domain control, which introduces spatial degrees of freedom in multimode lasers to support spatiotemporal bound states [361,363].

However, implementing these control mechanisms reliably requires addressing the inherent fragility of bound states. Unlike single solitons, SMs are characterized by shallow binding potentials, rendering them highly susceptible to environmental perturbations [73,371,372]. Thermal drifts reshape the interaction potential, while mechanical vibrations in non-polarization-maintaining fibers induce birefringence fluctuations, leading to phase slippage and dissociation [373,374]. Crucially, beyond the additive noise that creates timing jitter, multiplicative noise, stemming from pump power fluctuations, plays a distinct and often more destructive role[375]. Theoretical and experimental evidence suggests that the multiplicative noise, which scales with the field intensity, can modulate the binding strength and drive the system across bifurcation points, triggering spontaneous collapse [376]. Furthermore, the practical deployment of these states is hindered by the "reproducibility crisis" associated with multidimensional hysteresis and the coexistence of multiple attractors in dissipative cavities; consequently, restoring the same control parameters does not guarantee deterministic recovery of a specific molecular state [268,377].

To resolve this issue, architectural stabilization is the first line of defense. Switching to all-polarization-maintaining fiber is essential. This hardware foundation eliminates polarization drift and isolates the system from mechanical vibrations, creating a stable platform for the physical controls mentioned above [374,378]. Complementing this hardware is algorithmic control. Researchers now use genetic algorithms and machine learning to navigate the complex, multistable landscape of these lasers. These data-driven approaches automate optimization and enable "self-healing," effectively bridging the gap between the theoretical controllability and actual experimental reproducibility[312,379,380].

These programmable capabilities directly enable applications that use the internal degrees of freedom of soliton molecules. By controlling the relative phase, multilevel optical encoding has been demonstrated, such as a quaternary format[179]. Deep optomechanical potentials support all-optical bit storage with long retention times [381], while "soliton reactor" architectures extend this to analog simulation of many-body kinetics [307]. Furthermore, the dynamical control of intracavity soliton motion has been harnessed to realize programmable all-optical delay generators, enabling high-speed, scanning-free pump-probe spectroscopy with tunable nano- to picosecond delays [382]. On the hardware side, the observation of soliton molecules in microresonators offers a promising path toward integrated, chip-scale devices[383]. To scale up these systems, three main constraints remain: advancing switching speeds from thermal/acoustic time scales to GHz rates; developing compact, efficient decoding methods for $\Delta\tau$ and $\Delta\varphi$ (potentially, by dint of AI-assisted inference); and improving the integration and packaging. Looking ahead, distinct future directions emerge, including the use of SMs for precision sensing, exploiting their nonlinearity for neuromorphic computing [384], extending the control to mid-IR spectroscopy for molecular sensing [385], and shaping quantum frequency combs for high-dimensional quantum information processing [386,387].

In summary, the research landscape of soliton bound states is undergoing a fundamental shift from the observation of physical phenomena to the practical engineering of reliable tools. Future progress depends on transforming the current understanding of the complex dissipative dynamics in fiber lasers into precise capabilities. The objective is to achieve deterministic control, enabling the on-demand formation of bound states, with a specific focus on locking and manipulating the phase shift between pulses, which is the key parameter for many applications. Simultaneously, achieving reliability requires a rigorous defense against instability. It is essential to deepen the understanding of how laser-noise mechanisms and internal soliton interactions contribute to the timing jitter and phase drift. To suppress these fluctuations, advanced instrumentation, capable of real-time measurement (such as TS-DFT), should be combined with active-feedback strategies for implementing the dynamical error correction. Furthermore, the physical platform itself is evolving too. While fiber lasers have been serving as the primary testbed, research is rapidly expanding toward compact, chip-scale platforms, such as Kerr microresonators, semiconductor lasers, and integrated circuits. These platforms promise superior stability and the potential for mass integration, overcoming the environmental sensitivity of fiber systems. Driven by these advances, soliton molecules are expected to evolve from fundamental curiosities into versatile information carriers, driving next-generation stages in the high-dimensional coding, optical storage, and intelligent photonics.

## Declaration of Competing Interest

The authors declare that they have no known competing financial interests or personal relationships that could have appeared to influence the work reported in this paper.

## Acknowledgments

We thank Profs. Marcel Clerc and Mustapha Tlidi for the invitation to present a review article for the possible publication in Physics Reports. This work was supported in part by the National Key R&D Program of China (No. 2023YFF0715802), National Natural Science Foundation of China (No. 62575261, 62305299), and Zhejiang Provincial Natural Science Foundation of China (Nos. LZ24F050002, Z24F050002 and LQ23F050004). The work of BAM was supported, in part, by the Israel Science Foundation (grant No. 1695/22).

## List of Abbreviations

Throughout the whole paper, each abbreviation is defined the first time it appears. As further help, we list all the abbreviations here, in alphabetical order.

| | |
|---|---|
| AC | alternating-current |
| CGLE | Complex Ginzburg-Landau Equation |
| CNT | Carbon Nanotubes |
| CW | Continuous Wave |
| DM | Dispersion-Managed |
| EPCs | Electric Polarization Controllers |
| FWHM | Full Width at Half Maximum |
| GRIN | Graded-Index |
| GVD | Group-Velocity Dispersion |
| HML | Harmonic-Mode Locking |
| MMF | Multimode Fiber |
| NALM | Nonlinear Amplification Loop Mirrors |
| NLSE | Nonlinear Schrödinger Equation |
| NMI | Noise-Mediated Interaction |
| NPE | Nonlinear Polarization Evolution |
| OC | Optical Coupler |
| RO | Relaxation Oscillation |
| SESAM | Semiconductor Saturable Absorbers |
| SMC | Soliton Molecular Complexes |
| SMF | Single-Mode Fiber |
| SMs | Soliton Molecules |
| SPM | Self-Phase-Modulation |
| STDS | Spatiotemporal Dissipative Solitons |
| STIN | Step-Index |
| STML | Spatiotemporal Mode-Locking |
| TS-DFT | Time-Stretch Dispersive Fourier Transform |
| VDSs | Vector Dissipative Solitons |
| XPM | Cross-Phase Modulation |